\documentclass[twoside,twocolumn]{article}

\usepackage{xspace}
\usepackage{amsmath}
\usepackage{verbatim}
\usepackage{subfigure}
\usepackage{color}
\definecolor{pink}{RGB}{219, 48, 122}
\usepackage[normalem]{ulem}
\usepackage{soul}
\usepackage{paralist}
\usepackage{booktabs}
\usepackage{multirow}
\usepackage[capitalize]{cleveref}
\usepackage{graphics}
\usepackage{graphicx}

\newcommand\pwtcomment[1]{}
\newcommand\nccomment[1]{}
\newcommand\sjt[1]{}
\newcommand\sjtcomment[1]{}
\newcommand\kscomment[1]{}
\newcommand\npcomment[1]{}
\newcommand\kmcomment[1]{}
\newcommand\ks[1]{}
\newcommand\pout[1]{}

\newcommand{\bench}[1]{{\sffamily \mbox{#1}}}
\def\etsbench{\mbox{\bench{ets\_bench}}\xspace}
\def\bencherl{\mbox{BenchErl}\xspace}

\newcommand\wombat{\mbox{WombatOAM}\xspace}

\newcommand{\svgpath}{figures}
\newcommand{\executeiffilenewer}[3]{%
	\ifnum\pdfstrcmp{\pdffilemoddate{#1}}%
	{\pdffilemoddate{#2}}>0%
	{\immediate\write18{#3}}\fi%
}
\def\svgwidth{.49\textwidth}

\graphicspath{{\svgpath/}}

\usepackage{blindtext} 

\usepackage[sc]{mathpazo} 
\usepackage[T1]{fontenc} 
\usepackage{microtype} 

\usepackage[english]{babel} 

\usepackage[hmarginratio=1:1,top=32mm,columnsep=20pt]{geometry} 
\usepackage[hang, small,labelfont=bf,up,textfont=it,up]{caption} 
\usepackage{booktabs} 

\usepackage{lettrine} 

\usepackage{abstract} 

\usepackage{titlesec} 
\renewcommand\thesection{\Roman{section}} 
\renewcommand\thesubsection{\roman{subsection}} 
\titleformat{\section}[block]{\large\scshape\centering}{\thesection.}{1em}{} 
\titleformat{\subsection}[block]{\large}{\thesubsection.}{1em}{} 

\usepackage{fancyhdr} 
\usepackage{titling} 

\usepackage{hyperref} 

\pretitle{\begin{center}\Huge\bfseries} 
\posttitle{\end{center}} 
\title{Scaling Reliably: Improving the Scalability of the Erlang Distributed Actor Platform} 
\author{%
\textsc{Phil Trinder,
Natalia Chechina,
Nikolaos Papaspyrou,}
\and
\textsc{Konstantinos Sagonas,
Simon Thompson,
Stephen Adams,
Stavros Aronis,}
\and
\textsc{Robert Baker,
Eva Bihari,
Olivier Boudeville,
Francesco Cesarini,}
\and
\textsc{Maurizio Di Stefano,
Sverker Eriksson,
Viktoria Fordos,
Amir Ghaffari,}
\and
\textsc{Aggelos Giantsios,
Rockard Green,
Csaba Hoch,
David Klaftenegger,}
\and
\textsc{Huiqing Li,
Kenneth Lundin,
Kenneth MacKenzie,
Katerina Roukounaki,}
\and
\textsc{Yiannis Tsiouris,
Kjell Winblad} \\[1ex] 
\normalsize RELEASE EU FP7 STREP (287510) project \\ 
\normalsize \href{www.release-project.eu/}{www.release-project.eu/} 
}
\date{April 25, 2017} 
\renewcommand{\maketitlehookd}{%

\begin{abstract}
\noindent Distributed actor languages are an effective means of constructing
  scalable reliable systems,  and the Erlang
  programming language has a well-established and influential model. 
  While the Erlang model conceptually provides reliable scalability, it
  has some inherent scalability limits and these force developers to
  depart from the model at scale. This article establishes the
  scalability limits of Erlang systems, and reports the work of the EU RELEASE
  project to improve the scalability and understandability of the
  Erlang reliable distributed actor model.  

  We systematically study the scalability limits of Erlang, and then
  address the issues at the virtual machine, language and tool levels.
  More specifically: (1) We have evolved the Erlang virtual machine so that it
  can work effectively in large scale single-host multicore and NUMA
  architectures. We have made important changes and architectural
  improvements to the widely used Erlang/OTP release.  (2) We have
  designed and implemented Scalable Distributed (SD) Erlang libraries
  to address language-level scalability issues, and
  provided and validated a set of semantics for the new language
  constructs. 
  (3) To make large Erlang systems easier to deploy, monitor, and
  debug we have developed and made open source releases of five
  complementary tools, some specific to SD Erlang.

  Throughout the article we use two case studies to investigate the
  capabilities of our new technologies and tools: a distributed hash
  table based Orbit calculation and Ant Colony Optimisation (ACO).
  Chaos Monkey experiments show that two versions of ACO survive random process failure
  and hence that SD Erlang preserves the Erlang reliability
  model. While we report measurements on a range of NUMA and cluster
  architectures, the key scalability experiments are conducted on the
  Athos cluster with 256 hosts (6144 cores). Even for programs with no
  global recovery data to maintain, SD Erlang partitions the network
  to reduce network traffic and hence improves performance of the
  Orbit and ACO benchmarks above 80 hosts.  ACO measurements show that
  maintaining global recovery data dramatically limits scalability;
  however scalability is recovered by partitioning the recovery data. We exceed the established scalability limits of distributed Erlang, and do not reach the limits of SD Erlang for these benchmarks at this scale (256 hosts, 6144 cores).
\end{abstract}
}

\begin{document}

\maketitle


\section{Introduction}
\label{sec:introduction}
Distributed programming languages and frameworks are central to engineering large
scale systems, where key properties include  scalability and reliability. By scalability we mean that performance increases as hosts and cores are added, and by large scale we mean
architectures with hundreds of hosts and tens of thousands of
cores. Experience with high performance and data centre computing shows
that reliability is critical at these scales, e.g. host failures alone
account for around one failure per hour on commodity servers with
approximately 10$^{5}$ cores~\cite{BarrosoCH13-datacenter}. 
To be usable, programming languages employed on them must be supported by a suite of deployment, monitoring, refactoring and testing tools that work at scale. 

Controlling shared state is the only way to build reliable scalable
systems. State shared by multiple units of computation limits
scalability due to high synchronisation and communication
costs. Moreover shared state is a threat for reliability as failures
corrupting or permanently locking shared state may poison the entire
system.

Actor languages avoid shared state: actors or
processes have entirely local state, and only interact with each other by sending
messages~\cite{agha-86-overview}. Recovery is facilitated in this model, since actors, like
operating system processes, can fail independently without affecting
the state of other actors. Moreover an actor can \emph{supervise}
other actors, detecting failures and taking remedial action,
e.g. restarting the failed actor.

Erlang~\cite{armstrong-07-programming,cesarini-09-erlang} is a beacon
language for reliable scalable computing with a widely emulated
distributed actor model.  It has influenced the design of numerous
programming languages like Clojure~\cite{hickey-08-clojure} and
F\#~\cite{syme-15-expert}, and many languages have Erlang-inspired
actor frameworks, e.g.\ Kilim for Java~\cite{srinivasan-08-kilim},
Cloud Haskell~\cite{epstein-11-towards}, and Akka for C\#, F\# and
Scala~\cite{website-12-scala}. Erlang is widely used for building
reliable scalable servers, e.g. Ericsson's AXD301 telephone exchange
(switch)~\cite{ErlangAXD301}, the Facebook chat server, and the
Whatsapp instant messaging server~\cite{whatsapp-15-website}.

In Erlang, the actors are termed processes and are managed by a
sophisticated Virtual Machine on a single multicore or NUMA host,
while \emph{distributed Erlang} provides relatively transparent
distribution over networks of VMs on multiple hosts. Erlang is
supported by the Open Telecom Platform (OTP) libraries that capture
common patterns of reliable distributed computation, such as the
client-server pattern and process supervision. Any large-scale system
needs scalable persistent storage and, following the CAP
theorem~\cite{Gilbert2002}, Erlang uses and indeed implements
Dynamo-style NoSQL DBMS like Riak~\cite{Klophaus} and
Cassandra~\cite{Lakshman:2010}.

While the Erlang distributed actor model conceptually provides
reliable scalability, it has some inherent scalability limits, and
indeed large-scale distributed Erlang systems must depart from the
distributed Erlang paradigm in order to scale, e.g. not maintaining a
fully connected graph of hosts. The EU FP7 RELEASE project set out to
establish and address the scalability limits of the Erlang reliable
distributed actor model~\cite{RELEASEWeb}.

After outlining related work (Section~\ref{sec:context}) and the
benchmarks used throughout the article (Section~\ref{sec:benchmarks}) we
investigate the scalability limits of Erlang/OTP, seeking to identify
specific issues at the virtual machine, language and persistent storage levels
(Section~\ref{sec:erlang-scalability}).

We then report the RELEASE project work to address these
issues, working at the following three levels.
\begin{enumerate}
\item We have designed and implemented a set of Scalable Distributed
  (SD) Erlang libraries to address \emph{language-level} reliability
  and scalability issues.  An operational semantics is provided for
  the key new s\_group construct, and the implementation is validated
  against the semantics (Section~\ref{sec:language-scalability}).

\item We have \emph{evolved the Erlang virtual machine} so that it can
  work effectively in large-scale single-host multicore and NUMA
  architectures. We have improved the shared ETS tables, time
  management, and load balancing between schedulers. Most of these
  improvements are now included in the Erlang/OTP release, currently 
  downloaded  approximately 50K times each month (Section~\ref{sec:improving-vm-scalability}).

\item To facilitate the development of scalable Erlang systems, and to
  make them maintainable, we have developed three new tools: Devo, SDMon and \wombat, and enhanced two others: the visualisation tool Percept, and the refactorer Wrangler. 
  The tools support refactoring programs to make them more scalable,
  easier to deploy at large scale (hundreds of hosts), easier to monitor
  and visualise their behaviour.
  Most of these tools are freely available under open source licences;
  the \wombat deployment and monitoring tool is a commercial product
  (Section~\ref{sec:scalable-tools}).
\end{enumerate}

Throughout the article we use two benchmarks to investigate the
capabilities of our new technologies and tools. These are a
computation in symbolic algebra, more specifically an algebraic
`orbit' calculation that exploits a non-replicated distributed hash
table, and an Ant Colony Optimisation (ACO) parallel search program
(Section~\ref{sec:benchmarks}).

We report on the reliability and scalability implications of our new
technologies using Orbit, ACO, and other benchmarks. We use a Chaos
Monkey instance~\cite{bennett-12-chaos} that randomly kills processes
in the running system to demonstrate the reliability of the benchmarks
and to show that SD Erlang preserves the Erlang language-level reliability
model. While we report measurements on a range of NUMA and cluster
architectures as specified in Appendix~A, the key scalability
experiments are conducted on the Athos cluster with 256 hosts and 6144
cores. Having established scientifically the folklore limitations of
around 60 connected hosts/nodes for distributed Erlang systems in
Section 4, a key result is to show that the SD Erlang benchmarks exceed
this limit and do not reach their limits on the Athos cluster
(Section~\ref{sec:case-studies}).

\textbf{Contributions.} This article is the first systematic presentation
of the coherent set of technologies for engineering scalable reliable
Erlang systems developed in the RELEASE project.

Section~\ref{sec:erlang-scalability} presents the first scalability
study covering Erlang VM, language, and storage scalability. Indeed we
believe it is the first comprehensive study of any distributed actor
language at this scale (100s of hosts, and around 10K
cores). Individual scalability studies, e.g.~into Erlang VM
scaling~\cite{bencherl-12}, or language and storage scaling have
appeared before~\cite{ghaffari-13-scalable,ghaffari-14-investigating}.

At the language level the design, implementation and validation of the
new libraries (Section~\ref{sec:language-scalability}) have been
reported piecemeal~\cite{improving-jpdc-16,mackenzie-15-performance},
and are included here for completeness.

While some of the improvements made to the Erlang Virtual Machine
(Section~\ref{sec:vm-ets}) have been thoroughly reported in conference
publications~\cite{SH-EW12,ets_scalability-13,QDEuroPar,ScalableETS@Erlang-14,CA-ISPDC,CA-LCPC},
others are reported here for the first time
(Sections~\ref{sec:vm-scalability-time}, \ref{sec:vm-schedulers}).

In Section~\ref{sec:scalable-tools}, the WombatOAM and SD-Mon tools
are described for the first time, as is the revised Devo system and
visualisation.  The other tools for profiling, debugging and
refactoring developed in the project have previously been published
piecemeal~\cite{ASE12,percept2,PEPM2015,bakermulti},
but this is their first unified presentation.  

All of the performance results in Section~\ref{sec:case-studies} are
entirely new, although a comprehensive study of SD Erlang performance is
now available in a recent article by~\cite{evaluating-tpds-16}.


\section{Context}
\label{sec:context}
\subsection{Scalable Reliable Programming Models}
\label{subsec:progress-language}

There is a plethora of shared memory concurrent programming models
like PThreads or Java threads, and some models, like
OpenMP~\cite{chandra-01-parallel}, are simple and high level. However
synchronisation costs mean that these models generally do not scale
well, often struggling to exploit even 100 cores. Moreover, reliability mechanisms are
greatly hampered by the shared state: for example, a lock becomes
permanently unavailable if the thread holding it fails.

The High Performance Computing (HPC) community build large-scale (10$^{6}$
core) distributed memory systems using the \emph{de facto} standard MPI
communication libraries~\cite{snir-95-mpi}. Increasingly these are hybrid
applications that combine MPI with OpenMP. Unfortunately, MPI is not suitable
for producing general purpose concurrent software as it is too low level
with explicit message passing. Moreover, the most widely
used MPI implementations offer no fault recovery:\footnote{Some fault
  tolerance is provided in less widely used MPI implementations
  like~\cite{dewolfs-06-ft-mpi}.} if any part of the computation
fails, the entire computation fails. Currently the issue is addressed
by using what is hoped to be highly reliable computational and networking hardware, but
there is intense research interest in introducing reliability into HPC applications~\cite{gainaru2015errors}. 

Server farms use commodity computational and networking hardware, and
often scale to around 10$^{5}$ cores, where host failures are routine. They
typically perform rather constrained computations, e.g.~Big Data
Analytics, using reliable frameworks like Google
MapReduce~\cite{DBLP:journals/cacm/DeanG08} or Hadoop~\cite{DBLP:books/daglib/0029284}. The idempotent
nature of the analytical queries makes it relatively easy for the
frameworks to provide implicit reliability: queries are monitored and
failed queries are simply re-run. In contrast, actor languages like
Erlang are used to engineer reliable general purpose computation,
often recovering failed stateful computations.


\subsection{Actor Languages}
\label{sec:actor-langs}

The actor model of concurrency consists of independent processes
communicating by means of messages sent asynchronously between
processes. A process can send a message to any other process for which
it has the address (in Erlang the ``process identifier'' or pid), and
the remote process may reside on a different host. While the notion of
actors originated in AI~\cite{hewitt-73-universal}, it has been used
widely as a general metaphor for concurrency, as well as being
incorporated into a number of niche programming languages in the 1970s
and 80s. More recently it has come back to prominence through the rise
of not only multicore chips but also larger-scale distributed
programming in data centres and the cloud.

With built-in concurrency and data isolation, actors are a natural
paradigm for engineering reliable scalable general-purpose
systems~\cite{AghaPhD,hewitt-10-actor}. The model has two main
concepts: actors that are the unit of computation, and messages that
are the unit of communication. Each actor has an \emph{address-book}
that contains the addresses of all the other actors it is aware
of. These addresses can be either locations in memory, direct
physical attachments, or network addresses. In a pure actor language,
messages are the only way for actors to communicate.

After receiving a message an actor can do the following: (i) send
messages to another actor from its \emph{address-book}, (ii) create
new actors, or (iii) designate a behaviour to handle the next message
it receives. The model does not impose any restrictions in the order
in which these actions must be taken. Similarly, two messages sent
concurrently can be received in any order.
These features enable actor based systems to support indeterminacy and
quasi-commutativity, while providing locality, modularity, reliability
and scalability~\cite{hewitt-10-actor}.

Actors are just one message-based paradigm, and other languages and
libraries have related message passing paradigms. Recent  example languages include
Go~\cite{donovan2015go} and Rust~\cite{matsakis2014rust} that provide explicit channels, similar
to actor mailboxes. Probably the most famous message passing library
is MPI~\cite{snir-95-mpi}, with APIs for many languages and widely used on
clusters and High Performance Computers. It is, however, arguable that
the most important contribution of the actor model is the one-way
asynchronous communication~\cite{hewitt-10-actor}. Messages are not
coupled with the sender, and neither they are transferred
synchronously to a temporary container where transmission takes place,
e.g.~a buffer, a queue, or a mailbox. Once a message is sent, the
receiver is the only entity responsible for that message.

Erlang~\cite{armstrong-07-programming,cesarini-09-erlang} is the
pre-eminent programming language based on the actor
model,
having a history of use in production systems, initially with its
developer Ericsson and then more widely through open source adoption.
There are now actor frameworks for many  other languages; these include Akka for C\#, F\# and
Scala~\cite{website-12-scala}, CAF\footnote{\url{http://actor-framework.org}} for C++,
Pykka\footnote{http://pykka.readthedocs.org/en/latest/},
Cloud Haskell~\cite{epstein-11-towards},
PARLEY~\cite{parley-10-python} for Python, and Termite
Scheme~\cite{germain-06-concurrency}, and each of these is currently
under active use and development. Moreover, the recently defined Rust
language~\cite{matsakis2014rust} has a version of the
actor model built in, albeit in an imperative context.


\subsection{Erlang's Support for Concurrency}
\label{sec:erlang}

In Erlang, actors are termed \textit{processes}, and virtual machines are
termed \textit{nodes}. The key elements of the actor model are:
fast process creation and destruction;
lightweight processes, e.g.\ enabling 10$^{6}$ concurrent processes 
on a single host with 8GB RAM; fast asynchronous message passing with copying
semantics; process monitoring; strong
dynamic typing, and selective message reception.  

By default Erlang  processes are addressed by their \emph{process
identifier} (\textit{pid}), e.g.
\begin{verbatim}
Pong_PID = spawn(fun some_module:pong/0)
\end{verbatim}
spawns a process to execute the anonymous function given as argument to the \texttt{spawn} primitive, and binds \texttt{Pong\_PID} to the new process identifier. Here the new process will execute the \texttt{pong/0} function which is defined in \texttt{some\_module}. A subsequent call
\begin{verbatim}
Pong_PID ! finish
\end{verbatim}
sends the messaged \texttt{finish} to the process identified by
\texttt{Pong\_PID}. Alternatively, processes can be given names using
a call of the form:
\begin{verbatim}
register(my_funky_name, Pong_PID)
\end{verbatim}
which registers this process name in the node's \textit{process name table}
if not already present. Subsequently, these names can
be used to refer to or communicate with the corresponding processes
(e.g. send them a message):
\begin{verbatim}
my_funky_process ! hello.
\end{verbatim}

A \emph{distributed Erlang} system executes on multiple nodes, and the
nodes can be freely deployed across hosts, e.g.~they can be located on
the same or different hosts. To help make distribution transparent to
the programmer, when any two nodes connect they do so transitively,
sharing their sets of connections. Without considerable care this
quickly leads to a fully connected graph of nodes. A process may be
spawned on an explicitly identified node, e.g.
\begin{verbatim}
Remote_Pong_PID = spawn(some_node,
               fun some_module:pong/0).
\end{verbatim}
After this, the remote process can be addressed just as if it were local. It is a significant burden on the programmer to  identify the remote nodes in large systems, and we will return to this in Sections~\ref{sec:languagescaling}, \ref{subsubsec:design-semi-explicit-placement}.

\subsection{Scalability and Reliability in Erlang Systems}
\label{sec:scalability-reliability}

Erlang was designed to solve a particular set of problems, namely
those in building telecommunications' infrastructure, where systems
need to be scalable to accommodate hundreds of thousands of calls
concurrently, in soft real-time. These systems need to be
highly-available and reliable: i.e.\ to be robust in the case of
failure, which can come from software or hardware faults. Given the
inevitability of the latter, Erlang adopts the ``let it fail''
philosophy for error handling. That is, encourage programmers to embrace
the fact that a process may fail at any point, and have them rely on the
supervision mechanism, discussed shortly, to handle the failures.

\begin{figure}
\centering
\includegraphics[scale=0.38]{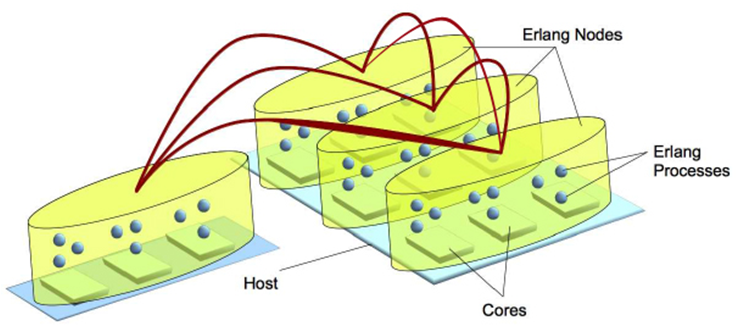}
\caption{Conceptual view of Erlang's concurrency, multicore support and distribution.}
\label{fig:cores-nodes}
\end{figure}

Figure~\ref{fig:cores-nodes} illustrates Erlang's support for
concurrency, multicores and distribution. Each Erlang node
is represented by a
yellow shape, and each rectangle represents a \emph{host} with an IP
address. Each red arc represents a \emph{connection} between Erlang nodes.
Each node can run on multiple cores, and exploit the inherent concurrency provided. This is done automatically by the VM, with no user intervention needed. Typically each core has an associated \emph{scheduler} that schedules processes; a new process will be spawned on the same core as the process that spawns it, but work can be moved to a different scheduler through a work-stealing allocation algorithm.
Each scheduler allows a process that is ready to compute at most a
fixed number of computation steps before switching to another. Erlang
\emph{built-in functions} or BIFs are implemented in C, and at the
start of the project were run to completion once scheduled, causing
performance and responsiveness problems if the BIF had a long execution time.

Scaling in Erlang is provided in two different ways. It is possible to scale \emph{within a single node} by means of the multicore virtual machine exploiting the concurrency provided by the multiple cores or NUMA nodes. It is also possible to scale \emph{across multiple hosts} using multiple distributed Erlang nodes.

Reliability in Erlang is multi-faceted. As in all actor languages
each process has private state, preventing a failed or failing process
from corrupting the state of other processes. Messages enable stateful
interaction, and contain a deep copy of the value to be shared, with
no references (e.g. pointers) to the senders' internal state. Moreover
Erlang avoids type errors by enforcing strong typing, albeit
dynamically~\cite{armstrong-10-erlang}.  
Connected nodes check liveness with heartbeats, and can be monitored
from outside Erlang, e.g. by an operating system process.

However, the most important way to achieve reliability is
\emph{supervision}, which allows a process to monitor the status of a
child process and react to any failure, for example by spawning a
substitute process to replace a failed process. Supervised processes
can in turn supervise other processes, leading to a \emph{supervision
  tree}.  The supervising and supervised processes may be in different
nodes, and on different hosts, and hence the supervision tree may span
multiple hosts or nodes.

To provide reliable distributed service registration, a global
namespace is maintained on every node, which maps \emph{process names}
to pids. It is this that we mean when we talk about a `reliable'
system: it is one in which a named process in a distributed system can
be restarted without requiring the client processes also to be
restarted (because the name can still be used for communication).

To see global registration in action, consider a pong server process
\begin{verbatim}
global:register_name(pong_server,
                     Remote_Pong_PID).
\end{verbatim}
Clients of the server can send messages to the registered name, e.g.
\begin{verbatim}
global:whereis_name(pong_server)!finish.
\end{verbatim}
If the server fails the supervisor can spawn a replacement server process with a new pid and register it with the same name (\texttt{pong\_server}). Thereafter client messages to the \texttt{pong\_server} will be delivered to the new server process. We return to discuss the scalability limitations of maintaining a global namespace in Section~\ref{sec:language-scalability}.


\subsection{ETS: Erlang Term Storage}
\label{sec:ETS}

Erlang is a pragmatic language and the actor model it supports is not
pure. Erlang processes, besides communicating via asynchronous message
passing, can also share data in public memory areas called \emph{ETS
  tables}.

The Erlang Term Storage (ETS) mechanism is a central component of
Erlang's implementation. It is used internally by many libraries and
underlies the in-memory databases. ETS tables are key-value stores:
they store Erlang tuples where one of the positions in the tuple
serves as the lookup key.
An ETS table has a \emph{type} that may be either \texttt{set},
\texttt{bag} or \texttt{duplicate\_bag}, implemented as a hash table,
or \texttt{ordered\_set} which is currently implemented as an AVL tree.
The main operations that ETS supports are table creation, insertion of
individual entries and atomic bulk insertion of multiple entries in a
table, deletion and lookup of an entry based on some key, and destructive
 update. The operations are implemented as C built-in functions in the Erlang VM. 

The code snippet below  creates a \texttt{set} ETS table keyed by the first element of the  entry; atomically inserts two elements with keys \texttt{some\_key} and \texttt{42}; updates the value associated with the table entry with key \texttt{42}; and then looks up this entry.
\begin{verbatim}
Table = ets:new(my_table,
          [set, public, {keypos, 1}]),
ets:insert(Table, [
            {some_key, an_atom_value},
            {42, {a,tuple,value}}]),
ets:update_element(Table, 42,
        [{2, {another,tuple,value}}]),
[{Key, Value}] = ets:lookup(Table, 42).
\end{verbatim}

ETS tables are heavily used in many Erlang applications. This is
partly due to the convenience of sharing data for some programming
tasks, but also partly due to their fast implementation. As a shared
resource, however, ETS tables induce contention and become a
scalability bottleneck, as we shall see in Section~\ref{sec:VMScaling}.


\section{Benchmarks for scalability and reliability}
\label{sec:benchmarks}
The two benchmarks that we use throughout this article are Orbit, that measures scalability without looking at reliability,
and Ant Colony Optimisation (ACO) that allows us to measure the impact
on scalability of adding global namespaces to ensure
reliability. The source code for the benchmarks, together
  with more detailed documentation, is available at
  \url{https://github.com/release-project/benchmarks/}.
The RELEASE project team also worked to improve the reliability and
scalability of other Erlang programs including a substantial
(approximately 150K lines of Erlang code) Sim-Diasca
simulator~\cite{simdiasca} and an Instant Messenger that is more
typical of Erlang applications~\cite{chechina-16-scalable} but we do
not cover these systematically here.

\subsection{Orbit}
\label{subsec:orbit-overview}

Orbit is a computation in symbolic algebra, which generalises a transitive closure computation~\cite{lubeck-01-enumerating}. To compute the \textit{orbit} for a given space $[0..X]$, a list of generators $g_1, g_2,..., g_n$ are applied on an initial vertex $x_{0}\in [0..X]$.  This creates new values $(x_{1}...x_{n}) \in [0..X]$, where $x_i = g_i(x_0)$. The generator functions are applied on the new values until no new value is generated.

Orbit is a suitable benchmark because it has a number of aspects that
characterise a class of real applications.  The core data structure it
maintains is a set and, in distributed environments is implemented as
a \emph{distributed hash table} (DHT), similar to the DHTs used in
replicated form in NoSQL database management systems.  Also, in
distributed mode, it uses standard peer-to-peer (P2P) techniques like a credit-based termination
algorithm~\cite{matocha-98-taxonomy}.  By choosing the orbit size, the
benchmark can be parameterised to specify smaller or larger
computations that are suitable to run on a single machine
(Section~\ref{sec:VMScaling}) or on many nodes (Section\ref{subsec:case-studies-orbit}).
Moreover it is only a few hundred lines of code.

\begin{figure}[t]
\centering
\includegraphics[trim={0 3.8cm 0 3.8cm},clip,width=0.5\textwidth]{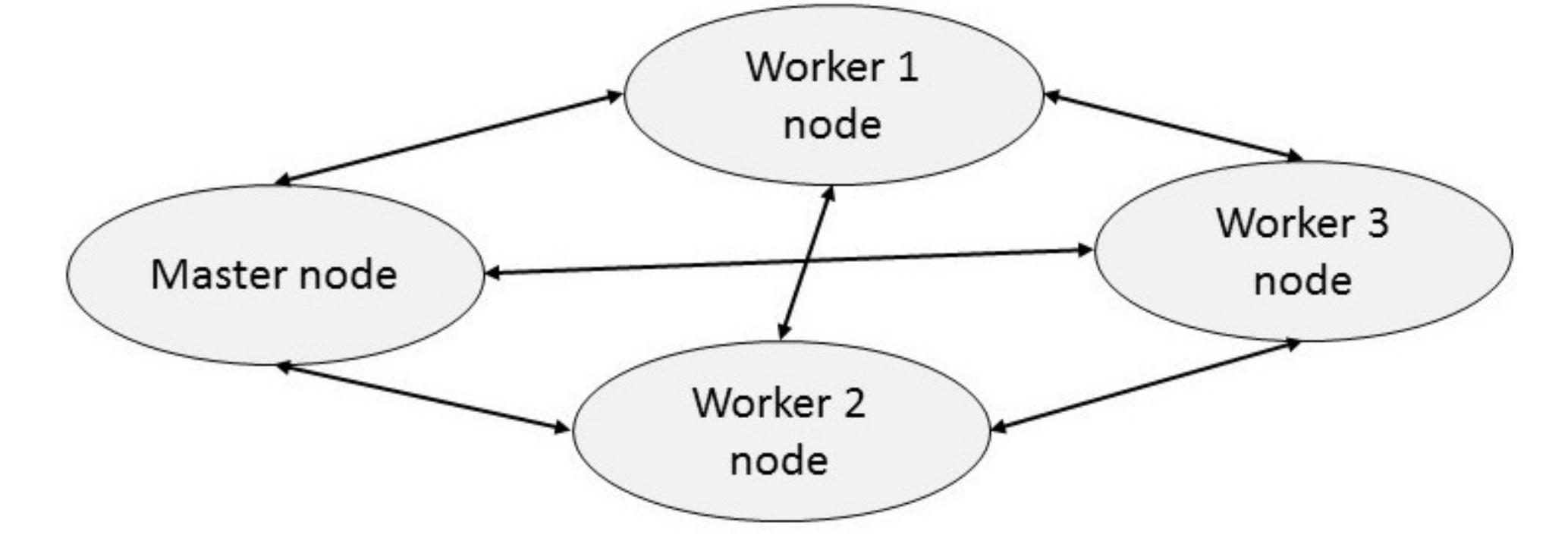}
\caption{Distributed Erlang Orbit (D-Orbit) architecture: workers are mapped to nodes.}
\label{fig:d-orbit-comm-model}
\end{figure}

As shown in Figure~\ref{fig:d-orbit-comm-model}, the computation is
initiated by a master which creates a number of workers.  In the
single node scenario of the benchmark, workers correspond to processes
but these workers can also spawn other processes to apply the
generator functions on a subset of their input values, thus
creating \emph{intra-worker parallelism}.  In the distributed version
of the benchmark, processes are spawned by the master node to worker
nodes, each maintaining a DHT fragment.  A newly spawned process gets
a share of the parent's credit, and returns this on completion.  The
computation is finished when the master node collects all credit,
i.e.\ all workers have completed.

\subsection{Ant Colony Optimisation (ACO)}
\label{subsec:aco-overview}

Ant Colony Optimisation~\cite{Dorigo-book} is a meta-heuristic which
has been applied to a large number of combinatorial optimisation
problems. For the purpose of this article, we have applied it to an
NP-hard scheduling problem known as the Single Machine Total Weighted
Tardiness Problem (SMTWTP)~\cite{McNaughton-Scheduling}, where a
number of jobs of given lengths have to be arranged in a single linear
schedule. The goal is to minimise the \textit{cost} of the schedule,
as determined by certain constraints.

\begin{figure}[!b]
\centering
\includegraphics[trim={0 5.1cm 0 3.2cm},clip,width=.5\textwidth]{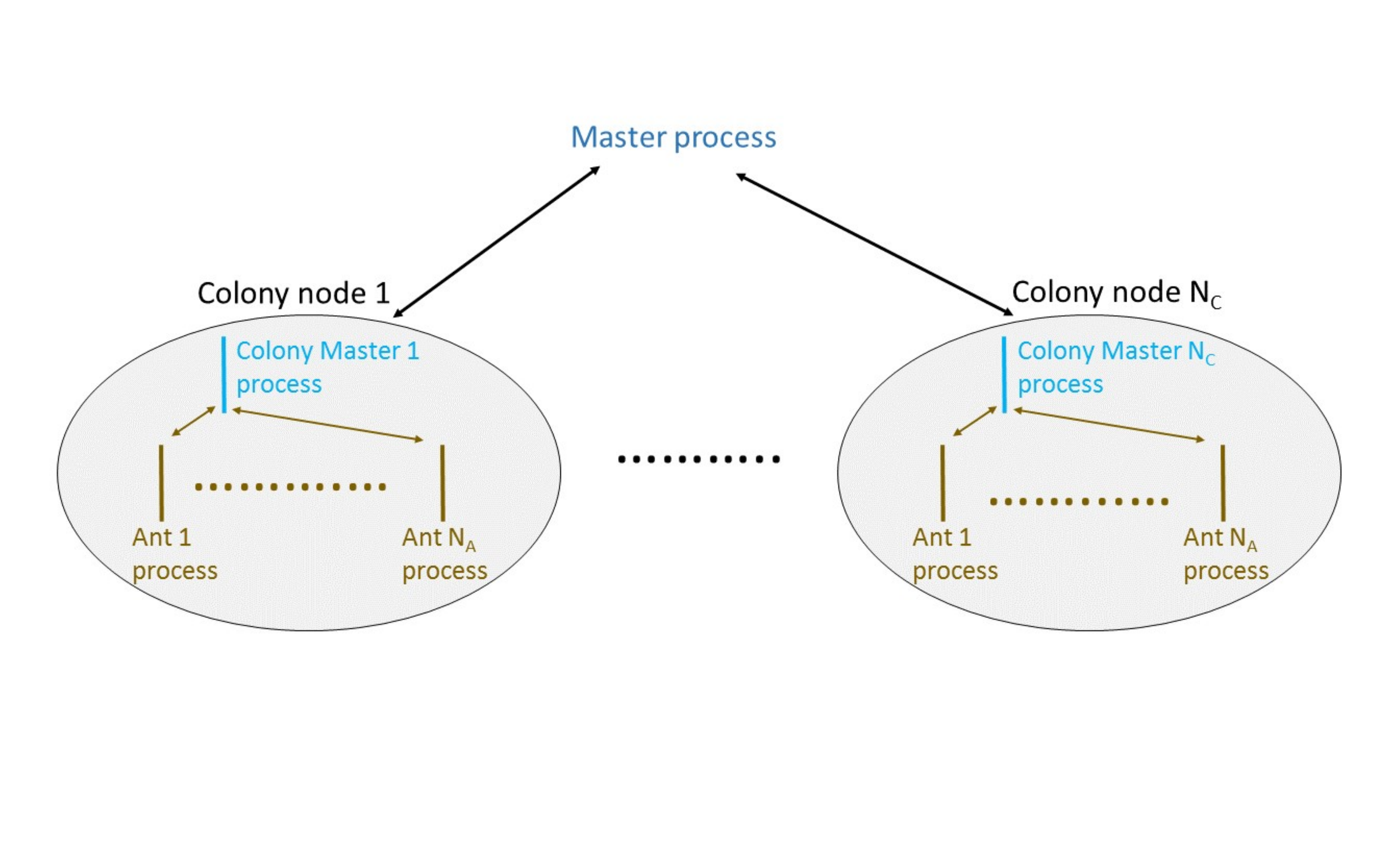}
\caption{Distributed Erlang Two-level Ant Colony Optimisation (TL-ACO) architecture.}
\label{fig:two-level-aco}
\end{figure}

\begin{figure*}[!t]
\centering
\includegraphics[trim={1.4cm 2.5cm 1cm 1cm},clip,width=0.55\textwidth,angle=-90]{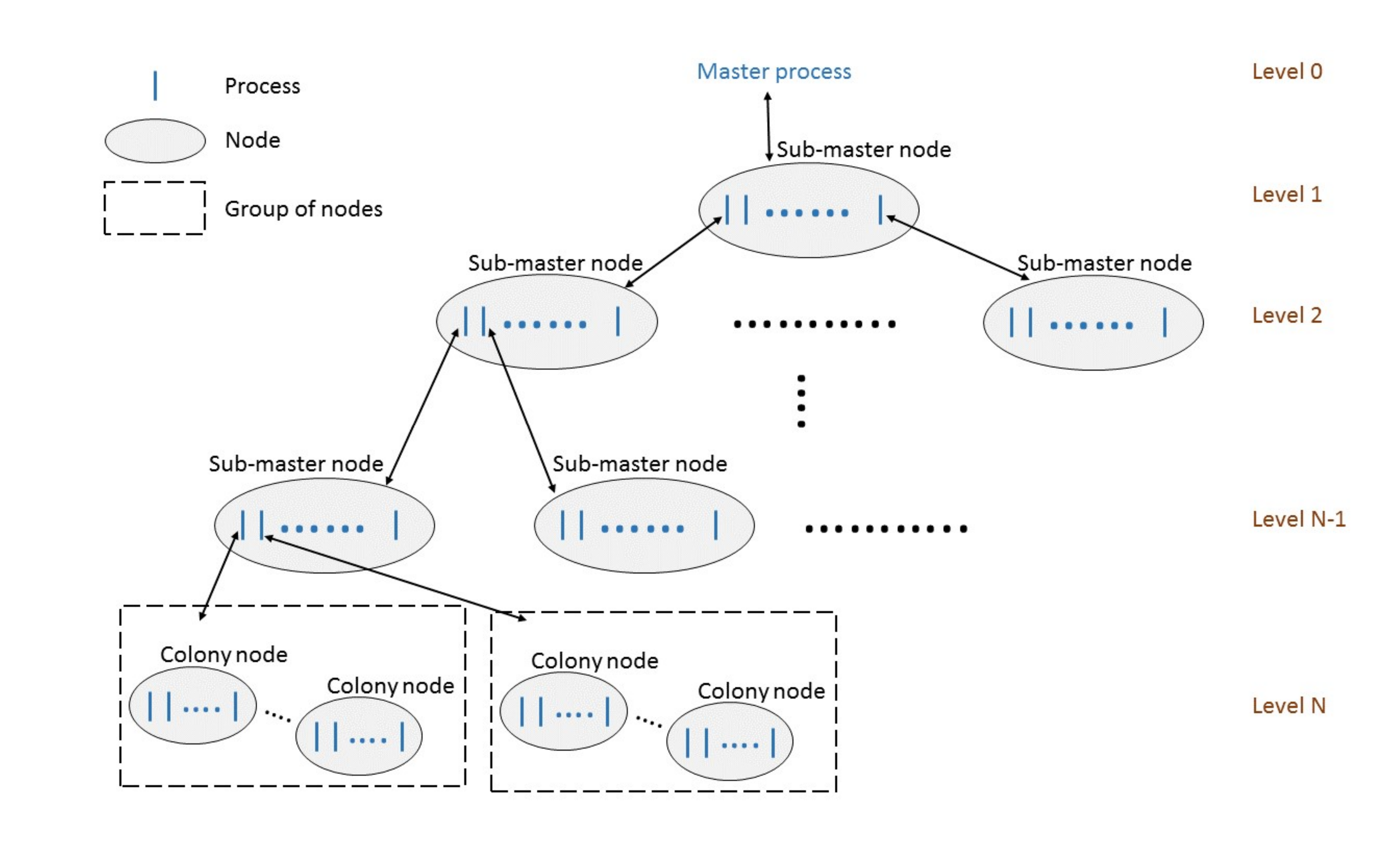}
\caption{Distributed Erlang Multi-level Ant Colony Optimisation (ML-ACO) architecture.}
\label{fig:multi-level-aco}
\end{figure*}

The ACO method is attractive from the point of view of distributed
computing because it can benefit from having multiple cooperating
colonies, each running on a separate compute node and consisting of
multiple ``ants''.  
Ants are simple computational agents which
concurrently compute possible solutions to the input problem guided by
shared information about good paths through the search space; there is
also a certain amount of stochastic variation which allows the ants to
explore new directions. 
Having multiple colonies increases the number of ants, thus increasing
the probability of finding a good solution.

We implement four distributed coordination patterns for the same
multi-colony ACO computation as follows. In each implementation, the
individual colonies perform some number of \textit{local iterations}
(i.e.~generations of ants) and then report their best solutions; the
globally-best solution is then selected and is reported to the
colonies, which use it to update their pheromone matrices. This
process is repeated for some number of \textit{global iterations}.

\textit{Two-level ACO (TL-ACO)} has a single master node that
collects the colonies' best solutions and distributes the overall best
solution back to the colonies. Figure~\ref{fig:two-level-aco} depicts
the process and node placements of the \textit{TL-ACO} in a cluster
with \textit{$N_C$} nodes. The master process spawns \textit{$N_C$}
colony processes on available nodes. 
In the next step, each colony process spawns \textit{$N_A$} ant
processes on the local node.  Each ant iterates $I_A$ times, returning
its result to the colony master. Each colony iterates $I_M$ times,
reporting their best solution to, and receiving the globally-best
solution from, the master process. We validated the implementation by applying TL-ACO to a number of
standard SMTWTP instances
~\cite{Crauwel-Potts-SMTWTP,Beasley-ORLIB,Geiger-New-SMTWTP-instances}, 
obtaining good results in all cases, and confirmed that the number of perfect solutions
increases as we increase the number of colonies.

\textit{Multi-level ACO (ML-ACO)}. In TL-ACO the master node receives
messages from \textit{all} of the colonies, and thus could become a
bottleneck. ML-ACO addresses this by having a tree of submasters
(Figure~\ref{fig:multi-level-aco}), with each node in the bottom level
collecting results from a small number of colonies. These are then fed
up through the tree, with nodes at higher levels selecting the best
solutions from their children. 

\textit{Globally Reliable ACO (GR-ACO)}. In ML-ACO if a single colony
fails to report back the system will wait indefinitely. GR-ACO adds
fault tolerance, supervising colonies so that a faulty colony can be
detected and restarted, allowing the system to continue execution.

\textit{Scalable Reliable ACO (SR-ACO)} also adds fault-tolerance, but
using supervision within our new s\_groups from
Section~\ref{subsubsec:design-s-groups}, and the architecture of SR-ACO is
discussed in detail there.


\section{Erlang Scalability Limits}
\label{sec:erlang-scalability}
This section investigates the scalability of Erlang at VM, language,
and persistent storage levels. An aspect we choose not to explore is
the security of large scale systems where, for example, one might
imagine providing enhanced security for systems with multiple clusters
or cloud instances connected by a Wide Area Network. We assume that
existing security mechanisms are used, e.g.\ a Virtual Private Network.


\subsection{Scaling Erlang on a Single Host}
\label{sec:VMScaling}

\npcomment{This section still does not address issues related to the scheduler.}

To investigate Erlang scalability we built \bencherl, an extensible
open source benchmark suite with a web interface.\footnote{Information
  about \bencherl is available at
  \url{http://release.softlab.ntua.gr/bencherl/}.}
\bencherl shows how an application's performance changes when
resources, like cores or schedulers, are added; or when options that
control these resources change:
\begin{itemize}
\item the \emph{number of nodes}, i.e.\ the number of Erlang VMs used,
  typically on multiple hosts;
\item the \emph{number of cores} per node;
\item the \emph{number of schedulers}, i.e.\ the OS threads that execute
  Erlang processes in parallel, and their \emph{binding} to the topology
  of the cores of the underlying computer node;
\item the \emph{Erlang/OTP release and flavor}; and
\item the \emph{command-line arguments} used to start the Erlang nodes.
\end{itemize}

Using \bencherl, we investigated the scalability of an initial set of
twelve benchmarks and two substantial Erlang applications using a
single Erlang node (VM) on machines with up to 64 cores, including
the Bulldozer machine specified in Appendix~A.
This set of experiments, reported by~\cite{bencherl-12}, confirmed
that some programs scaled well in the most recent Erlang/OTP release
of the time (R15B01) but also revealed VM and language level
scalability bottlenecks.

\begin{figure}
  \centering
  \includegraphics[width=0.49\textwidth,height=.23\textheight]{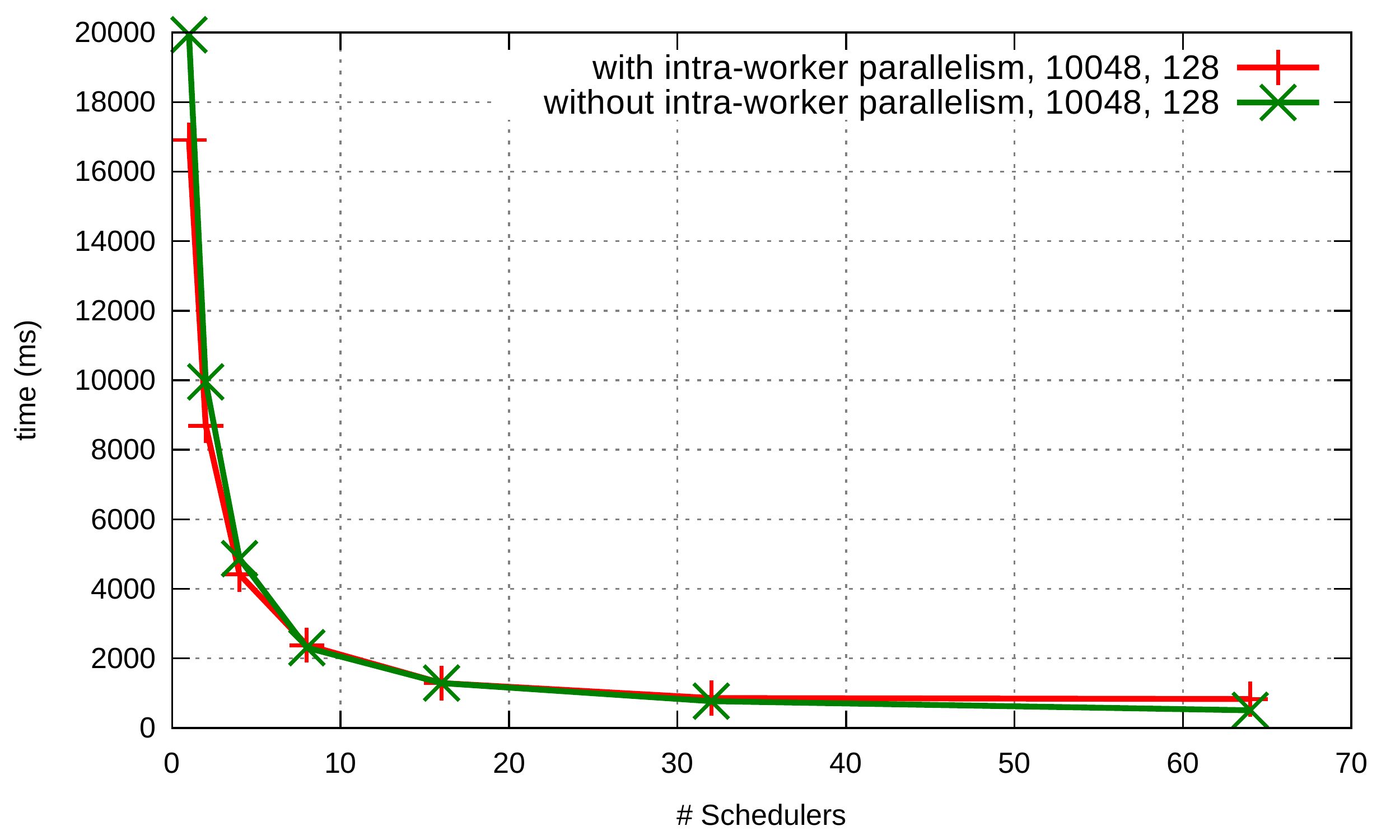}
  \includegraphics[width=0.49\textwidth,height=.23\textheight]{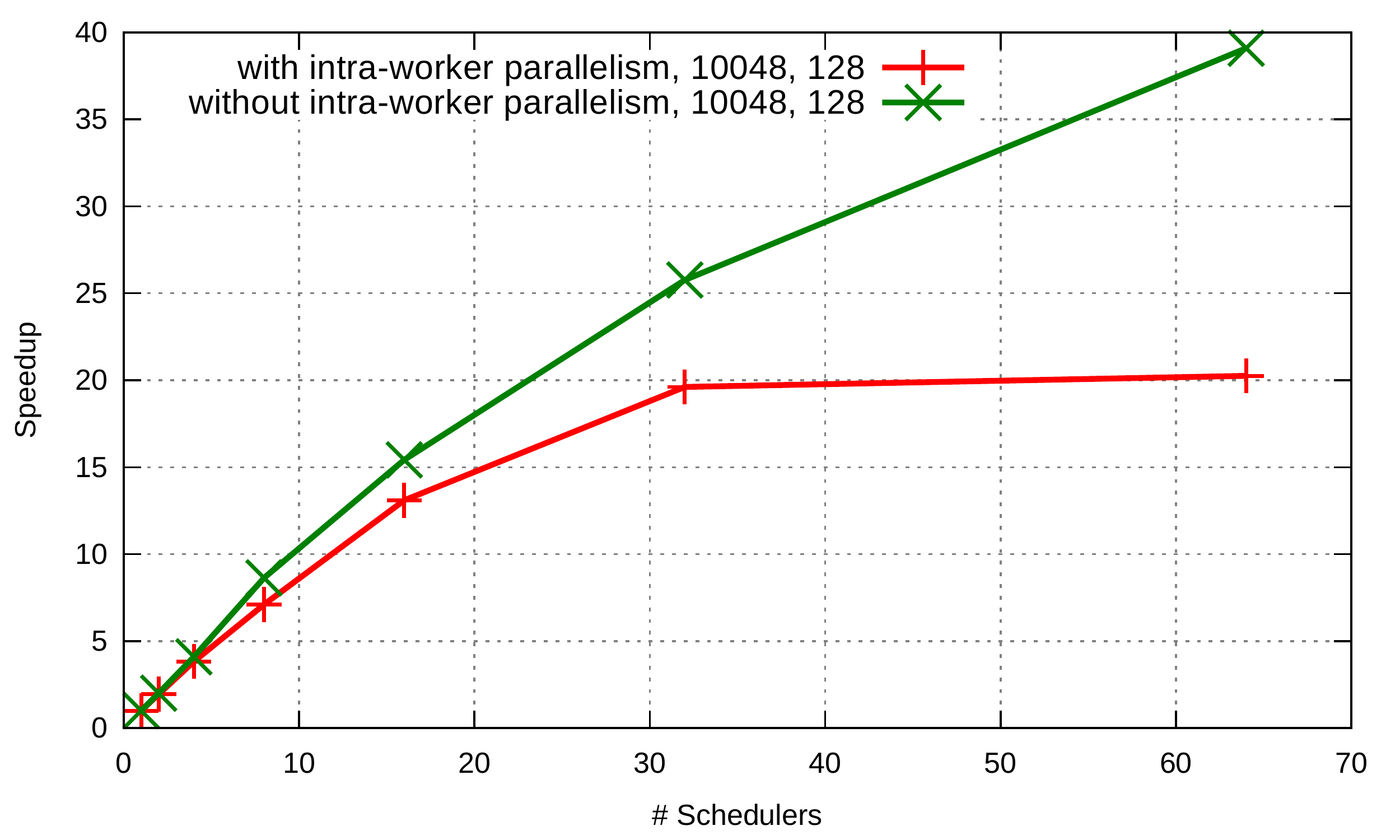}
  \caption{Runtime and speedup of two configurations of the Orbit benchmark using Erlang/OTP~R15B01.
    \label{fig:orbit}}
\end{figure}

Figure~\ref{fig:orbit} shows runtime and speedup curves for the Orbit benchmark
where master and workers run on a single Erlang node in configurations
with and without intra-worker parallelism.  In both configurations the
program scales.  Runtime continuously decreases as we add more
schedulers to exploit more cores. The speedup of the benchmark without
intra-worker parallelism, i.e.\ without spawning additional processes
for the computation (green curve), is almost linear up to 32 cores but
increases less rapidly from that point on; we see a similar but more
clearly visible pattern for the other configuration (red curve) where
there is no  performance improvement beyond 32 schedulers. This is due
to the asymmetric characteristics of the machine's micro-architecture,
which consists of modules that couple two conventional x86
out-of-order cores that share the early pipeline stages, the floating
point unit, and the L2 cache with the rest of the module~\cite{Bulldozer}.

\begin{figure}[!t]
  \centering
  \includegraphics[width=0.49\textwidth,height=.23\textheight]{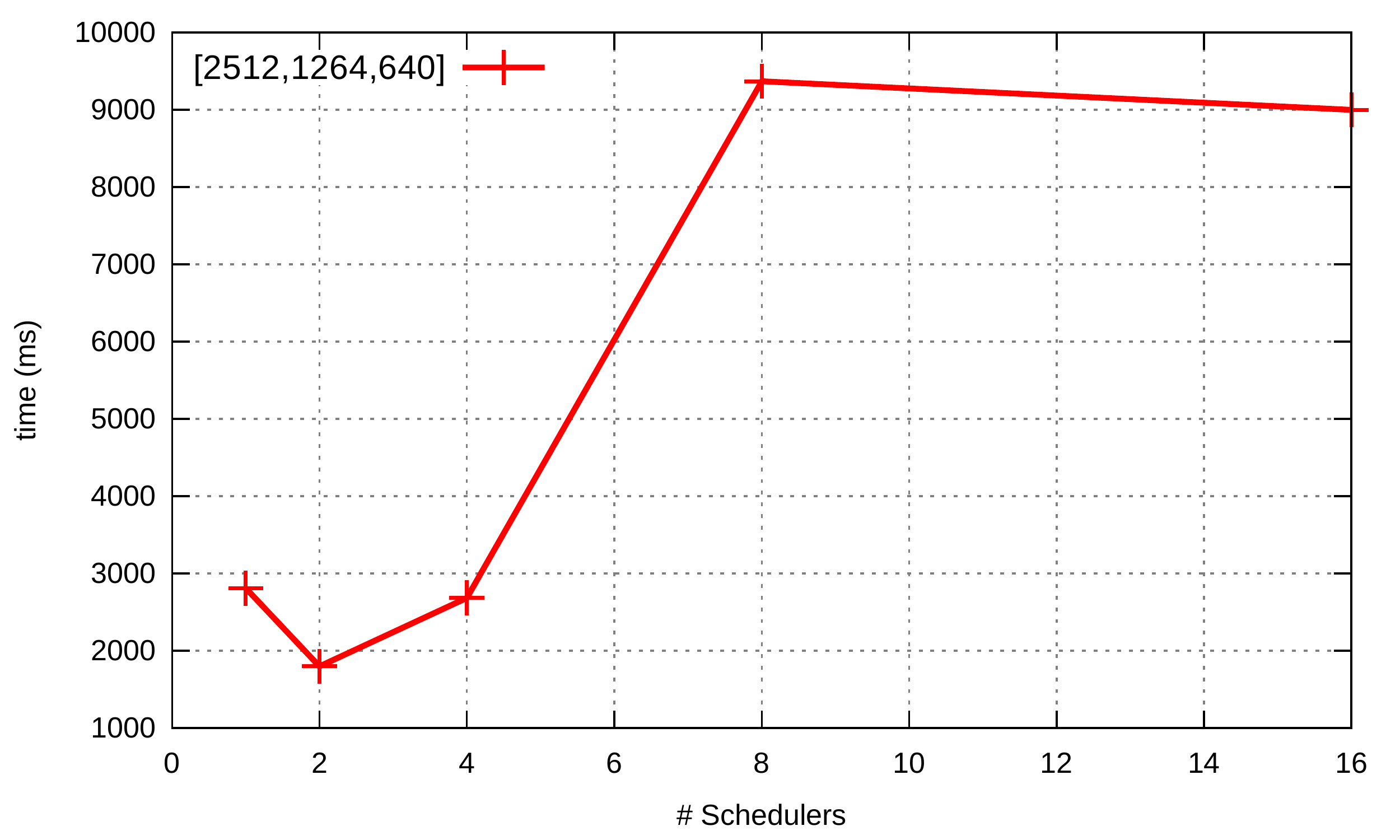}
  \includegraphics[width=0.49\textwidth,height=.23\textheight]{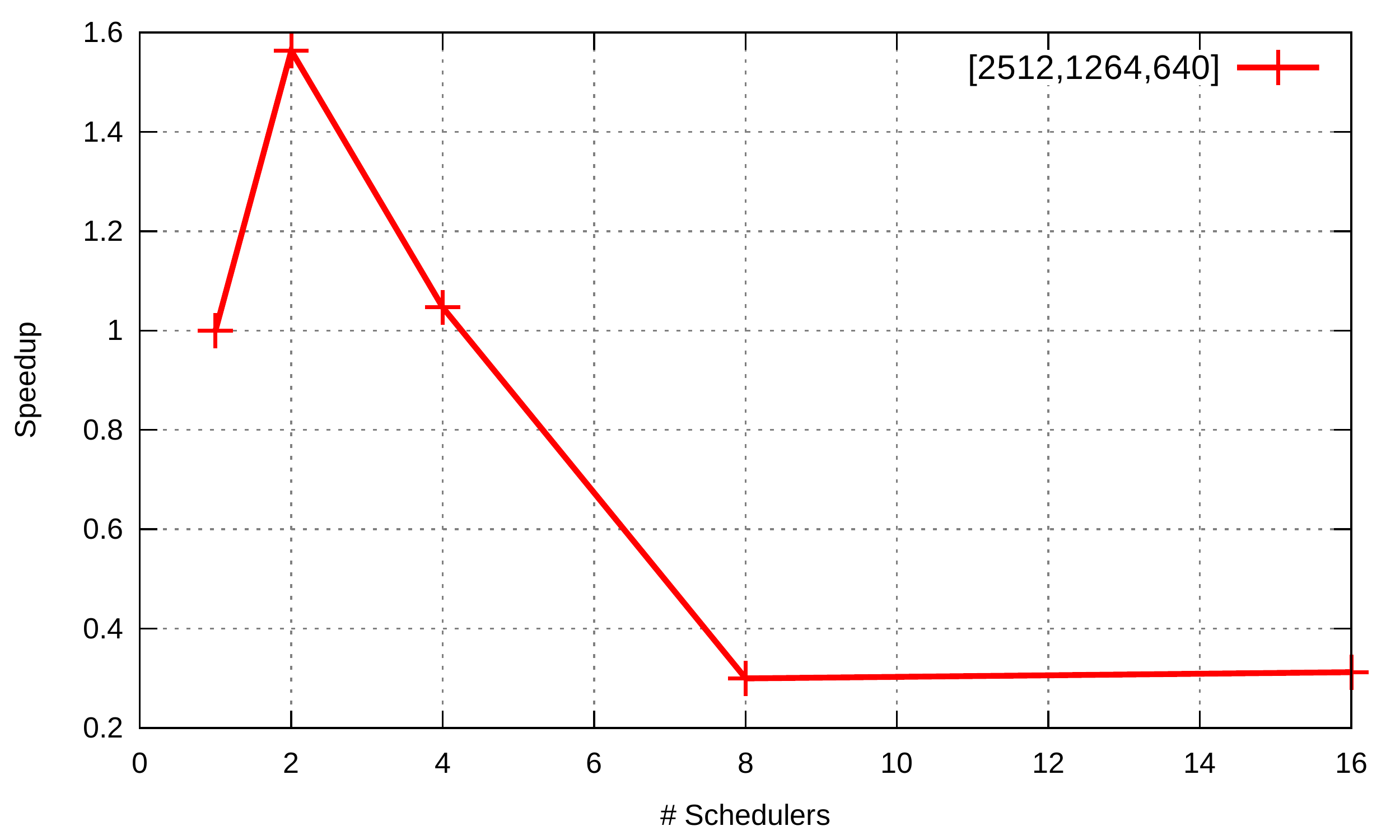}
  \caption{Runtime and speedup of the \bench{ets\_test} \bencherl benchmark using Erlang/OTP~R15B01. 
    \label{fig:ets_test}}
\end{figure}

Some other benchmarks, however, did not scale well or experienced
significant slowdowns when run in many VM schedulers (threads).
For example the \bench{ets\_test} benchmark has multiple processes
accessing a shared ETS table.  Figure~\ref{fig:ets_test} shows runtime
and speedup curves for \bench{ets\_test} on a 16-core (eight cores
with hyperthreading) Intel Xeon-based machine.  It shows that runtime
increases beyond two schedulers, and that the program exhibits a
slowdown instead of a speedup.

For many benchmarks there are obvious reasons for poor scaling like
limited parallelism in the application, or contention for shared
resources.  The reasons for poor scaling  are less obvious for other
benchmarks, and it is exactly these we have chosen to study in detail
in subsequent work~\cite{bencherl-12,ets_scalability-13}.

A simple example is the \bench{parallel} \bencherl benchmark, that
spawns some~$n$ processes, each of which creates a list of~$m$
timestamps and, after it checks that each timestamp in the list is
strictly greater than the previous one, sends the result to its parent.
\begin{figure}[!t]
  \centering
  \includegraphics[width=0.49\textwidth,height=.23\textheight]{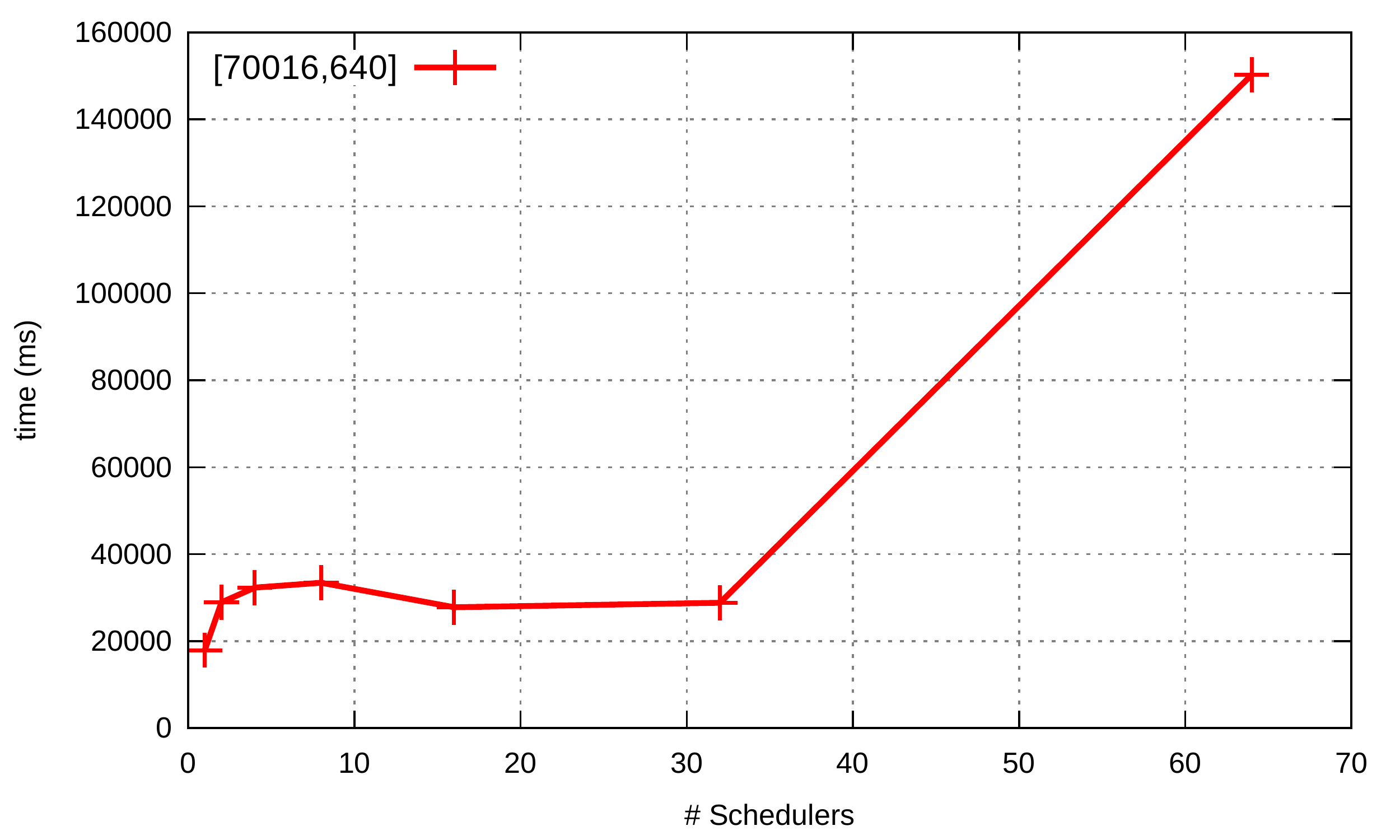}
  \hfill
  \includegraphics[width=0.49\textwidth,height=.23\textheight]{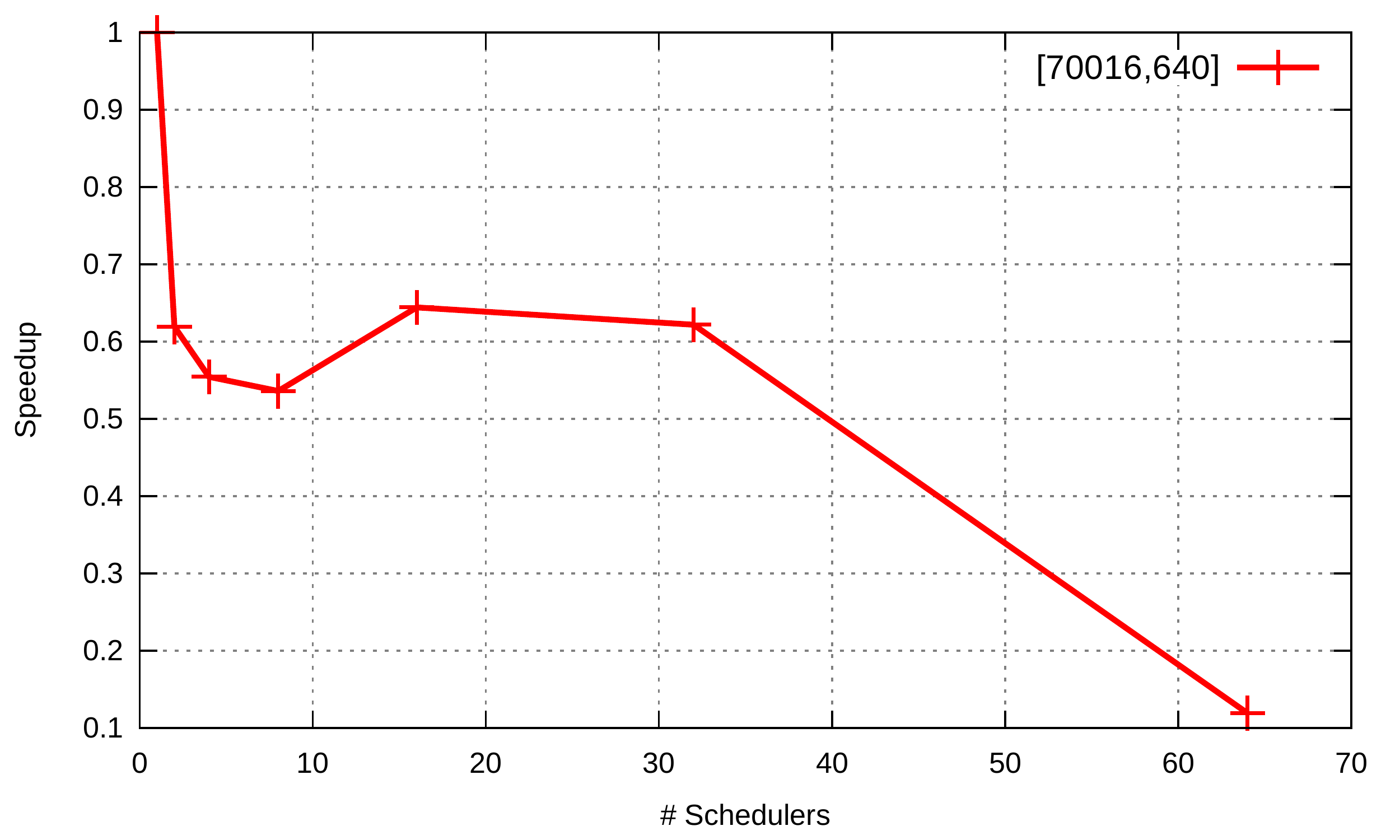}
  \caption{Runtime and speedup of the \bencherl benchmark called \bench{parallel} using Erlang/OTP~R15B01.
    \label{fig:parallel}}
\end{figure}
Figure~\ref{fig:parallel} shows that up to eight cores each
additional core leads to a slowdown, thereafter a small speedup is
obtained up to 32 cores, and then again a slowdown. A small aspect of
the benchmark, easily overlooked, explains the poor scalability. The
benchmark creates timestamps using the \verb|erlang:now/0| built-in
function, whose implementation acquires a \emph{global lock} in order
to return a \emph{unique timestamp}. That is, two calls
to \verb|erlang:now/0|, even from different processes are guaranteed
to produce monotonically increasing values.  This lock is precisely
the bottleneck in the VM that limits the scalability of this
benchmark. We describe our work to address VM timing issues
in Figure~\ref{sec:vm-scalability-time}.

\paragraph{Discussion} Our investigations identified contention for
shared ETS tables, and for commonly-used shared resources like timers,
as the key VM-level scalability issues.
Section~\ref{sec:improving-vm-scalability} outlines how we addressed
these issues in recent Erlang/OTP releases.

\npcomment{Revise if/when schedulers are included.}


\subsection{Distributed Erlang Scalability}
\label{sec:languagescaling}

\paragraph{Network Connectivity Costs} When any normal distributed
Erlang nodes communicate, they share their connection sets and this
typically leads to a fully connected graph of nodes. So a
system with $n$ nodes will maintain $O(n^{2}$) connections, and these
are relatively expensive TCP connections with continual maintenance
traffic. This design aids transparent distribution as there is no need
to discover nodes, and the design works well for small numbers of nodes.
However at emergent server architecture scales, i.e.\  hundreds of nodes,
this design becomes very expensive and system architects must switch
from the default Erlang model, e.g. they need to start using hidden
nodes that do not share connection sets.

We have investigated the scalability limits imposed by network
connectivity costs using several Orbit calculations on two large
clusters: Kalkyl and Athos as specified in Appendix~A.  The Kalkyl
results are discussed by~\cite{improving-jpdc-16}, and
Figure~\ref{fig:orbit-2-5-speedup} in Section~\ref{subsec:case-studies-orbit} shows
representative results for distributed Erlang computing orbits with 2M
and 5M elements on Athos. In all cases performance degrades beyond some
scale (40 nodes for the 5M orbit, and 140 nodes for the 2M orbit).
Figure~\ref{fig:network-traffic-sent} illustrates the additional network traffic
induced by the fully connected network. It allows a comparison between
the number of packets sent in a fully connected network (ML-ACO) with
those sent in a network partitioned using our new s\_groups (SR-ACO).

\begin{figure*}[!t]
  \centering
  \includegraphics[keepaspectratio,width=\textwidth]{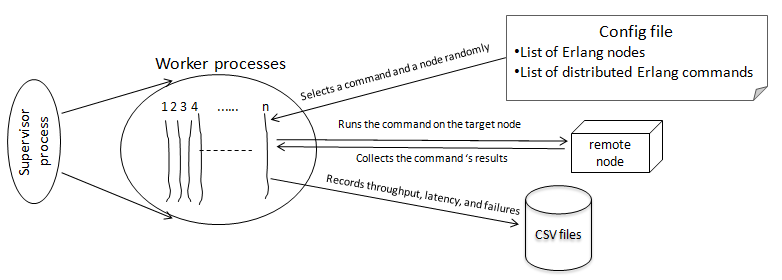}
  \caption{DE-Bench's internal workflow.}
  \label{fig:de_bench_design}
\end{figure*}

\paragraph{Global Information Costs} Maintaining global information
is known to limit the scalability of distributed systems, and
crucially the process namespaces used for reliability are global.  To
investigate the scalability limits imposed on distributed Erlang by
such global information we have designed and implemented
\emph{DE-Bench}, an open source, parameterisable and scalable
peer-to-peer benchmarking
framework~\cite{ghaffari-14-investigating,Ghaffari2014DEBenchCode}. DE-Bench
measures the throughput and latency of distributed Erlang commands on
a cluster of Erlang nodes, and the design is influenced by the Basho
Bench benchmarking tool for Riak~\cite{website-14-basho-bench}. Each
DE-Bench instance acts as a peer, providing scalability and
reliability by eliminating central coordination and any single point
of failure. 

To evaluate the scalability of distributed Erlang, we measure how
adding more hosts increases the throughput, i.e.~the total number of
successfully executed distributed Erlang commands per experiment.
Figure~\ref{fig:de_bench_design} shows the parameterisable internal workflow
of DE-Bench. There are three classes of commands in DE-Bench:%
\begin{inparaenum}[(i)]
\item \emph{Point-to-Point (P2P) commands}, where a function with
  tunable argument size and computation time is run on a remote node,
  include \texttt{spawn}, \texttt{rpc}, and synchronous calls to
  server processes, i.e. \texttt{gen\_server} or \texttt{gen\_fsm}.
\item \emph{Global commands}, which entail synchronisation across all
  connected nodes, such as \texttt{global:register\_name} and
  \texttt{global:unregister\_name}.
\item \emph{Local commands}, which are executed independently by a single
  node, e.g.~ \texttt{whereis\_name}, a look up in the local name table. 
\end{inparaenum}

The benchmarking is conducted on 10 to 100 host configurations
of the Kalkyl cluster (in steps of 10) and measures
the throughput of successful commands per second over 5 minutes. There
is one Erlang VM on each host and one DE-Bench instance on
each VM. The full paper~\cite{ghaffari-14-investigating} investigates
the impact of data size, and computation time in P2P calls both
independently and in combination, and the scaling properties of the
common Erlang/OTP generic server processes \texttt{gen\_server} and
\texttt{gen\_fsm}.

\begin{figure*}[!t]
  \centering
  \includegraphics[keepaspectratio,width=0.75\textwidth]{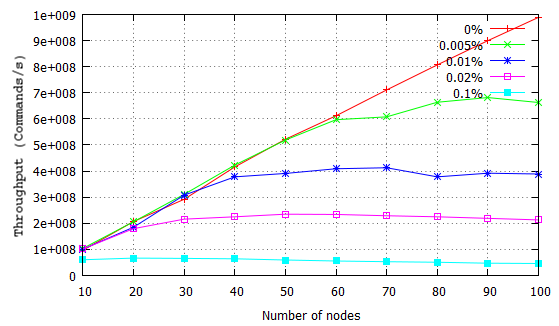}
  \caption{Scalability vs.\ percentage of global commands in Distributed Erlang.}
  \label{fig:frequency_of_global_operation}
\end{figure*}

\begin{figure*}
  \centering
  \includegraphics[trim={0 3pt 3pt 0},clip,keepaspectratio,width=0.75\textwidth] {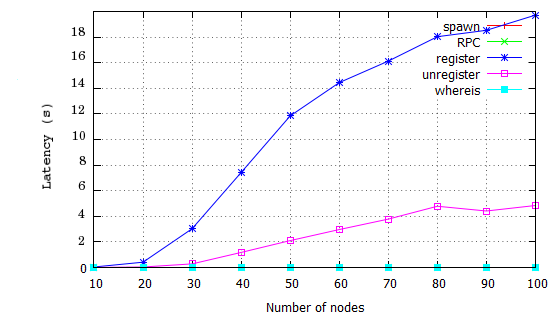}
  \caption{Latency of commands as the number of Erlang nodes increases.}
  \label{fig:latency}
\end{figure*}

Here we focus on the impact of different proportions of global
commands, mixing global with P2P and local commands.
Figure~\ref{fig:frequency_of_global_operation} shows that even a low
proportion of global commands limits the scalability of distributed
Erlang, e.g.~just 0.01\% global commands limits scalability to around
60 nodes.
Figure~\ref{fig:latency} reports the latency of all commands and shows that,
while the latencies for P2P and local commands are stable at scale,
the latency of the global commands increases dramatically with scale.
Both results illustrate that the impact of global operations on
throughput and latency in a distributed Erlang system is severe.

\paragraph{Explicit Placement} While network connectivity and global
information impact \emph{performance} at scale, our investigations
also identified explicit process placement as a \emph{programming}
issue at scale.  Recall from Section~\ref{sec:erlang} that distributed
Erlang requires the programmer to identify an explicit Erlang node (VM)
when spawning a process. Identifying an appropriate node becomes a
significant burden for large and dynamic systems. The problem is
exacerbated in large distributed systems where (1) the hosts may not
be identical, having different hardware capabilities or different
software installed; and (2) communication times may be non-uniform: it may
be fast to send a message between VMs on the same host, and slow if
the  VMs are on different  hosts in a large distributed system.

These factors make it difficult to deploy applications, especially in
a scalable and \textit{portable} manner.  Moreover while the
programmer may be able to use platform-specific knowledge to decide
where to spawn processes to enable an application to run efficiently,
if the application is then deployed on a different platform, or if the
platform changes as hosts fail or are added, this becomes outdated.

\paragraph{Discussion}
Our investigations confirm three language-level scalability
limitations of Erlang from developer folklore.%
\begin{inparaenum}[(1)]
\item Maintaining a \emph{fully connected network} of Erlang nodes limits scalability, for example  Orbit is typically limited to just 40 nodes.
\item \emph{Global operations}, and crucially the global operations required for reliability, i.e.\ to maintain a \emph{global namespace}, seriously limit the scalability of distributed Erlang systems.
\item Explicit \emph{process placement} makes it hard to built performance portable applications for large architectures.
\end{inparaenum}
These issues cause designers of reliable large scale systems in Erlang to depart from the standard Erlang model, e.g.\ using techniques like hidden nodes and storing pids in data structures.
In Section~\ref{sec:language-scalability} we develop language technologies to address these issues.


\subsection{Persistent Storage}
\label{sec:storagescaling}

\npcomment{I think this section should be beefed up, to resemble the previous ones.}

Any large scale system needs reliable scalable persistent storage, and
we have studied the scalability limits of Erlang persistent storage
alternatives~\cite{ghaffari-13-scalable}.  We envisage a typical large
server having around 10$^{5}$ cores on around 100 hosts.
We have reviewed the requirements for scalable and available
persistent storage and evaluated four popular Erlang DBMS against
these requirements. For a target scale of around 100 hosts, Mnesia and
CouchDB are, unsurprisingly, not suitable. However, Dynamo-style NoSQL
DBMS like Cassandra and Riak have the potential to be.

We have investigated the current scalability limits of the Riak NoSQL
DBMS using the Basho Bench benchmarking framework on a cluster with up
to 100 nodes and independent disks. We found that that the scalability
limit of Riak version 1.1.1 is 60 nodes on the Kalkyl cluster. The
study placed into the public scientific domain what was previously
well-evidenced, but anecdotal, developer experience.

We have also shown that resources like memory, disk, and network do not
limit the scalability of Riak. By instrumenting the \texttt{global} and
\texttt{gen\_server} OTP libraries we identified a specific Riak remote
procedure call that fails to scale. We outline how later releases of
Riak are refactored to eliminate the scalability bottlenecks.

\paragraph{Discussion} We conclude that Dynamo-like NoSQL DBMSs have
the potential to deliver reliable persistent storage for Erlang at our
target scale of approximately 100 hosts. Specifically an Erlang
Cassandra interface is available and Riak 1.1.1 already provides
scalable and available persistent storage on 60 nodes.  Moreover the
scalability of Riak is much improved in subsequent versions.
\npcomment{This section seems to end with ``all was good when we
  started''.}  \pwtcomment{That's just the point. We investigated, and
  found there was nothing to fix.}



\section{Improving Language Scalability}
\label{sec:language-scalability}
\newcommand{\fgfh}[2]{\big({\it #1}, {\it #2}\big)}   
\newcommand{\gr}[3]{\big({\it #1}, {\it #2}, {\it #3}\big)}   
\newcommand{\grset}[3]{\big\{({\it #1}, {\it #2}, {\it #3})\big\}}   
\newcommand{\inputset}[2]{{\it #1} \in {\it #2}}
\newcommand{\listcomprehension}[2]{\setbig{{\it #1}\ |\ {\it #2}}}
\newcommand{\nd}[4]{\big({\it #1}, {\it #2}, {\it #3}, {\it #4}\big)}   
\newcommand{\nioplusnis}{\{{\it ni}\} \oplus nis}   
\newcommand{\ns}[2]{({\it #1}, {\it #2})}   
\newcommand{\nsset}[2]{\{({\it #1}, {\it #2})\}}   

\newcommand{\setsmall}[1]{\{#1\}}
\newcommand{\setbig}[1]{\big\{#1\big\}}

\newcommand{\state}[4]{\big({\it #1}, {\it #2}, {\it #3}, {\it #4}\big)}   
\newcommand{\stateinit}{({\it grs, fgs, fhs, nds})}   

\newcommand{\statesetting}[3]{#1 & \in \setsmall{#2} \equiv \setbig{#3}}
\newcommand{\statesettinga}[3]{#1 & \in \setsmall{#2} \equiv \setsmall{#3}}

\newcommand{\statesettinglong}[4]{\mathit{#1} & \in \mathit{#2} \equiv \setsmall{\mathit{#3}} \equiv \setbig{\mathit{#4}}}
\newcommand{\statesettingshort}[2]{{#1} & \in {\setsmall{#2}}}

\newcommand{\transstateinit}{({\it grs}, & {\it fgs, fhs, nds})}

\newcommand{\transition}[5]{({\it #1}, {\it #2}, {\it #3}) & \longrightarrow ({\it #4}, {\it #5})}
\newcommand{\transitionhead}[2]{\big(\transstateinit, {\it #1}({\it #2}), {\it ni}\big) & }
\newcommand{\transitionbodyifresultfun}[4]{\longrightarrow\ & \big({\it #1}, {\it #2}) & {\it if\ } {\it #3}({\it #4}\big)}
\newcommand{\transitionbodyotherwise}[2]{\longrightarrow\ & \big({\it #1}, {\it #2}\big) & {\it otherwise}}
\newcommand{\transitionwhere}{{\it where} &}
\newcommand{\transitionmatchfun}[3]{& {\it #1} \equiv {\it #2}({\it #3})}
\newcommand{\transitionmatchgrs}{& {\it \grset{s}{\nioplusnis}{ns} \oplus grs'} \equiv {\it grs}}

\newcommand{\auxhead}[2]{{\it #1} & ({\it #2})}
\newcommand{\auxmatch}[5]{{\it #1}({\it #2})=\exists{\it #3}\ .\ {\it #4} \equiv {\it #5}}
\newcommand{\auxbody}[1]{={\it #1}}

\newcommand{\transitionk}[5]{({\it #1}, {\it #2}, {\it #3}) \longrightarrow ({\it #4}, {\it #5})}


This section outlines the Scalable Distributed (SD) Erlang
libraries~\cite{improving-jpdc-16} we have designed and implemented to
address the distributed Erlang scalability issues identified
in Section~\ref{sec:languagescaling}.  SD Erlang introduces two concepts to
improve scalability. \emph{S\_groups} partition the set of nodes in an
Erlang system to reduce network connectivity and partition global data
(Section~\ref{subsubsec:design-s-groups}). \emph{Semi-explicit placement}
alleviates the issues of explicit process placement in large
heterogeneous networks
(Section~\ref{subsubsec:design-semi-explicit-placement}). The two features
are independent and can be used separately or in combination.  We
overview SD Erlang in Section~\ref{subsec:sd-erlang-design}, and outline
s\_group semantics and validation
in Sections~\ref{subsec:sd-erlang-semantics}, \ref{subsec:sd-erlang-semantics-validation}
respectively.

\subsection{SD Erlang Design}
\label{subsec:sd-erlang-design}

\subsubsection{S\_groups}
\label{subsubsec:design-s-groups}
reduce both the number of connections a node maintains, and the size
of name spaces, i.e.\ they minimise global information. Specifically
names are registered on, and synchronised between, only the nodes
within the s\_group. An s\_group has the following parameters: a name,
a list of nodes, and a list of registered names.
A node can belong to many s\_groups or to none. If a node belongs to
no s\_group it behaves as a usual distributed Erlang node.

\begin{table*}[!t]%
\caption{Summary of \texttt{s\_group} functions.\label{tab:functions-summary}}{%
  \renewcommand{\arraystretch}{1.3}
  \begin{tabular}{@{}p{.5\textwidth}@{}p{.5\textwidth}@{}}\toprule
    \textbf{Function} & \textbf{Description} \\
    \midrule
    \texttt{new\_s\_group(SGroupName, Nodes)} & Creates a new s\_group consisting of some nodes. \\

    \texttt{delete\_s\_group(SGroupName)} & Deletes an s\_group. \\

    \texttt{add\_nodes(SGroupName, Nodes)} & Adds a list of nodes to an s\_group. \\

    \texttt{remove\_nodes(SGroupName, Nodes)} & Removes a list of nodes from an s\_group. \\

    \texttt{s\_groups()} & Returns a list of all s\_groups known to the node. \\

    \texttt{own\_s\_groups()} & Returns a list of s\_group tuples the node belongs to. \\

    \texttt{own\_nodes()} & Returns a list of nodes from all s\_groups the node belongs to. \\
    \texttt{own\_nodes(SGroupName)} & Returns a list of nodes from the given s\_group. \\

    \texttt{info()} & Returns s\_group state information. \\

    \texttt{register\_name(SGroupName, Name, Pid)} & Registers a name in the given s\_group. \\

    \texttt{re\_register\_name(SGroupName, Name, Pid)} & Re-registers a name (changes a registration) in a given s\_group.\\

    \texttt{unregister\_name(SGroupName, Name)} & Unregisters a name in the given s\_group. \\

    \texttt{registered\_names($\{$node,Node$\}$)} & Returns a list of all registered names on the given node. \\
    \texttt{registered\_names($\{$s\_group,SGroupName$\}$)} & Returns a list of all registered names in the given s\_group.  \\

    \texttt{whereis\_name(SGroupName, Name)} & Return the pid of a name registered in the given \\
    \texttt{whereis\_name(Node, SGroupName, Name)} & s\_group. \\

    \texttt{send(SGroupName, Name, Msg)} & Send a message to a name registered in the given \\
    \texttt{send(Node, SGroupName, Name, Msg)} & s\_group. \\
\bottomrule
\end{tabular}}
\end{table*}%

The \texttt{s\_group} library defines the functions shown
in Table~\ref{tab:functions-summary}. Some of these functions manipulate
s\_groups and provide information about them, such as creating s\_groups
and providing a list of nodes from a given s\_group. The remaining
functions manipulate names registered in s\_groups and provide
information about these names. For example, to register a process,
\texttt{Pid}, with name \texttt{Name} in s\_group \texttt{SGroupName}
we use the following function. The name will only be registered if the
process is being executed on a node that belongs to the given
s\_group, and neither \texttt{Name} nor \texttt{Pid} are already
registered in that group.

\begin{verbatim}
s_group:register_name(SGroupName, Name,
                      Pid) -> yes | no
\end{verbatim}

\begin{figure*}[!t]
\centering
\includegraphics[width=0.75\textwidth]{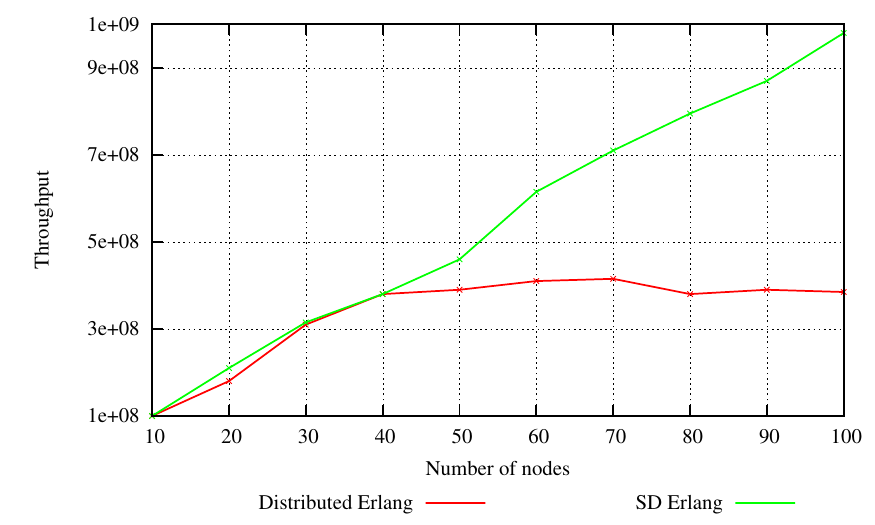}
\caption{Global operations in Distributed Erlang vs.\ SD Erlang.}
\label{fig:d-sd-erlang-global-operations}
\end{figure*}

To illustrate the impact of s\_groups on scalability we repeat the
global operations experiment from Section~\ref{sec:languagescaling}
(Figure~\ref{fig:frequency_of_global_operation}). In the SD Erlang
experiment we partition the set of nodes into s\_groups each
containing ten nodes, and hence 
the names are replicated and synchronised on just ten nodes, and not on
all nodes as in distributed Erlang. The results
in Figure~\ref{fig:d-sd-erlang-global-operations} show that with 0.01\%
of global operations throughput of distributed Erlang stops growing at
40 nodes while throughput of SD Erlang continues to grow linearly.

\begin{figure*}
  \centering
  \includegraphics[trim={0 5pt 0 15pt},clip,width=.98\textwidth]{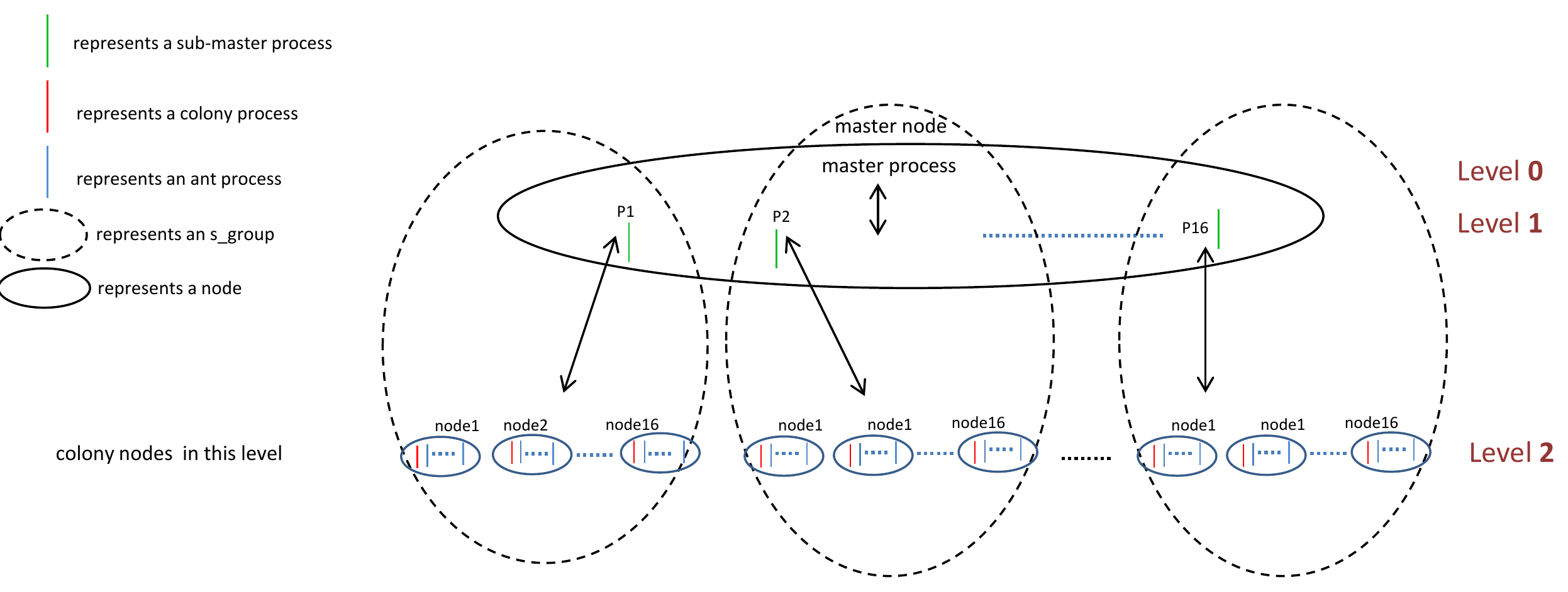}
  \caption{SD Erlang ACO (SR-ACO) architecture.}
  \label{fig:sd-aco-comm-model}
\end{figure*}

The connection topology of s\_groups is extremely flexible: they may
be organised into a hierarchy of arbitrary depth or branching, e.g.\
there could be multiple levels in the tree of s\_groups;
see Figure~\ref{fig:sd-aco-comm-model}. Moreover it is not necessary to
create an hierarchy of s\_groups, for example, we have constructed an
Orbit implementation using a ring of s\_groups.  

Given such a flexible way of organising distributed systems, key
questions in the design of an SD Erlang system are the following.
\textit{How should s\_groups be structured?} Depending on
the reason the nodes are grouped -- reducing the number of
connections, or reducing the namespace, or both -- s\_groups can be
freely structured as a tree, ring, or some other topology. \textit{How
  large should the s\_groups be?} Smaller s\_groups mean more
inter-group communication, but the synchronisation of the s\_group
state between the s\_group nodes constrains the maximum size of
s\_groups.  We have not found this constraint to be a serious
restriction. For example many s\_groups are either relatively small,
e.g.\ 10-node, internal or terminal elements in some topology,
e.g. the leaves and nodes of a tree. \textit{How do nodes from different
  s\_groups communicate?}  While any two nodes can communicate in an
SD Erlang system, to minimise the number of connections communication
between nodes from different s\_groups is typically routed via gateway
nodes that belong to both s\_groups.  \textit{How do we avoid single
  points of failure?}  For reliability, and to minimise communication
load, multiple gateway nodes and processes may be required. 

Information to make these design choices is provided by the tools
in Section~\ref{sec:scalable-tools} and by benchmarking. A further
challenge is how to systematically refactor a distributed Erlang
application into SD Erlang, and this is outlined
in Section~\ref{subsec:refactoring-for-scalability}. A detailed
discussion of distributed system design and refactoring in SD Erlang
provided in a recent article~\cite{evaluating-tpds-16}.

\begin{figure*}
  \centering
  \includegraphics[trim={0 1.7cm 0 1.7cm},clip,width=0.85\textwidth]{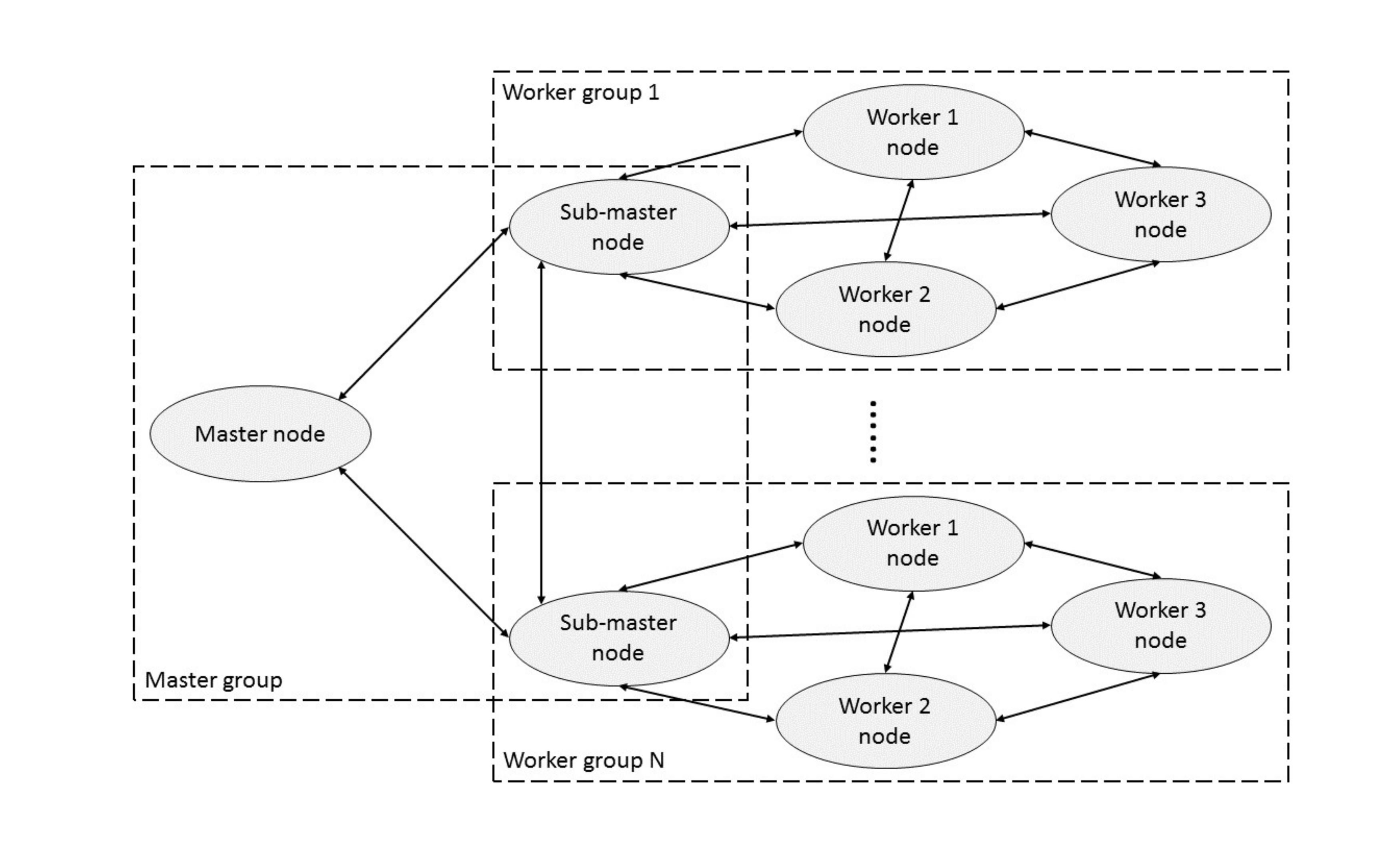}
    \caption{SD Erlang (SD-Orbit) architecture.}
    \label{fig:sd-orbit-comm-model}
\end{figure*}

We illustrate typical SD Erlang system designs by showing refactorings
of the Orbit and ACO benchmarks from Section~\ref{sec:benchmarks}. In
both distributed and SD Erlang the computation starts on the
\textit{Master} node and the actual computation is done on the
\textit{Worker} nodes. In the distributed Erlang version all nodes are
interconnected, and messages are transferred directly from the sending
node to the receiving node (Figure~\ref{fig:d-orbit-comm-model}). In
contrast, in the SD Erlang version nodes are grouped into s\_groups,
and messages are transferred between different s\_groups via
\textit{Sub-master} nodes (Figure~\ref{fig:sd-orbit-comm-model}). 

A fragment of code that
creates an s\_group on a \textit{Sub-master} node is as follows:
\begin{verbatim}
create_s_group(Master,
               GroupName, Nodes0) ->
case s_group:new_s_group(GroupName,
                         Nodes0) of
{ok, GroupName, Nodes} -> 
           Master ! {GroupName, Nodes};
_ -> io:format("exception: message")
end.
\end{verbatim}

Similarly, we introduce s\_groups in the GR-ACO benchmark from
Section~\ref{subsec:aco-overview}  to create Scalable Reliable ACO (SR-ACO);
see Figure~\ref{fig:sd-aco-comm-model}. 
Here, apart from reducing the number of connections, s\_groups also
reduce the global namespace information. That is, instead of
registering the name of a pid globally, i.e.\ with all nodes, the
names is registered only on all nodes in the s\_group with
\texttt{s\_group:register\_name/3}. 

A comparative performance evaluation of distributed Erlang and SD
Erlang Orbit and ACO is presented in Section~\ref{sec:case-studies}.

\subsubsection{Semi-Explicit Placement}
\label{subsubsec:design-semi-explicit-placement}

Recall from Section~\ref{sec:erlang} that distributed Erlang spawns a
process onto an explicitly named Erlang node, e.g.
\begin{verbatim}
spawn(some_node,fun some_module:pong/0).
\end{verbatim}
and also recall the portability and programming effort issues
associated with such explicit placement in large scale systems
discussed in Section~\ref{sec:languagescaling}.

To address these issues we have developed a semi-explicit placement library
that enables the programmer to select nodes on which to spawn
processes based on run-time information about the properties of the
nodes. For example, if a process performs a lot of computation one
would like to spawn it on a node with considerable computation power, or
if two processes are likely to communicate frequently then it would be
desirable to spawn them on the same node, or nodes with a fast interconnect.

We have implemented two Erlang libraries to support semi-explicit
placement \cite{mackenzie-15-performance}. The first deals with
\textit{node attributes}, and describes properties of individual
Erlang VMs and associated hosts, such as total and currently available
RAM, installed software, hardware configuration, etc. The second deals
with a notion of \textit{communication distances} which models the
communication times between nodes in a distributed system.%
\sjt{It's not clear that this is runtime information here, in the
  sense that it \emph{must} be calculated and held at runtime. I had
  the impression that these distances were pre- calculated.}
\pwtcomment{Kenneth please confirm. My understanding is that
  communication distances are static, although load is dynamic.}
\kmcomment{The communication distances are calculated from a static
  description of the network structure (read from a configuration file
  when the library is loaded). Attributes can be either static
  (specified in advance or calculated once at load-time) or dynamic
  (calculated anew every time they're queried).  Maybe we should miss
  out/modify the sentence referring to run-time information.}%
Therefore, instead of specifying a node we can use the
\texttt{attr:choose\_node/1} function to define the target node, i.e.
\begin{verbatim}
spawn(attr:choose_node(Params),
      fun some_module:pong/0).
\end{verbatim}

\cite{mackenzie-15-performance} report an investigation into the
communication latencies on a range of NUMA and cluster architectures,
and demonstrate the effectiveness of the placement libraries using the
ML-ACO benchmark on the Athos cluster.


\subsection{S\_group Semantics}
\label{subsec:sd-erlang-semantics}
For precise specification, and as a basis for validation, we provide a
small-step operational semantics of the s\_group
operations~\cite{improving-jpdc-16}. Figure~\ref{eq:sd-erlang-state}
defines the state of an SD Erlang system and associated abstract
syntax variables. The abstract syntax variables on the left are
defined as members of sets, denoted $\{\}$, and these in turn may
contain tuples, denoted $()$, or further sets. In particular $nm$ is a
process name, $p$ a pid, $ni$ a node\_id, and $nis$ a set of
node\_ids. The state of a system is modelled as a four tuple
comprising a set of $s\_group$s, a set of $free\_group$s, a set of
$free\_hidden\_group$s, and a set of $node$s. Each type of group is
associated with nodes and has a namespace. An $s\_group$ additionally
has a name, whereas a $free\_hidden\_group$ consists of only one node,
i.e.~a hidden node simultaneously acts as a node and as a group,
because as a group it has a namespace but does not share it with any
other node. Free normal and hidden groups have no names, and are
uniquely defined by the nodes associated with them. Therefore, group
names, $gr\_names$, are either $NoGroup$ or a set of
$s\_group\_name$s. A $namespace$ is a set of $name$ and process id,
$pid$, pairs and is replicated on all nodes of the associated group.

A $node$ has the following four parameters: $node\_id$ identifier, $node\_type$ that can be either hidden or normal, $connections$, and $group\_names$, i.e.~names of groups the node belongs to. The node can belong to either a list of s\_groups or one of the free groups. The type of the free group is defined by the node type. Connections are a set of $node\_id$s.

\begin{figure*}[!t]
\begin{align*}
\statesetting{\stateinit}{state}{\state{\setsmall{s\_group}}{\setsmall{free\_group}}{\setsmall{free\_hidden\_group}}{\setsmall{node}}}\\
\statesettinglong{gr}{grs}{s\_group}{\gr{s\_group\_name}{\setsmall{node\_id}}{namespace}}\\
\statesettinglong{fg}{fgs}{free\_group}{ \fgfh{\setsmall{node\_id}}{namespace} }\\
\statesettinglong{fh}{fhs}{free\_hidden\_group}{ \fgfh{node\_id}{namespace} }\\
\statesettinglong{nd}{nds}{node}{ \nd{node\_id}{node\_type}{connections}{gr\_names} }\\
\statesetting{gs}{gr\_names}{NoGroup, \setsmall{s\_group\_name}}\\
\statesetting{ns}{namespace}{ \nsset{name}{pid} }\\
\statesetting{cs}{connections}{\setsmall{node\_id}}\\
\statesettinga{nt}{node\_type}{Normal, Hidden}\\
\statesettingshort{s}{NoGroup, s\_group\_name}
\end{align*}
\caption{SD Erlang state~\cite{improving-jpdc-16}.}
\label{eq:sd-erlang-state}
\end{figure*}

Transitions in the semantics have the form
$
  \transitionk{state}{command}{ni}{state'}{value}
$
meaning that executing $command$ on node $ni$ in $state$ returns
$value$ and transitions to $state'$.

The semantics is presented in more detail by~\cite{improving-jpdc-16}, but we illustrate it here with the \texttt{s\_group:registered\_names/1} function from Section~\ref{subsubsec:design-s-groups}. The function returns a list of names registered in s\_group $s$ if node $ni$ belongs to the s\_group, an empty list otherwise (Figure~\ref{eq:sd-erlang-semantics}).
Here $\oplus$ denotes disjoint set union; \textit{IsSGroupNode}
returns true if node $ni$ is a member of some s\_group $s$, false
otherwise; and \textit{OutputNms} returns a set of process names
registered in the $ns$ namespace of s\_group $s$.

\begin{figure*}[!t]
\begin{align*}
\transitionhead{\mathrm{s\_group:registered\_names}}{\mathrm{s}}\\
\transitionbodyifresultfun{\stateinit}{nms}{IsSGroupNode}{ni,s,grs}\\
\transitionbodyotherwise{\stateinit}{\setsmall{}}\\
\transitionwhere\\
\transitionmatchgrs\\
\transitionmatchfun{nms}{OutputNms}{s,ns}
\end{align*}
\begin{align*}
\auxmatch{IsSGroupNode}{ni, s, grs}{nis, ns, grs'}{\grset{s}{\nioplusnis}{ns} \oplus grs'}{grs}
\end{align*}
\begin{align*}
\auxhead{OutputNms}{s, ns} \auxbody{ \listcomprehension{(s, nm)}{ \inputset{ \ns{nm}{p} }{ns} } }
\end{align*}
\caption{SD Erlang Semantics of \texttt{s\_group:registered\_names/1} Function}
\label{eq:sd-erlang-semantics}
\end{figure*}


\subsection{Semantics Validation}
\label{subsec:sd-erlang-semantics-validation}

\begin{figure}[!b]
\centering
\includegraphics[width=0.48\textwidth]{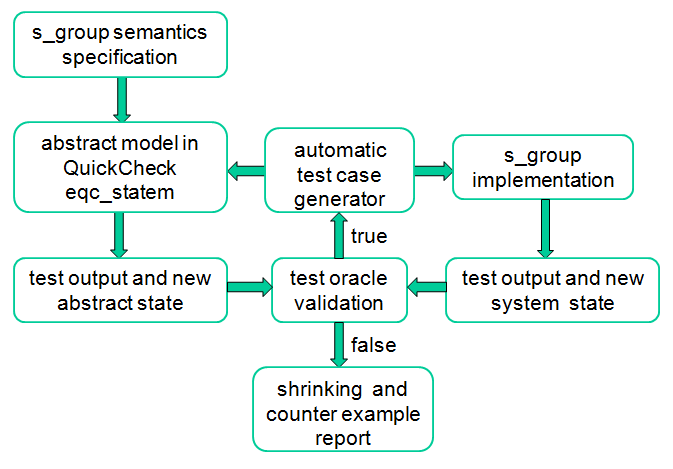}
\caption{Testing s\_groups using QuickCheck.}
\label{fig:semantics-validation}
\end{figure}

As the semantics is concrete it can readily be made \emph{executable}
in Erlang, with lists replacing sets throughout. Having an executable
semantics allows users to engage with it, and to understand how the
semantics behaves \emph{vis \`a vis} the library, giving them an
opportunity to assess the correctness of the library against the
semantics.

Better still, we can automatically assess how the system behaves in
comparison with the (executable) semantics by executing them in
lockstep, guided by the constraints of which operations are possible
at each point. We do that by building an abstract state machine model
of the library. We can then generate random sequences (or traces)
through the model, with appropriate library data generated too. This
random generation is supported by the QuickCheck property-based
testing system~\cite{quickCheck,TestingTelecoms}.

The architecture of the testing framework is shown in Figure~\ref{fig:semantics-validation}. First an abstract state machine embedded as an ``\texttt{eqc\_statem}'' module is derived from the executable semantic specification.
The state machine defines the abstract state representation and the transition
from one state to another when an operation is applied.  Test case and data generators are then defined to control the test case generation; this includes the automatic generation of eligible s\_group operations and the input data to those operations. Test oracles are encoded as the postcondition for s\_group operations.

During testing, each test command is applied to both the abstract
model and the  s\_group library. The application of the
test command to the abstract model takes the abstract model from its
current state to a new state as described by the transition functions;
whereas the application of the test command to the library leads
the system to a new actual state. The actual state information is
collected from each node in the distributed system, then merged and
normalised to the same format as the abstract state
representation. For a test to be successful, after the execution of a
test command, the test oracles specified for this command should be
satisfied. Various test oracles can be defined for s\_group
operations; for instance one of the generic constraints that applies
to all the s\_group operations is that after each s\_group operation,
the normalised system state should be equivalent to the abstract
state.

Thousands of tests were run, and three kinds of errors --- which have
subsequently been corrected --- were found. Some errors in the library
implementation were found, including one error due to the
synchronisation between nodes, and the other related to the
\texttt{remove\_nodes} operation, which erroneously raised an
exception. We also found a couple of trivial errors in the semantic
specification itself, which had been missed by manual
examination. Finally, we found some situations where there were
inconsistencies between the semantics and the library implementation,
despite their states being equivalent: an example of this was in the
particular values returned by functions on certain errors.
Overall, the automation of testing boosted our confidence in the
correctness of the library implementation and the semantic specification.
This work is reported in more detail by~\cite{improved-semantics}.


\section{Improving VM Scalability}
\label{sec:improving-vm-scalability}
\sjt{A general comment on this section. It contains statements of (a) the status quo ante and (b) improvements. In principle the (a) parts could go into the ``context'' part -- section 2 -- and the improvements into here. On the other hand, if that's a chore, we could put forward references into Section 2 explaining that some of the background is explained in the body of the paper, namely at blah ...}
\npcomment{I think that the structure of the paper is complicated enough and it would become more so if we moved things around or added more forward references.  The only ``status quo ante'' here is some information about timers and schedulers.  It is not quantitative, so it cannot go to Section 4, nor is it so important to become new subsection(s) in Section 2.  I suggest that we leave it.}
\pwtcomment{I agree with Nikos}

This section reports the primary VM and library improvements we have
designed and implemented to address the scalability and reliability
issues identified in Section~\ref{sec:VMScaling}.

\subsection{Improvements to Erlang Term Storage}
\label{sec:vm-ets}
Because ETS tables are so heavily used in Erlang systems, they are a
focus for scalability improvements. We start by describing their
redesign, including some improvements that pre-date our RELEASE
project work, i.e.\ those prior to Erlang/OTP R15B03. These historical
improvements are very relevant for a scalability study and form the
basis for our subsequent changes and improvements.  At the point when
Erlang/OTP got support for multiple cores (in release R11B), there was a single
reader-writer lock for each ETS table. Optional fine grained locking
of hash-based ETS tables (i.e. \texttt{set}, \texttt{bag} or
\texttt{duplicate\_bag} tables) was introduced in Erlang/OTP~R13B02-1,
adding 16 reader-writer locks for the hash buckets.  Reader groups to
minimise read synchronisation overheads were introduced in
Erlang/OTP~R14B. The key observation is that a single count of the
multiple readers must be synchronised across many cache lines,
potentially far away in a NUMA system.  Maintaining reader counts in
multiple (local) caches makes reads fast, although writes must now
check every reader count. In Erlang/OTP R16B the number of bucket
locks, and the default number of reader groups, were both upgraded
from 16 to 64.

We illustrate the scaling properties of the ETS concurrency options
using the \etsbench BenchErl benchmark on the Intel NUMA machine
with 32 hyperthreaded cores specified in Appendix~A.  The \etsbench
benchmark inserts 1M items into the table, then records the time to
perform 17M operations, where an operation is either a lookup, an
insert, or a delete. The experiments are conducted on a hash-based
(\texttt{set}) ETS table with different percentages of update
operations, i.e.\ insertions or deletions.

\begin{figure}[!t]
\centering
{\def\svgwidth{.49\textwidth}%
	\executeiffilenewer{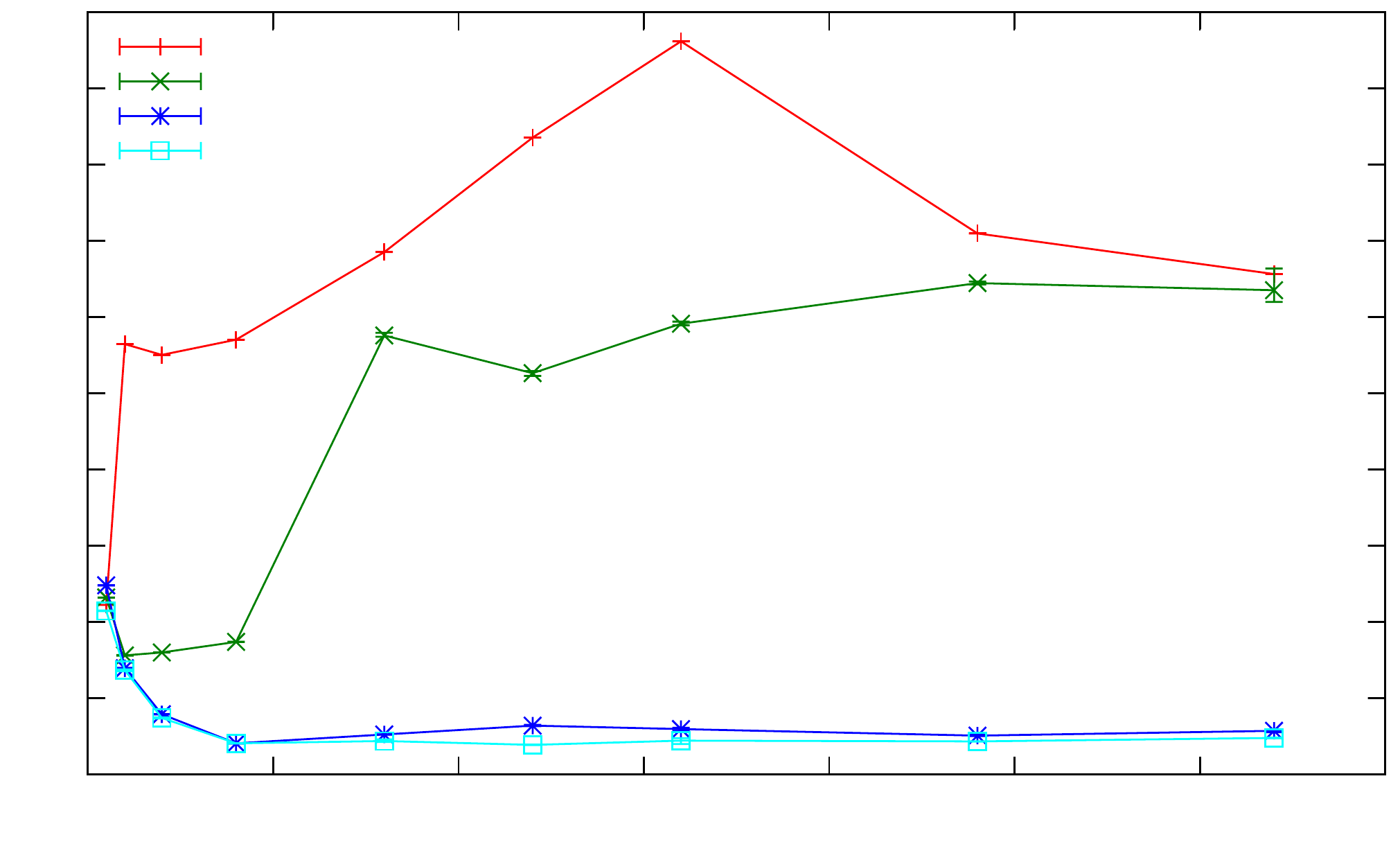}{\svgpath/different_versions_mixed_90_per_set.pdf}%
	{inkscape -z -D --file=\svgpath/different_versions_mixed_90_per_set.svg %
	--export-pdf=\svgpath/different_versions_mixed_90_per_set.pdf --export-latex}%
	{\scriptsize\input{\svgpath/different_versions_mixed_90_per_set.pdf_tex}}%
}\\%
{\def\svgwidth{.49\textwidth}%
	\executeiffilenewer{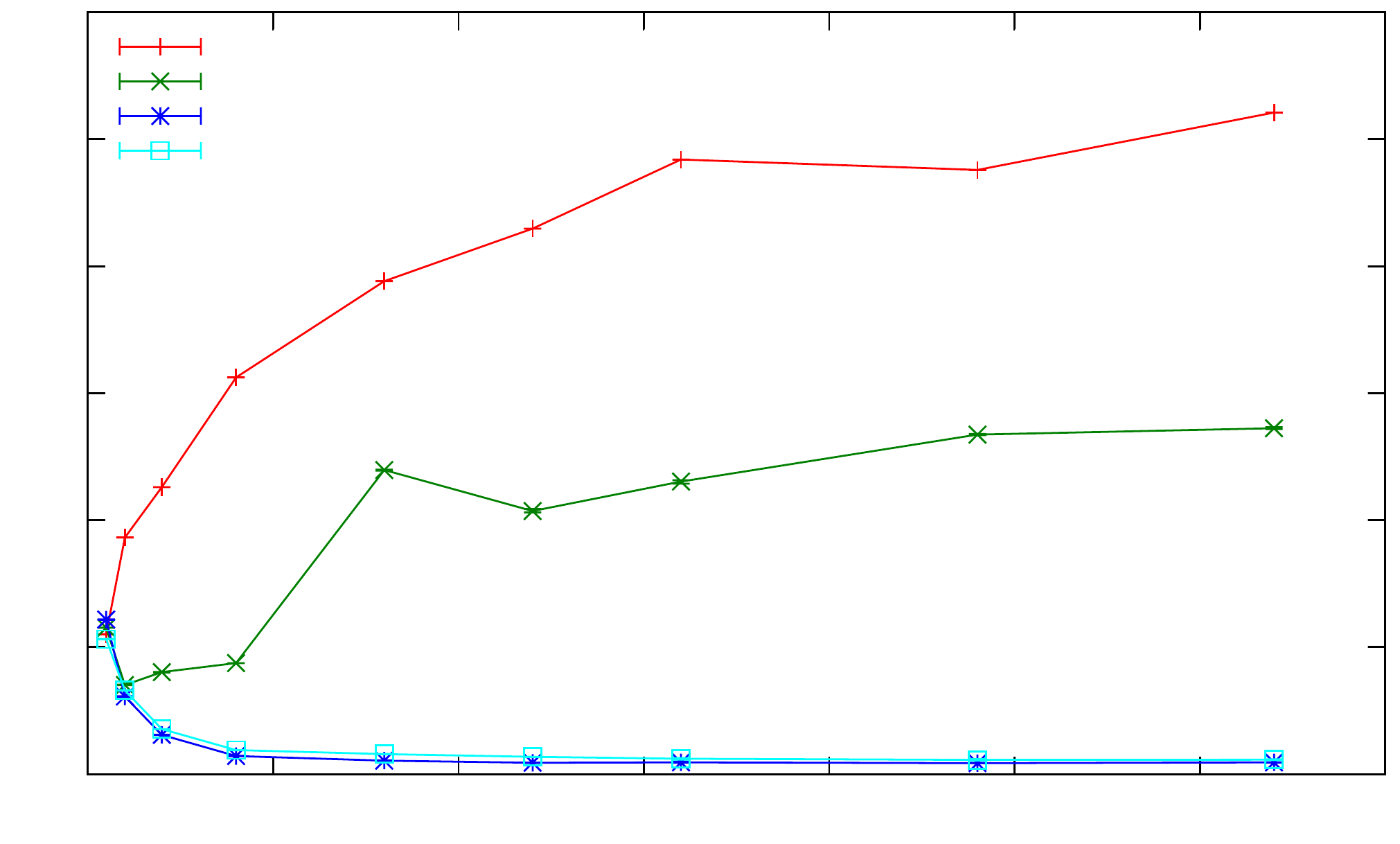}{\svgpath/different_versions_mixed_99_per_set.pdf}%
	{inkscape -z -D --file=\svgpath/different_versions_mixed_99_per_set.svg %
	--export-pdf=\svgpath/different_versions_mixed_99_per_set.pdf --export-latex}%
	{\scriptsize\input{\svgpath/different_versions_mixed_99_per_set.pdf_tex}}%
}
  \caption{Runtime of 17M ETS operations in Erlang/OTP releases: 10\% updates (left) and 1\% updates (right). The y-axis shows time in seconds (lower is better) and the x-axis is number of OS threads (VM schedulers).}
  \label{fig:releases_set}
\end{figure}

\begin{figure}[!t]
\centering
{\def\svgwidth{.49\textwidth}%
	\executeiffilenewer{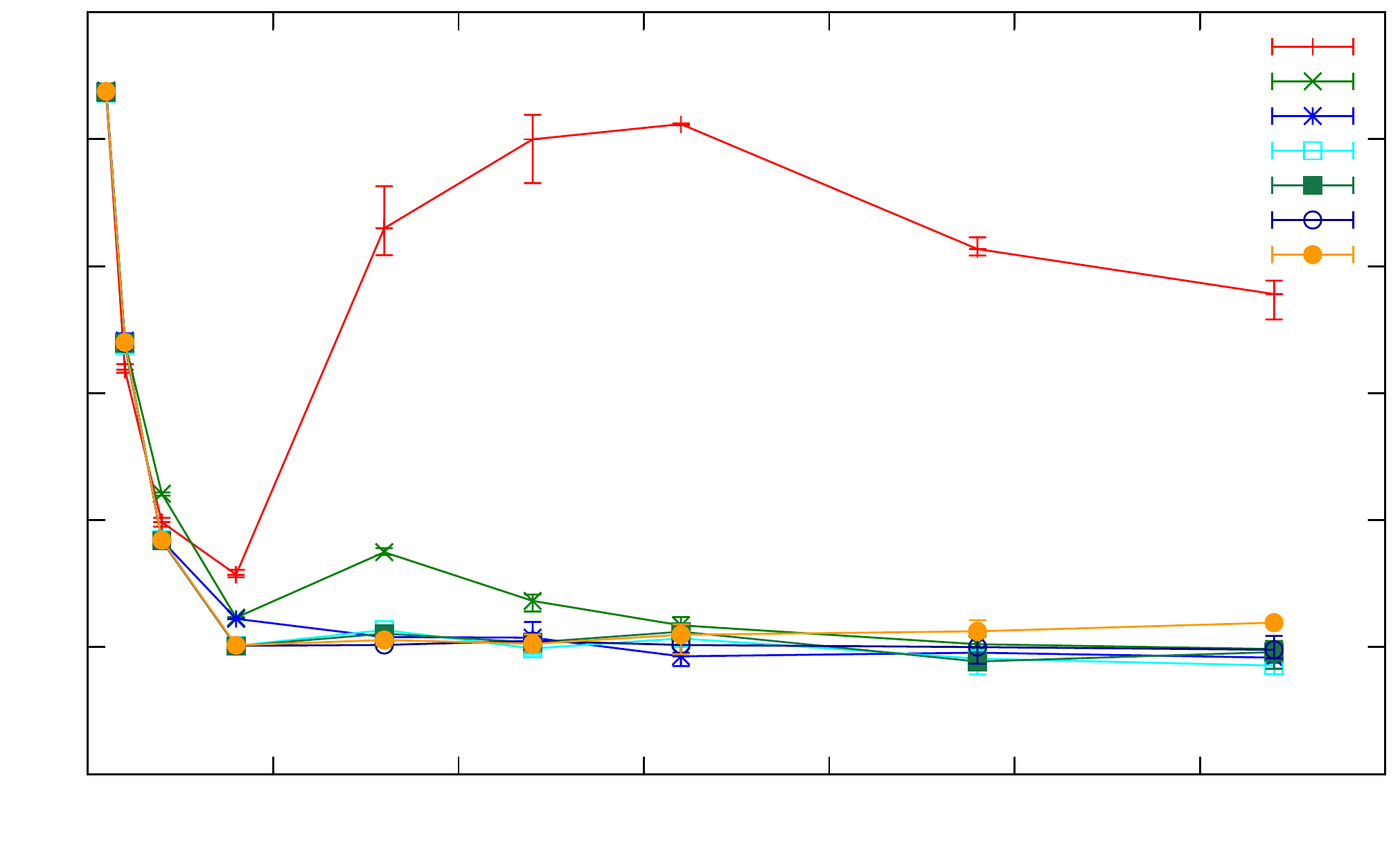}{\svgpath/rg_set_90.pdf}%
	{inkscape -z -D --file=\svgpath/rg_set_90.svg %
	--export-pdf=\svgpath/rg_set_90.pdf --export-latex}%
	{\scriptsize\input{\svgpath/rg_set_90.pdf_tex}}%
}\\%
{\def\svgwidth{.49\textwidth}%
	\executeiffilenewer{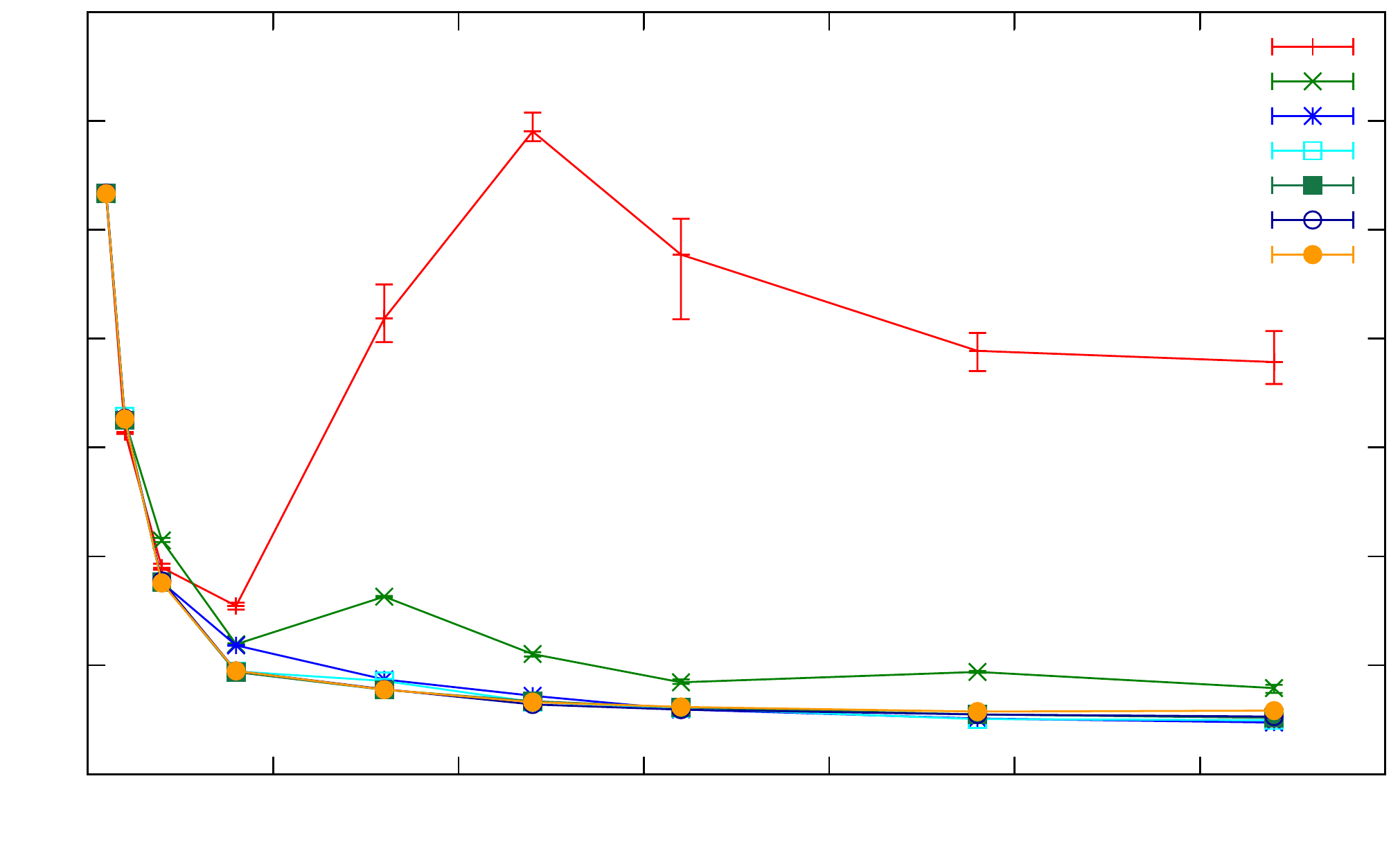}{\svgpath/rg_set_99.pdf}%
	{inkscape -z -D --file=\svgpath/rg_set_99.svg %
	--export-pdf=\svgpath/rg_set_99.pdf --export-latex}%
	{\scriptsize\input{\svgpath/rg_set_99.pdf_tex}}%
}
  \caption{Runtime of 17M ETS operations with varying numbers of reader groups: 10\% updates (left) and 1\% updates (right). The y-axis shows runtime in seconds (lower is better) and the x-axis is number of schedulers.}
  \label{fig:rg_set}
\end{figure}

Figure~\ref{fig:releases_set} shows the runtimes in seconds of 17M
operations in different Erlang/OTP versions, varying the number of
schedulers (x-axis), reflecting how the scalability of ETS tables has
improved in more recent Erlang/OTP releases.  Figure~\ref{fig:rg_set} shows
the runtimes in seconds of 17M operations on an ETS table with
different numbers of reader groups, again varying the number of
schedulers.  We see that one reader group is not sufficient with 10\%
updates, nor are two with 1\% updates.  Beyond that, different numbers
of reader groups have little impact on the benchmark performance
except that using 64 groups with 10\% updates slightly degrades performance.

We have explored four other extensions or redesigns in the ETS
implementation for better scalability.
\begin{inparaenum}
\item
  Allowing more programmer control over the number of bucket locks in
  hash-based tables, so the programmer can reflect the number of
  schedulers and the expected access pattern.
\item
  Using contention-adapting trees to get better scalability
  for \texttt{ordered\_set} ETS tables as described
  by~\cite{ScalableETS@Erlang-14}.
\item
  Using queue delegation locking to improve scalability~\cite{QDEuroPar}.
\item
  Adopting schemes for completely eliminating the locks in the meta
  table.
\end{inparaenum}
A more complete discussion of our work on ETS can be found
in the papers by~\cite{ScalableETS@Erlang-14} and~\cite{QDEuroPar}.

\begin{figure}[!t]
\centering
  \includegraphics[width=.49\textwidth]{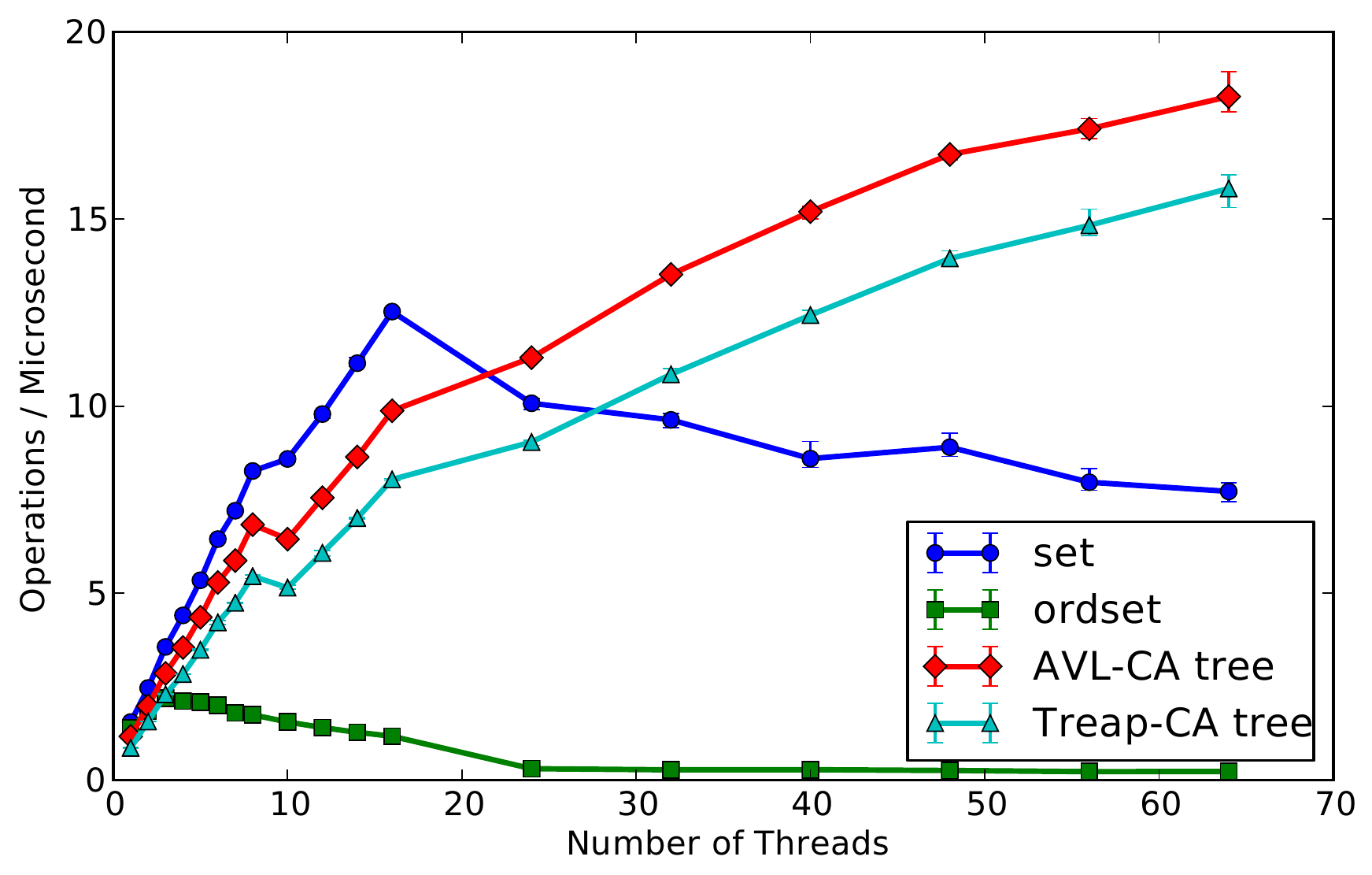}\hspace{0em}\\%
  \includegraphics[width=.49\textwidth]{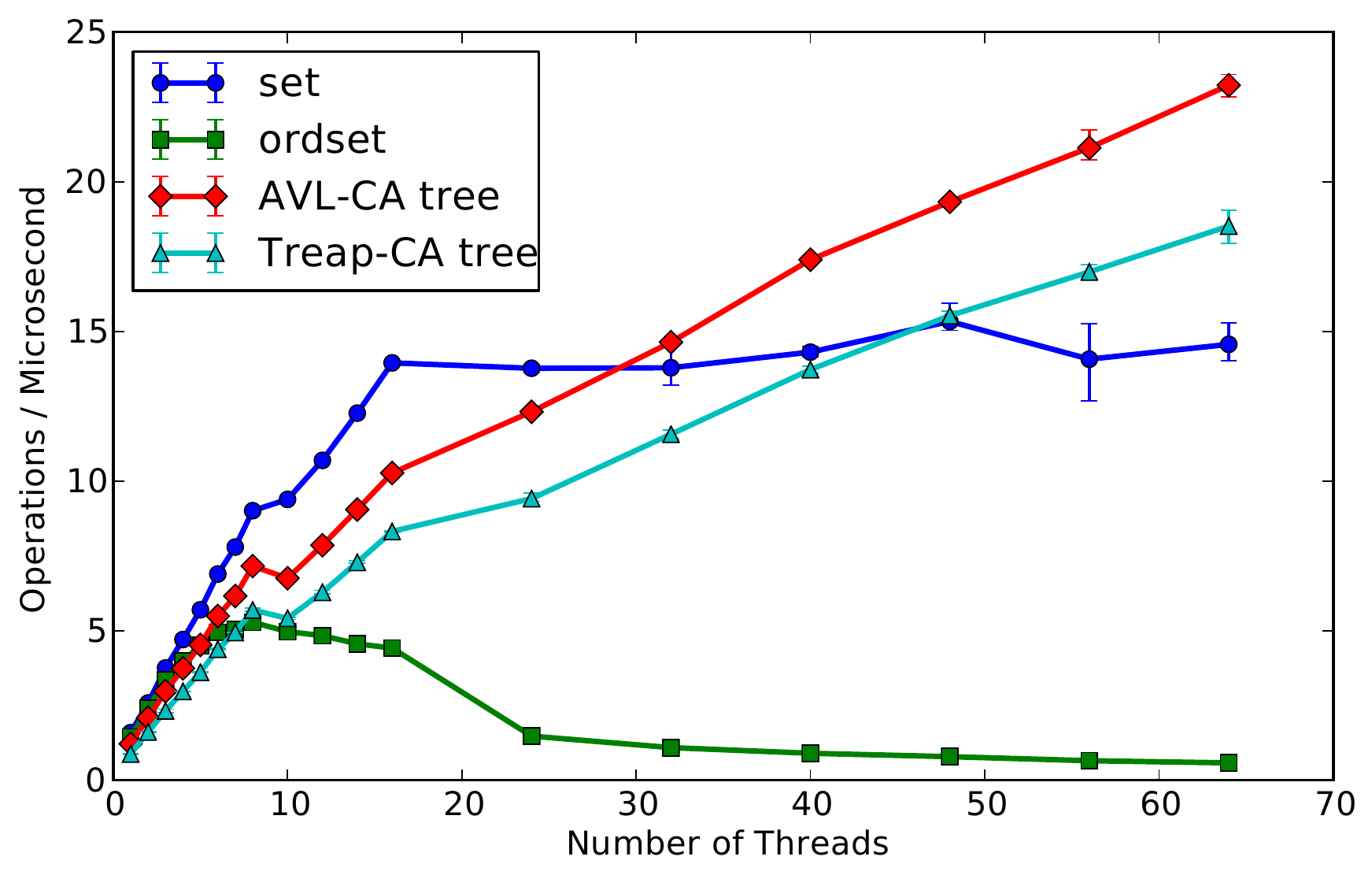}\hspace{0em}%
  \caption{Throughput of CA tree variants: 10\% updates (left) and 1\% updates (right).}
  \label{fig:CAtrees}
\end{figure}

Here we outline only our work on contention-adapting (CA) trees. A CA
tree monitors contention in different parts of a tree-shaped data
structure, introducing routing nodes with
locks in response to high contention, and removing them in response to
low contention. For experimental purposes two variants of the CA tree
have been implemented to represent \texttt{ordered\_set}s in the
virtual machine of Erlang/OTP~17.0.  One extends the existing AVL
trees in the Erlang VM, and the other uses a \emph{Treap} data
structure~\cite{Treap}.  Figure~\ref{fig:CAtrees} compares the
throughput of the CA tree variants with that of \texttt{ordered\_set}
and \texttt{set} as the number of schedulers increases. It is
unsurprising that the CA trees scale so much better than an
\texttt{ordered\_set} which is protected by a single readers-writer lock.
It is more surprising that they also scale better than \texttt{set}.
This is due to hash tables using fine-grained locking at a fixed
granularity, while CA trees can adapt the number of locks to the
current contention level, and also to the parts of the key range where
contention is occurring.

\subsection{Improvements to Schedulers}
\label{sec:vm-schedulers}

In the Erlang VM, a scheduler is responsible for executing multiple
processes concurrently, in a timely and fair fashion, making optimal
use of hardware resources.
The VM implements preemptive multitasking with soft real-time guarantees.
Erlang processes are normally scheduled on a reduction count basis
where one reduction is roughly equivalent to a function call. Each
process is allowed to execute until it either blocks waiting for
input, typically a message from some other process, or until it has
executed its quota of reductions.

The Erlang VM is usually started with one scheduler per logical core
(SMT-thread) available on the host machine, and schedulers are
implemented as OS threads.  When an Erlang process is spawned it is
placed in the run queue of the scheduler of its parent, and it waits
on that queue until the scheduler allocates it a slice of core time.
Work stealing is used to balance load between cores, that is an idle
scheduler may \emph{migrate} a process from another run queue.
Scheduler run queues are visualised in Figure~\ref{sec:dtrace}.

The default load management mechanism is load compaction that aims to
keep as many scheduler threads as possible fully loaded with work,
i.e.\ it attempts to ensure that scheduler threads do not run out of work.
We have developed a new optional \emph{scheduler utilisation balancing}
mechanism that is available from Erlang/OTP~17.0.
The new mechanism aims to balance scheduler utilisation between
schedulers; that is, it will strive for equal scheduler utilisation on all
schedulers.

The scheduler utilisation balancing mechanism has no performance impact on
the system when not enabled. On the other hand, when enabled, it results in
changed timing in the system; normally there is a small overhead due to
measuring of utilisation and calculating balancing information, which
depends on the underlying primitives provided by the operating system. 

The new balancing mechanism results in a better distribution of
processes to schedulers, reducing the probability of core contention.
Together with other VM improvements, such as interruptable BIFs and
garbage collection, it results in lower latency and improved
responsiveness, and hence reliability, for soft real-time
applications.  \pwtcomment{Is this reliability or responsiveness, or
both?}

\subsection{Improvements to Time Management}
\label{sec:vm-scalability-time}

Soon after the start of the RELEASE project, time management in the
Erlang VM became a scalability bottleneck for many applications, as
illustrated by the \bench{parallel} benchmark in Figure~\ref{sec:VMScaling}.
The issue came to prominence as other, more severe, bottlenecks were
eliminated.  This subsection motivates and outlines the improvements
to time management that we made; these were incorporated into
Erlang/OTP 18.x as a new API for time and time warping. The old API is
still supported at the time of writing, but its use is deprecated.

The original time API provides the \texttt{erlang:now/0} built-in that
returns ``Erlang system time'' or time since Epoch with micro second
resolution.  This time is the basis for all time internally in the
Erlang VM. 

Many of the scalability problems of \texttt{erlang:now/0} stem from
its specification, written at a time when the Erlang VM was not
multi-threaded, i.e.\ SMT-enabled. The documentation promises that
values returned by it are strictly increasing and many applications
ended up relying on this. For example applications often employ
\texttt{erlang:now/0} to generate unique integers.

Erlang system time should align with the operating system's
view of time since Epoch or ``OS system time''. However, while OS
system time can be freely changed both forwards and backwards, Erlang
system time cannot, without invalidating the strictly increasing value
guarantee. The Erlang VM therefore contains a mechanism that slowly
adjusts Erlang system time towards OS system time if they do not align.

One problem with time adjustment is that the VM deliberately presents
time with an inaccurate frequency; this is required to align Erlang
system time with OS system time smoothly when these two have deviated,
e.g. in the case of clock shifts when leap seconds are inserted or deleted.
Another problem is that Erlang system time and OS system time can
differ for very long periods of time.  In the new API, we resolve this
using a common OS technique~\cite{High-resolution-timer-API}, i.e. a
monotonic time that has its zero point at some unspecified point in
time. Monotonic time is not allowed to make leaps forwards and
backwards while system time is allowed to do this.  Erlang system time
is thus just a dynamically varying offset from Erlang monotonic time.

\paragraph{Time Retrieval}
Retrieval of Erlang system time was previously protected by a global
mutex, which made the operation thread safe, but scaled poorly.
Erlang system time and Erlang monotonic time need to run at the same
frequency, otherwise the time offset between them would not be constant.
In the common case, monotonic time delivered by the operating system
is solely based on the machine's local clock and cannot be changed,
while the system time is adjusted using the Network Time Protocol (NTP).
That is, they will run with different frequencies. Linux is an
exception with a monotonic clock that is NTP adjusted and runs with
the same frequency as system time~\cite{Linux-timers}.
To align the frequencies of Erlang monotonic time and Erlang system
time, we adjust the frequency of the Erlang monotonic clock.  This is
done by comparing monotonic time and system time delivered by the OS,
and calculating an adjustment.
To achieve this scalably, one VM thread calculates the time adjustment
to use at least once a minute. If the adjustment needs to be changed,
new adjustment information is published and used to calculate Erlang
monotonic time in the future.

When a thread needs to retrieve time, it reads the monotonic time
delivered by the OS and the time adjustment information previously
published and calculates Erlang monotonic time.
To preserve monotonicity it is important that all threads that read
the same OS monotonic time map this to exactly the same Erlang
monotonic time. This requires synchronisation on updates to the
adjustment information using a readers-writer (RW) lock. This RW lock
is write-locked only when the adjustment information is changed. This
means that in the vast majority of cases the RW lock will be
\mbox{read-locked}, which allows multiple readers to run concurrently.
To prevent bouncing the lock cache-line we use a bespoke reader
optimised RW lock implementation where reader threads notify about
their presence in counters on separate cache-lines. The concept  is
similar to the reader indicator algorithm described
by~\cite[Fig.~11]{QDL-TPDS} and alternatives include the ingress-egress
counter used by~\cite{NUMA-aware@PPoPP-13} and the SNZI algorithm
of~\cite{SNZI@PODC-07}.

\paragraph{Timer Wheel and BIF Timer}
The timer wheel contains all timers set by Erlang processes. The
original implementation was protected by a global mutex and scaled
poorly. To increase concurrency, each scheduler thread has been assigned
its own timer wheel that is used by processes executing on the scheduler.

The implementation of timers in Erlang/OTP uses a built in function
(BIF), as most low-level operations do.
Until Erlang/OTP 17.4, this BIF was also protected by a global mutex.
Besides inserting timers into the timer wheel,
the BIF timer implementation also maps timer references to a
timer in the timer wheel.
To improve concurrency, from  Erlang/OTP 18 we provide scheduler-specific
BIF timer servers as Erlang processes. These keep information about
timers in private ETS tables and only insert one timer at the time
into the timer wheel.

\paragraph{Benchmarks}
We have measured several benchmarks on a 16-core Bulldozer machine with
eight dual CPU AMD Opteron 4376 HEs.\footnote{See \S 2.5.4
  of the RELEASE project Deliverable 2.4
  (\url{http://release-project.eu/documents/D2.4.pdf}).} 
We present three of them here.

The first micro benchmark compares the execution time of an Erlang
\texttt{receive} with that of a \texttt{receive\ after} that specifies
a timeout and provides a default value. The \texttt{receive\ after}
sets a timer when the process blocks in the \texttt{receive}, and
cancels it when a message arrives.  In Erlang/OTP 17.4 the total
execution time with standard timers is 62\% longer than without
timers. Using the improved implementation in Erlang/OTP 18.0, total
execution time with the optimised timers is only 5\% longer than
without timers.

The second micro benchmark repeatedly checks the system time, calling
the \mbox{built-in} \texttt{erlang:now/0} in Erlang/OTP 17.4, and calling
both \texttt{erlang:monotonic\_time/0} and
\texttt{erlang:time\_offset/0} and adding the results in Erlang/OTP
18.0. In this machine, where the VM uses 16 schedulers by default,
the 18.0 release is more than 69 times faster than the 17.4 release.

\begin{figure}[!t]
  \centering
  \includegraphics[width=0.49\textwidth,height=.23\textheight]{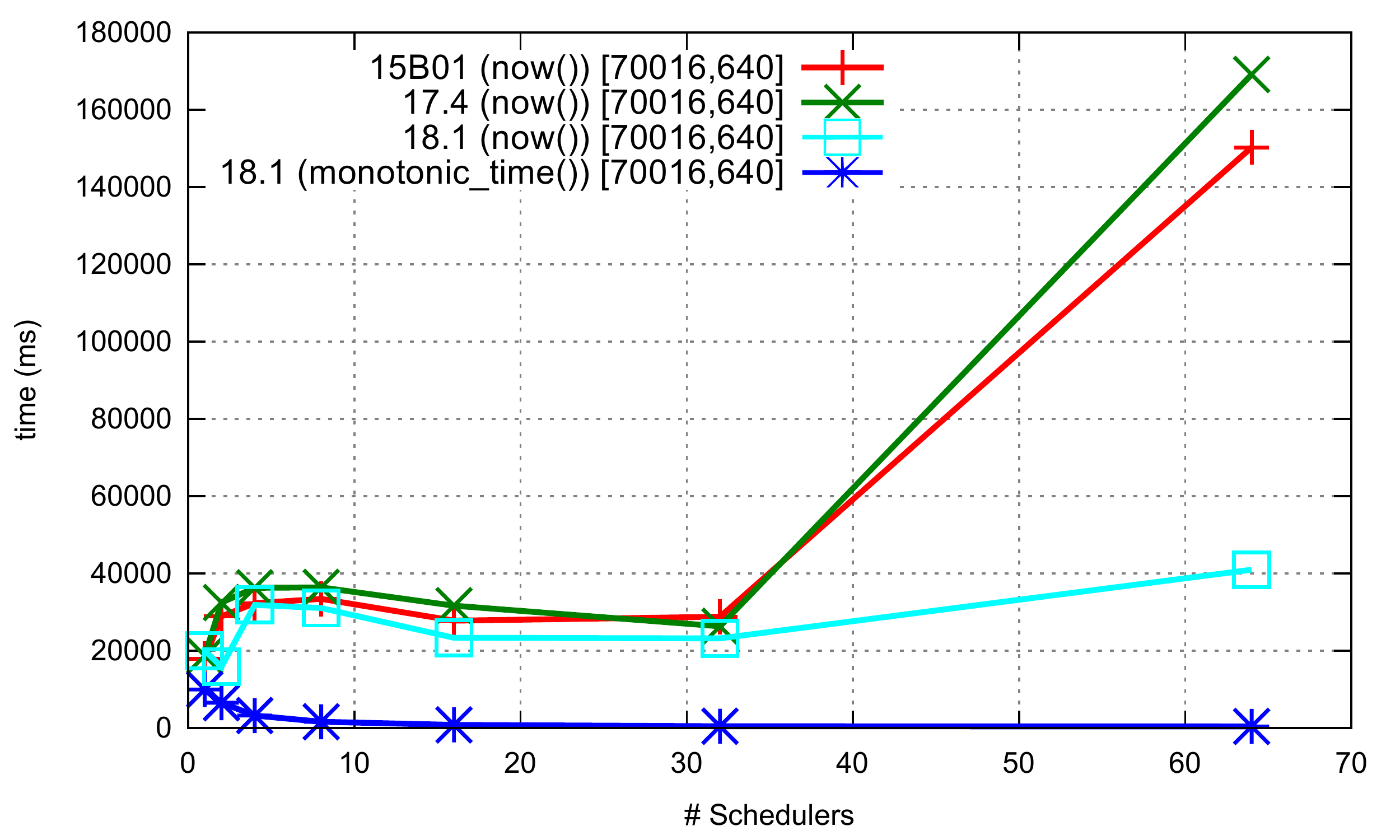}
  \hfill
  \includegraphics[width=0.49\textwidth,height=.23\textheight]{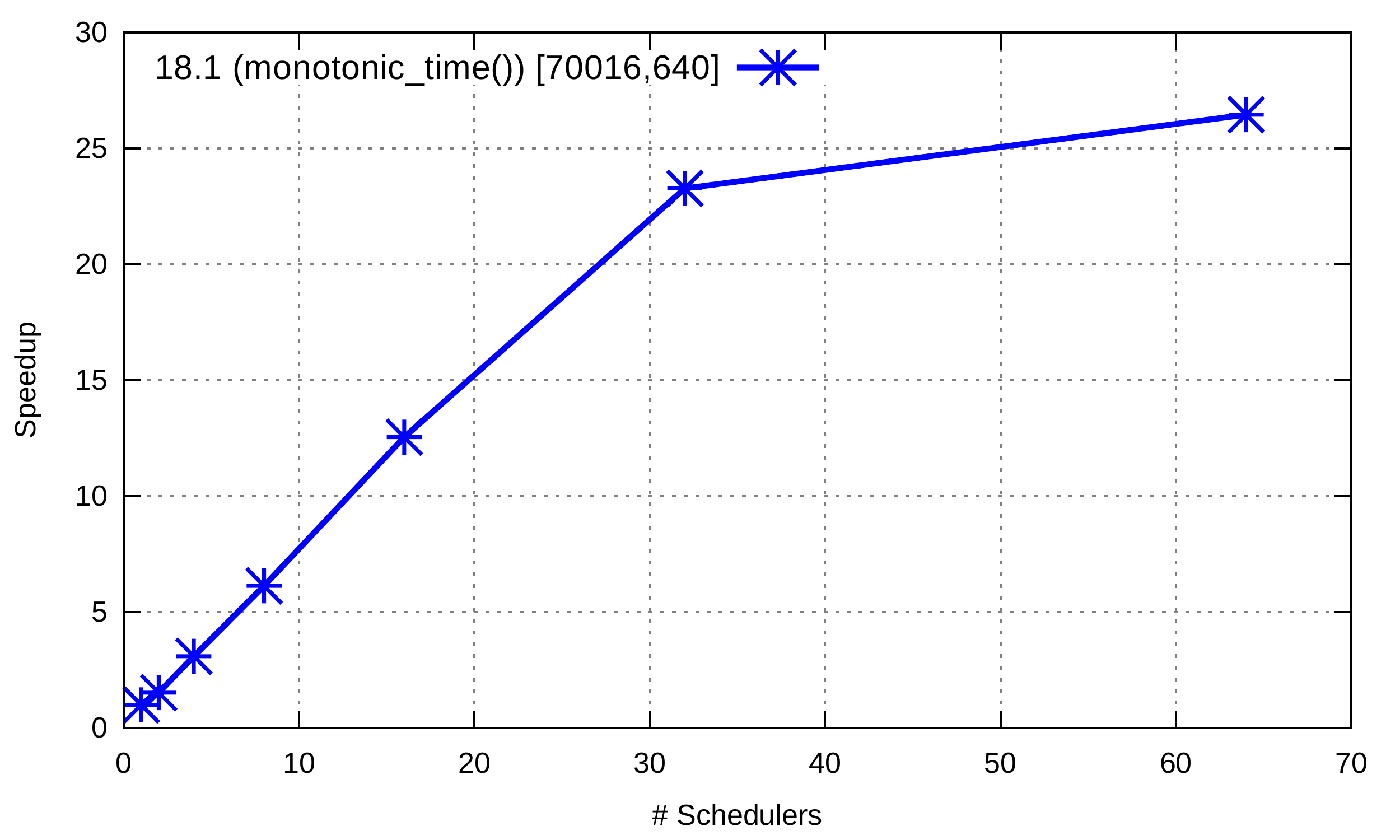}
  \caption{\bencherl{} \bench{parallel} benchmark using
    \texttt{erlang:monotonic\_time/0} or \texttt{erlang:now/0} in
    different Erlang/OTP releases: runtimes (left) and speedup
    obtained using \texttt{erlang:monotonic\_time/0} (right).}
\label{fig:parallel-OTPs}  
\end{figure}

The third benchmark is the \bench{parallel} \bencherl benchmark
from Section~\ref{sec:VMScaling}.
Figure~\ref{fig:parallel-OTPs} shows the results of executing  the
original version of this benchmark, which uses \texttt{erlang:now/0}
to create monotonically increasing unique values, using three Erlang/OTP
releases: R15B01, 17.4, and~18.1. We also measure a version of the
benchmark in Erlang/OTP~18.1 where the call to \texttt{erlang:now/0}
has been substituted with a call to \texttt{erlang:monotonic\_time/0}.
The graph on its left shows that:
\begin{inparaenum}
\item the performance of time management has remained roughly
  unchanged between Erlang/OTP releases prior to~18.0;
\item the improved time management in Erlang/OTP~18.x 
  make time management less likely to be a scalability bottleneck even
  when using \texttt{erlang:now/0}, and
\item the new time API (using \texttt{erlang:monotonic\_time/0} and
  friends) provides a scalable solution.
\end{inparaenum}
The graph on the right side of Figure~\ref{fig:parallel-OTPs} shows the
speedup that the modified version of the \bench{parallel} benchmark
achieves in Erlang/OTP~18.1.


\section{Scalable Tools}
\label{sec:scalable-tools}
This section outlines five tools developed in the
RELEASE project to support scalable Erlang systems. Some tools were
developed from scratch, like Devo, SDMon and \wombat, while others extend
existing tools, like Percept and Wrangler. These include tooling to
transform programs to make them more scalable, to deploy them for
scalability, to monitor and visualise them. Most of the tools are
freely available under open source licences (Devo, Percept2, SD-Mon, Wrangler); while WombatOAM is a commercial product. The tools have been used for profiling and refactoring the
ACO and Orbit benchmarks from Section~\ref{sec:benchmarks}.

The Erlang tool ``ecosystem'' consists of small stand-alone tools for
tracing, profiling and debugging Erlang systems that can be used
separately or together as appropriate for solving the problem at
hand, rather than as a single, monolithic, super-tool. The tools
presented here have been designed to be used as part of that
ecosystem, and to complement already available functionality rather
than to duplicate it. The Erlang runtime system \pwtcomment{How is this different from the VM?} has built-in support
for tracing many types of events, and this infrastructure forms the
basis of a number of tools for tracing and profiling. Typically the tools build on or
specialise the services offered by the Erlang virtual machine, through
a number of built-in functions. Most recently, and since the RELEASE
project was planned, the
Observer\footnote{\url{http://www.erlang.org/doc/apps/observer/}}
application gives a comprehensive overview of many of these data on a
node-by-node basis.

As actor frameworks and languages (see Section~\ref{sec:actor-langs}) have only recently become widely
adopted commercially, their tool support remains relatively immature and  generic in
nature. That is, the tools support the language itself, rather than its distinctively concurrent aspects. Given the widespread use of Erlang, tools developed for it
point the way for tools in other actor languages and frameworks.
For example, just as many Erlang tools use  tracing support provided
by the Erlang VM, so can other actor frameworks, e.g.\  Akka can use the
Kamon\footnote{\url{http://kamon.io}} JVM monitoring system.
Similarly, tools for other actor languages or frameworks could use data derived
through OS-level tracing frameworks DTrace\footnote{\url{http://dtrace.org/blogs/about/}} and SystemTap\footnote{\url{https://sourceware.org/systemtap/wiki}} probes as we show in this section for Erlang, provided that the
host language has tracing hooks into the appropriate infrastructure.


\subsection{Refactoring for Scalability}
\label{subsec:refactoring-for-scalability}
Refactoring~\cite{OpdykeThesis,Fowler99,thompson-li-jfp} is the
process of changing how a program works without changing what it
does. This can be done for readability, for testability, to prepare it
for modification or extension, or --- as is the case here --- in order
to improve its scalability.  Because refactoring involves the
transformation of source code, it is typically performed using machine
support in a refactoring tool. There are a number of tools that
support refactoring in Erlang: in the RELEASE project we have chosen
to extend
Wrangler\footnote{\url{http://www.cs.kent.ac.uk/projects/wrangler/}}~\cite{li2008refactoring};
other tools include Tidier~\cite{Tidier} and
RefactorErl~\cite{RefactorErl}.

\paragraph{Supporting API Migration}

The SD Erlang libraries modify Erlang's \texttt{global\_group} library, becoming the new \texttt{s\_group} library; as a result, Erlang programs using \texttt{global\_group} will have to be refactored to use \texttt{s\_group}. This kind of API migration problem is not uncommon, as software evolves and this often changes the API of a library. Rather than simply extend Wrangler with a refactoring to perform this particular operation, we instead added a  framework for the automatic generation of API migration refactorings from a user-defined adaptor module.

Our approach to automatic API migration works in this way: when an API function's interface is
changed, the author of this API function implements an \emph{adaptor function}, defining calls
to the old API in terms of the new. From this definition we automatically generate the refactoring
that transforms the client code to use the new API; this refactoring can also be supplied by the API
writer to clients on library upgrade, allowing users to upgrade their code automatically. The refactoring works by generating a set of rules that  ``fold in'' the adaptation to the client code, so that the resulting code works directly with the new API. More details of the design choices underlying the work and the technicalities of the implementation can be found in a paper by~\cite{ASE12}.

\paragraph{Support for Introducing Parallelism}

We have introduced support for parallelising explicit list operations (\texttt{map} and \texttt{foreach}), for process introduction to complete a computationally intensive task in parallel, for introducing a worker process to deal with call handling in an Erlang ``generic server'' and to parallelise a tail recursive function. We discuss these in turn now; more details and practical examples of the refactorings appear in a conference paper describing that work~\cite{PEPM2015}.

Uses of \texttt{map} and \texttt{foreach} in list processing are among of the most obvious places where parallelism can be introduced. We have added a small library to Wrangler, called \texttt{para\_lib}, which provides parallel implementations of \texttt{map} and \texttt{foreach}. The transformation from an explicit use of sequential \texttt{map/foreach} to the use of their parallel counterparts is very straightforward, even manual refactoring would not be a problem. However a \texttt{map/foreach} operation could also be implemented differently using recursive functions, list comprehensions, etc.; identifying this kind of implicit \texttt{map/foreach} usage can be done using Wrangler's code inspection facility, and a refactoring that turns an implicit \texttt{map/foreach} to an explicit \texttt{map/foreach} can also be specified
using Wrangler's rule-based transformation API.

If the computations of two non-trivial tasks do not depend on each other, then they can be executed in parallel.
The \emph{Introduce a New Process} refactoring implemented in Wrangler can be used to spawn a new process to execute a task in parallel with its parent process. The  result of the new process is sent back to the parent process, which will then consume it when needed. In order not to block other computations that do not depend on the result returned by the new process, the \texttt{receive} expression is placed immediately before the point where the result is needed.

While some tail-recursive list processing functions can be refactored to an explicit map operation,
many cannot due to data dependencies. For instance, an example might perform a recursion over a list while accumulating results in an
 accumulator variable. In such a situation it is possible to ``float out'' some of the computations into parallel computations. This can only be done when certain dependency constraints are satisfied, and these are done by program slicing, which is discussed below.

\paragraph{Support for Program Slicing}
Program slicing is a general technique of program analysis for extracting the part of a program, also called
the \emph{slice},  that influences or is influenced by a given point of interest, i.e.\  the \emph{slicing criterion}. Static program slicing is generally based on program dependency including both control dependency and data dependency. \emph{Backward intra-function slicing} is used by some of the refactorings described above; it is also useful in general, and made available to
 end-users under Wrangler's Inspector menu~\cite{PEPM2015}.

Our work can be compared with that in PaRTE\footnote{\url{http://paraphrase-enlarged.elte.hu/downloads/D4-3_user_manual.pdf}}~\cite{Boz2015}, a tool developed in another EU project that also re-uses the Wrangler front end. This work concentrates on skeleton introduction, as does some of our work, but we go further in using static analysis and slicing in transforming programs to make them suitable for introduction of parallel structures.


\begin{figure*}[!t]
\begin{center}
\includegraphics[trim={0 15pt 0 8pt},clip,width=0.9\textwidth]{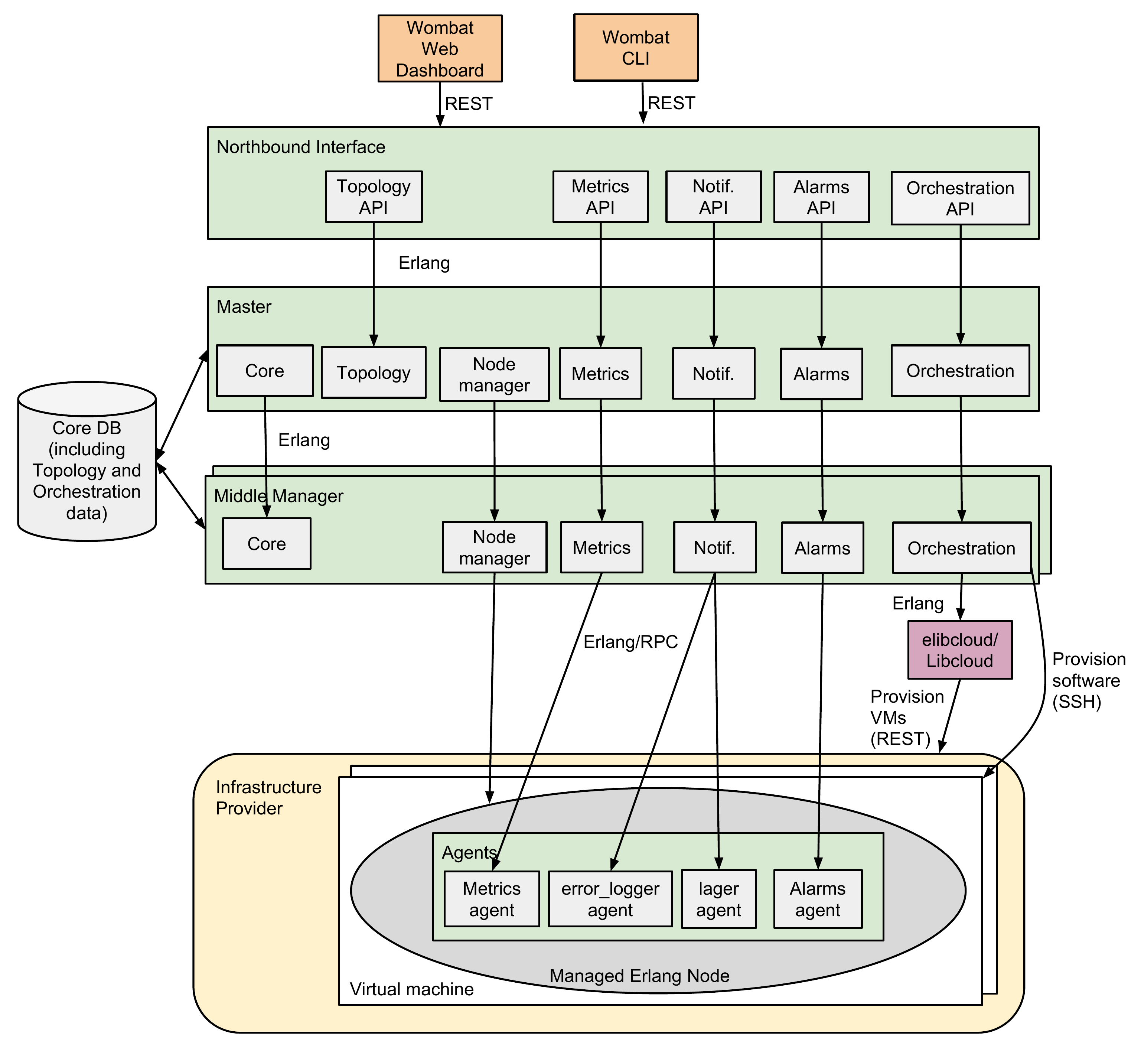}
\caption{The architecture of \wombat.}
\label{fig:wombat-arch}
\end{center}
\end{figure*}

\subsection{Scalable Deployment}
\label{sec:scalable-deployment}

We have developed the \wombat tool\footnote{\wombat (\url{https://www.erlang-solutions.com/products/wombat-oam.html}) is a
  commercial tool available from Erlang Solutions~Ltd.} to
provide a deployment, operations and maintenance framework for large-scale
Erlang distributed systems. These systems typically consist of a
number of Erlang nodes executing on different hosts. These hosts
may have different hardware or operating systems, be physical or
virtual, or run different versions of Erlang/OTP. Prior to the
development of \wombat, deployment of systems would use scripting in
Erlang and the shell, and this is the state of the art for other actor
frameworks; it would be possible to adapt the \wombat approach to
these frameworks in a straightforward way.

\paragraph{Architecture}
The architecture of \wombat is summarised in Figure~\ref{fig:wombat-arch}. Originally the system had problems addressing full scalability because of the role played by the central \emph{Master} node; in its current version an additional layer of \emph{Middle Managers} was introduced to allow the system to scale easily to thousands of deployed nodes. As the diagram shows, the ``northbound'' interfaces to the web dashboard and the command-line are provided through RESTful connections to the Master. The operations of the Master are delegated to the Middle Managers that engage directly with the managed nodes. Each managed node runs a collection of \emph{services} that collect metrics, raise alarms and so forth; we describe those now.

\paragraph{Services}
\wombat is designed to collect, store and display various kinds of information and event from running Erlang systems, and these data are accessed and managed through an AJAX-based Web Dashboard; these include the following.
\begin{description}
\item[Metrics]
\wombat supports the collection of some hundred metrics --- including, for instance, numbers of processes on a node and the message traffic in and out of the node --- on a regular basis from the Erlang VMs running on each of the hosts. It can also collect metrics defined by users within other metrics collection frameworks, such as Folsom\footnote{Folsom collects and publishes metrics through an Erlang API: \url{https://github.com/boundary/folsom}}, and interface with other tools such as graphite\footnote{\url{https://graphiteapp.org}} which can log and display such information. The metrics can be displayed as histograms covering different windows, such as the last fifteen minutes, hour, day, week or month.
\item[Notifications]
As well as metrics, it can support \emph{event-based} monitoring through the collection of notifications from running nodes. Notifications, which are \emph{one time events} can be generated using the Erlang System Architecture Support Libraries, \texttt{SASL}, which is part of the standard distribution, or the \texttt{lager} logging framework\footnote{\url{https://github.com/basho/lager}}, and will be displayed and logged as they occur.
\item[Alarms]
Alarms are more complex entities. Alarms have a state: they can be \emph{raised}, and once dealt with they can be \emph{cleared}; they also have identities, so that the same alarm may be raised on the same node multiple times, and each instance will need to be dealt with separately. Alarms are generated by \texttt{SASL} or \texttt{lager}, just as for notifications.
\item[Topology]
The Topology service handles adding, deleting and discovering nodes. It also monitors whether they are accessible, and if not, it notifies the other services, and periodically tries to reconnect. When the nodes are available again, it also notifies the other services. It doesn't have a middle manager part, because it doesn't talk to the nodes directly: instead it asks the Node manager service to do so.
\item[Node manager]
This service maintains the connection to all managed nodes via the Erlang distribution protocol. If it loses the connection towards a node, it periodically tries to reconnect. It also maintains the states of the nodes in the database (e.g. if the connection towards a node is lost, the Node manager changes the node state to DOWN and raises an alarm). The Node manager doesn't have a REST API, since the node states are provided via the Topology service's REST API.
\item[Orchestration]
This service can deploy new Erlang nodes on already running machines. It can also provision new virtual machine instances using several cloud providers, and deploy Erlang nodes on those instances. For communicating with the cloud providers, the Orchestration service uses an external library called \mbox{Libcloud},\footnote{The unified cloud API~\cite{libcloud}: \url{https://libcloud.apache.org}} for which Erlang Solutions has written an open source Erlang wrapper called \texttt{elibcloud}, to make Libcloud easier to use from \wombat. Note that \wombat Orchestration doesn't provide a platform for writing Erlang applications: it provides infrastructure for deploying them.
\end{description}

\paragraph{Deployment} The mechanism consists of the following five steps.

\begin{enumerate}
\item \textit{Registering a provider.} \wombat provides the same interface for different cloud providers which support the OpenStack standard or the Amazon EC2 API. \wombat also provides the same interface for using a fixed set of machines. In \wombat's backend, this has been implemented as two driver modules: the \emph{elibcloud} driver module which uses the \texttt{elibcloud} and Libcloud libraries to communicate with the cloud providers, and the \emph{SSH} driver module that keeps track of a fixed set of machines.

\item \textit{Uploading a release.} The release can be either a proper Erlang release archive or a set of Erlang modules. The only important aspect from \wombat's point of view is that start and stop commands should be explicitly specified. This will enable \wombat start and stop nodes when needed.

\item \textit{Defining a node family.} The next step is creating the node family, which is the entity that refers to a certain release, contains deployment domains that refer to certain providers, and contains other information necessary to deploy a node.

\item \textit{Defining a deployment domain.} At this step a deployment domain is created that specifies\begin{inparaenum}[(i)] \item which providers should be used for provisioning machines; \item the username that will be used when \wombat connects to the hosts using \emph{SSH}.\end{inparaenum}

\item \textit{Node deployment.} To deploy nodes a \wombat user needs only to specify the number of nodes and the node family these nodes belong to. Nodes can be dynamically added to, or removed from, the system depending on the needs of the application. The nodes are started, and \wombat is ready to initiate and run the application.
\end{enumerate}


\subsection{Monitoring and Visualisation}

A key aim in designing new monitoring and visualisation tools and adapting existing ones was to provide support for systems running on parallel and distributed hardware. Specifically, in order to run modern Erlang systems, and in particular SD Erlang systems, it is necessary to understand both their single host (``multicore'') and multi-host (``distributed'') nature. That is, we need to be able to understand how systems run on the Erlang multicore virtual machine, where the scheduler associated with a core manages its own run queue, and processes migrate between the queues through a work stealing algorithm; at the same time, we have to understand the dynamics of a distributed Erlang program, where the user explicitly spawns processes to  nodes.\footnote{This is in contrast with the multicore VM, where programmers have no control over where processes are spawned; however, they still need to gain insight into the behaviour of their programs to tune performance.}

\npcomment{Although we are out of space on this section, we have a couple of nice figures in D5.1 and D5.2 that we could use, showing the architecture of our tool for offline/online monitoring.}

\subsubsection{Percept2}

\begin{figure*}[!t]
\centering
\includegraphics[width=0.9\textwidth]{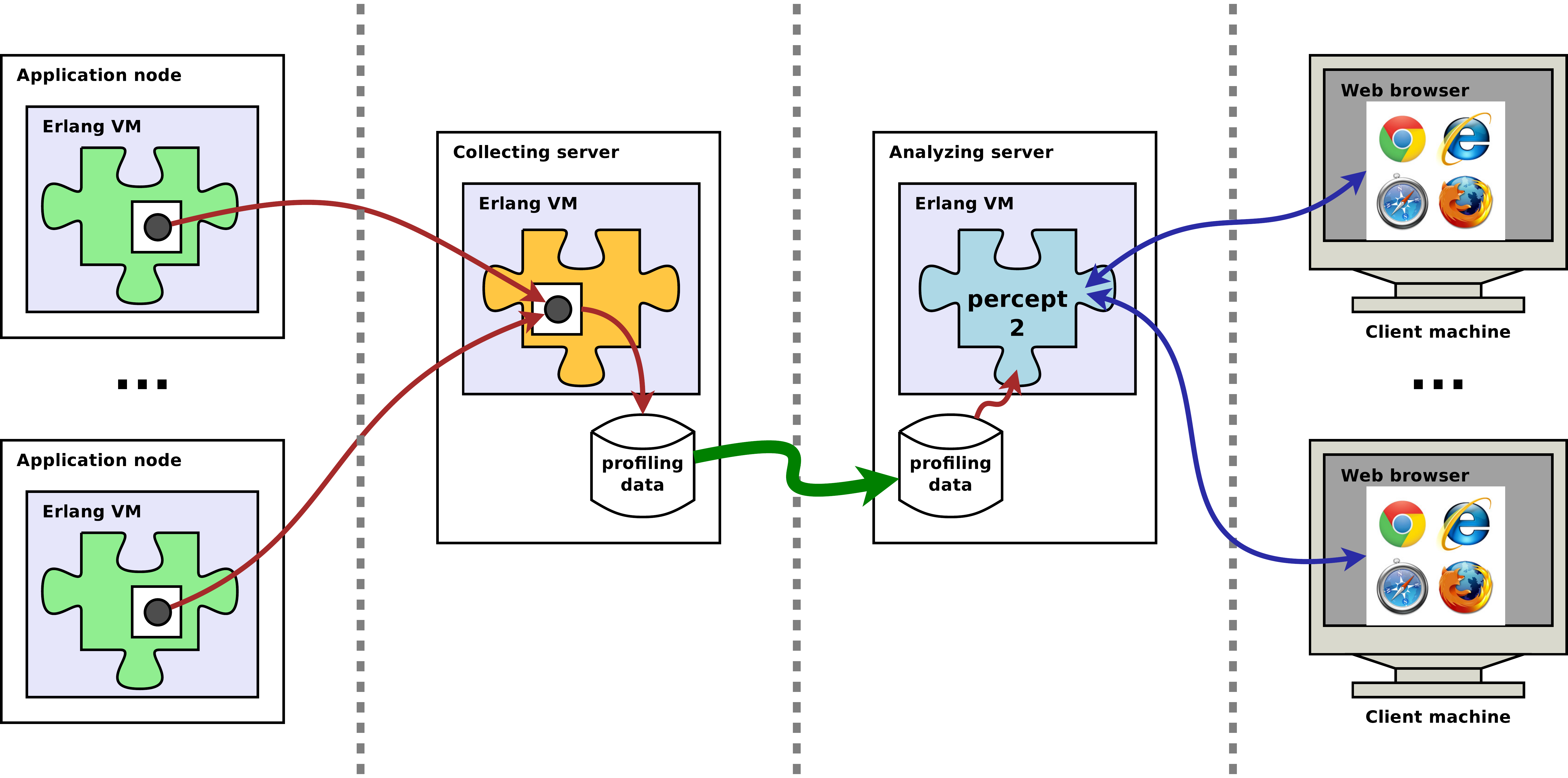}
\caption{Offline profiling of a distributed Erlang application
  using Percept2.
  \label{fig:percept-offline-trace}
}
\end{figure*}

Percept2\footnote{\url{https://github.com/RefactoringTools/percept2}}
builds on the existing Percept tool to provide \emph{post hoc} offline
analysis and visualisation of Erlang systems.  Percept2 is designed to
allow users to visualise and tune the parallel performance of Erlang
systems on a single node on a single manycore host. It visualises
Erlang application level concurrency and identifies concurrency
bottlenecks. Percept2 uses Erlang built in tracing and profiling to
monitor  process states, i.e. waiting, running, runnable,
free and exiting. A waiting or suspended process is considered an
inactive  and a running or runnable process is considered
active. As a program runs with Percept, events are collected
and stored to a file. The file is then analysed, with the results
stored in a RAM database, and this data is viewed through a web-based
interface.  The process of offline profiling for a distributed Erlang
application using Percept2 is shown in Figure~\ref{fig:percept-offline-trace}.

Percept generates an application-level zoomable concurrency graph, showing the number of active processes at each point during profiling; dips in the graph represent low concurrency. A lifetime bar for each process is also included, showing the points during its lifetime when the process was active, as well as other per-process information.

\begin{figure*}[!t]
\centering
\includegraphics[width=0.8\textwidth]{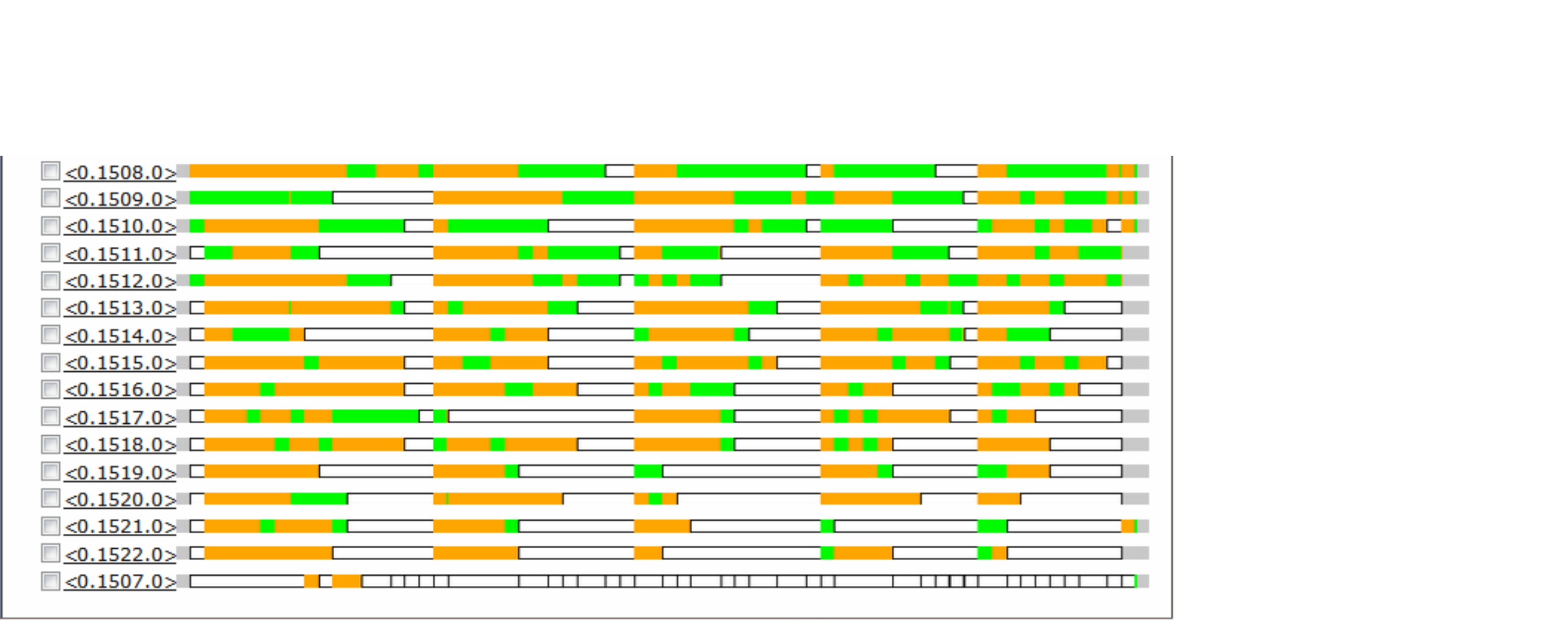}
\caption{Percept2: showing processes running and runnable (with 4 schedulers/run queues).}
\label{Percept2-pic}
\end{figure*}

Percept2 extends Percept in a number of ways --- as detailed by \cite{percept2} --- including most importantly:
\begin{itemize}
\item
Distinguishing between running and runnable time for each process: this is apparent in the process runnability comparison as shown in Figure~\ref{Percept2-pic}, where orange represents runnable and green represents running. This shows very clearly where potential concurrency is not being exploited.
\item
Showing scheduler activity: the number of active schedulers at any time during the profiling.
\item
Recording more information during execution, including the migration history of a process between run queues; statistics  about message passing between processes: the number of messages and the average message size sent/received by a process; the accumulated runtime per-process: the accumulated time when a process is in a running state.
\item
Presenting the process tree: the hierarchy structure indicating the parent-child relationships between processes.
\item
Recording dynamic function call graph/count/time: the hierarchy structure showing the calling relationships between functions during the program run, and the amount of time spent in a function.
\item
Tracing of s\_group activities in a distributed system. Unlike global group, s\_group allows dynamic changes to the s\_group structure of a distributed Erlang system. In order to support SD Erlang, we have also extended Percept2 to allow the profiling of s\_group related activities, so that the dynamic changes to the s\_group structure of a distributed Erlang system can be captured.
\end{itemize}
We have also improved on Percept as follows.
\begin{itemize}
\item
Enabling finer-grained control of what is profiled. The profiling of port activities, schedulers activities, message passing, process migration, garbage collection and s\_group activities can be enabled/disabled, while the profiling of process runnability (indicated by the ``proc'' flag) is always enabled.
\item
Selective function profiling of processes. In Percept2, we have built a version of fprof, which does not measure a function's own execution time, but measures everything else that fprof measures. Eliminating measurement of a function's own execution time gives users the freedom of not profiling all the function calls invoked during the program execution. For example, they can choose to profile only functions defined in their own applications' code, and not those in libraries.
\item
Improved dynamic function callgraph. With the dynamic function callgraph, a user is able to understand the causes of certain events, such as heavy calls of a particular function, by examining the region around the node for the function, including the path to the root of the graph. Each edge in the callgraph is annotated with the number of times the target function is called by the source function as well as further information.
\end{itemize}
Finally, we have also improved the scalability of Percept in three ways. First we have parallelised the processing of trace files so that multiple data files can be processed at the same time. We have also compressed the representation of call graphs, and cached the history of generated web pages. Together these make the system more responsive and more scalable.

\subsubsection{DTrace/SystemTap tracing}
\label{sec:dtrace}
DTrace provides dynamic tracing support for various flavours of Unix, including BSD and Mac OS X, and SystemTap does the same for Linux; both allow the monitoring
of live, running systems with minimal overhead.  They can be used by
administrators, language developers and application developers alike to
examine the behaviour of applications, language implementations and the
operating system during development or even on live production systems.  In
comparison to other similar tools and instrumentation frameworks, they are
relatively lightweight,  do not require special recompiled versions of
the software to be examined, nor special post-processing tools to
create meaningful information from the data gathered. Using these probs,
it is possible to identify bottlenecks both in the VM itself and in
applications.

\begin{figure}[t!]
\centering
\includegraphics[width=0.5\textwidth]{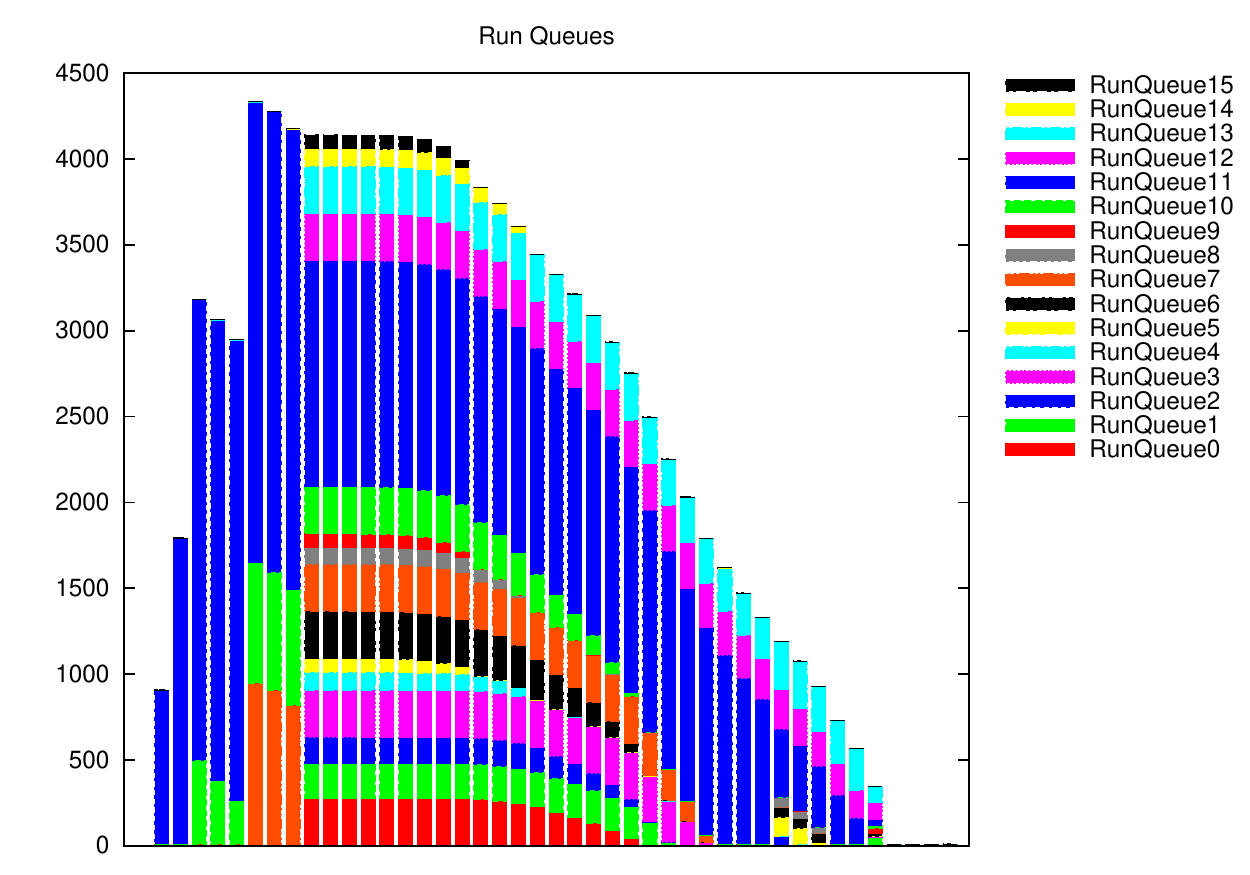}
\caption{DTrace: runqueue visualisation of \bench{bang} benchmark with 16 schedulers (time on X-axis).}
\label{fig:stacked-bargraph}
\end{figure}

VM bottlenecks are identified using a large number of probes inserted into
Erlang/OTP's VM, for example to explore scheduler run-queue lengths.
These probes can be used to measure the number of processes per
scheduler, the number of processes moved during work stealing, the number of
attempts to gain a run-queue lock, how many of these succeed immediately,
etc.  Figure~\ref{fig:stacked-bargraph} visualises the results of such
monitoring; it shows how the size of run queues vary during execution of the
\bench{bang} BenchErl benchmark on a VM with 16 schedulers.

Application bottlenecks are identified by an alternative
back-end for Percept2, based on DTrace/SystemTap instead of the Erlang
built-in tracing mechanism. The implementation re-uses
the existing Percept2 infrastructure as far as possible. It uses a
different mechanism for collecting information about Erlang programs, a
different format for the trace files, but the same storage
infrastructure and presentation facilities.

\subsubsection{Devo}
\label{sec:devo}
Devo\footnote{\url{https://github.com/RefactoringTools/devo}}~\cite{bakermulti}
is designed to provide real-time online visualisation of both the
low-level (single node, multiple cores) and high-level (multiple
nodes, grouped into s\_groups) aspects of Erlang and SD Erlang
systems. Visualisation is within a browser, with web sockets providing
the connections between the JavaScript visualisations and the running
Erlang systems, instrumented through the trace tool builder (\texttt{ttb}).

\begin{figure}
\centering
\includegraphics[width=0.48\textwidth]{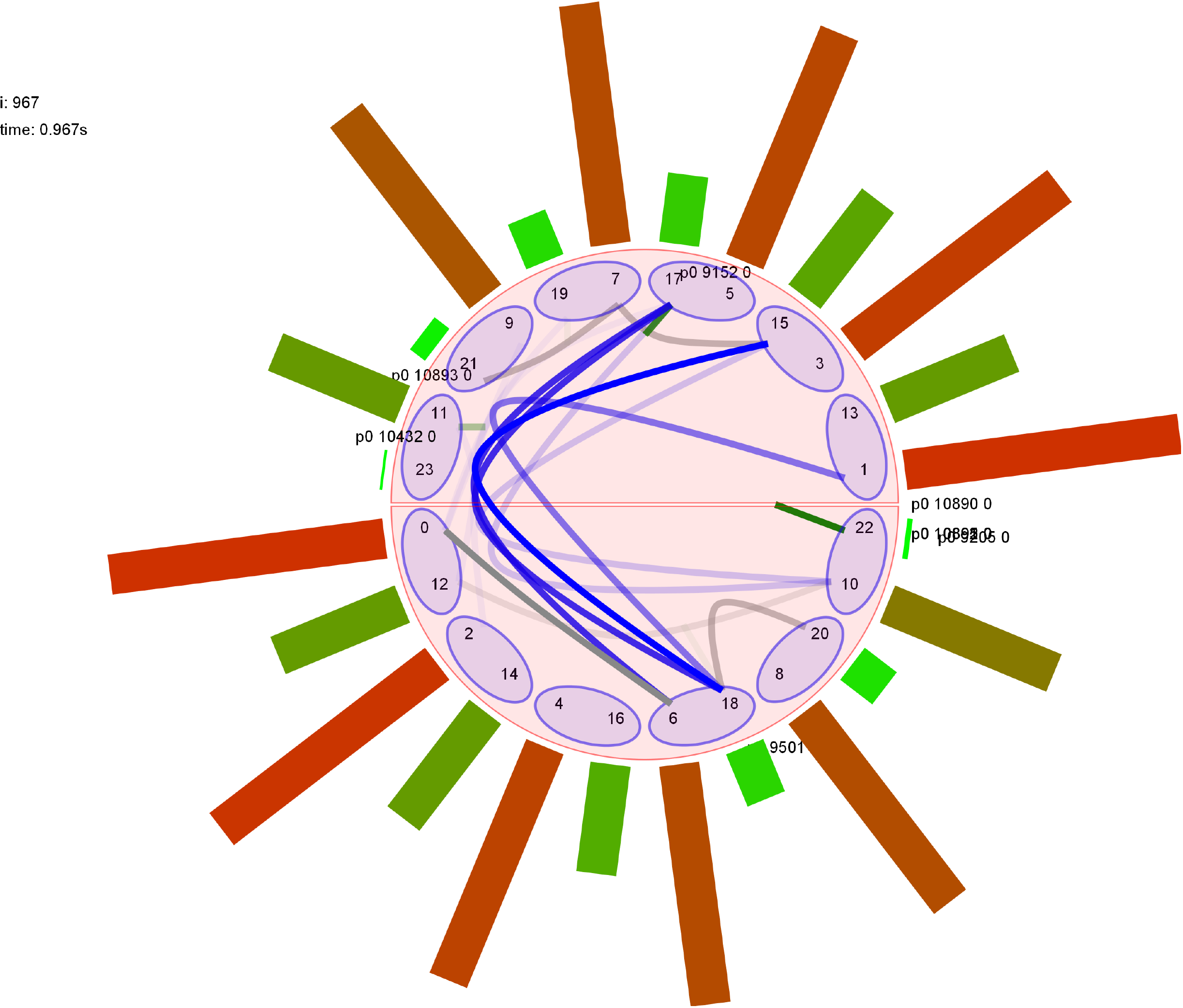}\
\includegraphics[width=0.48\textwidth]{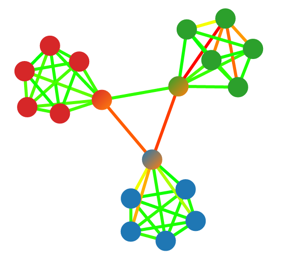}
\caption{Devo: low and high-level visualisations of SD-Orbit.}
\label{Devo-pic}
\end{figure}

Figure~\ref{Devo-pic} shows visualisations from devo in both modes. On the left-hand side a single compute node is shown. This consists of two physical chips (the upper and lower halves of the diagram) with six cores each; with hyperthreading this gives twelve virtual cores, and hence 24  run queues in total. The size of these run queues is shown by both the colour and height of each column, and process migrations are illustrated by (fading) arc between the queues within the circle. A green arc shows migration on the same physical core, a grey one on the same chip, and blue shows migrations between the two chips.

On the right-hand side of Figure~\ref{Devo-pic} is a visualisation of SD-Orbit in action. Each node in the graph represents an Erlang node, and colours (red, green, blue and orange) are used to represent the s\_groups to which a node belongs. As is evident, the three nodes in the central triangle belong to multiple groups, and act as routing nodes between the other nodes. The colour of the arc joining two nodes represents the current intensity of communication between the nodes (green quiescent; red busiest).

\begin{figure*}
\centering
\includegraphics[width=0.7\textwidth]{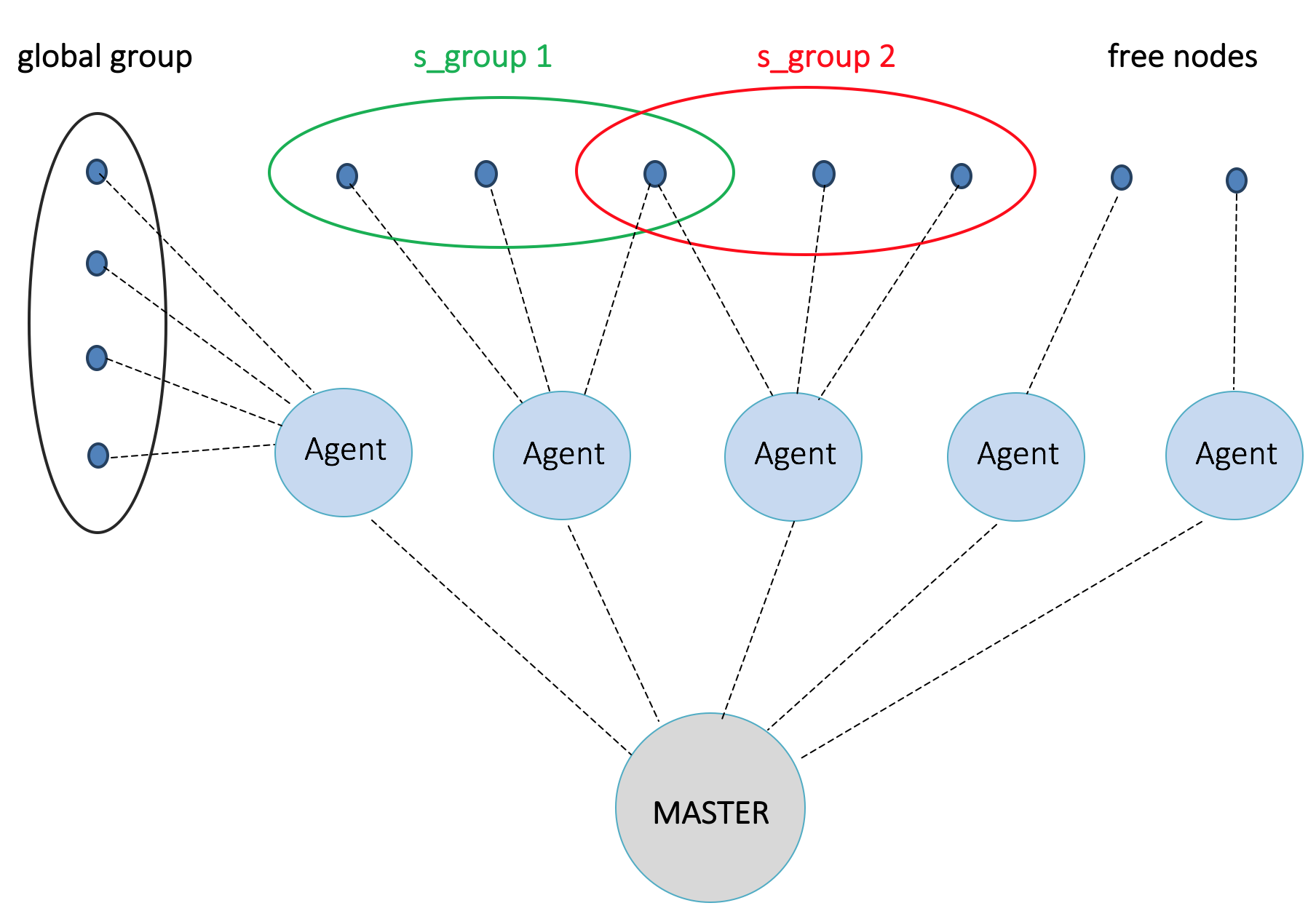}
\caption{SD-Mon architecture.}
\label{SD-Mon-pic}
\end{figure*}

\subsubsection{SD-Mon}
\label{sec:sd-mon}
SD-Mon is a tool specifically designed for monitoring SD-Erlang systems. This purpose is accomplished by means of a \emph{shadow network} of agents, that collect data from a running system. An example deployment is shown in Figure~\ref{SD-Mon-pic}, where blue dots represent nodes in the target system and the other nodes make up the \mbox{SD-Mon} infrastructure.
The network is deployed on the basis of a configuration file describing the network architecture in terms of hosts, Erlang nodes, global group and s\_group partitions. Tracing to be performed on monitored nodes is also specified within the configuration file.

\begin{figure}
\begin{center}
\includegraphics[width=0.45\textwidth]{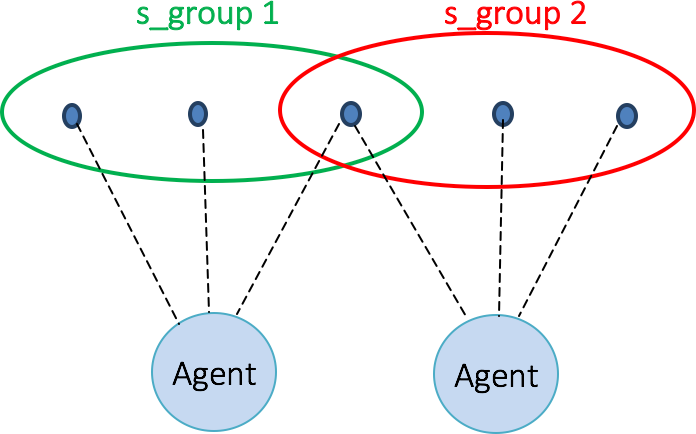}\\
\vspace{0.3cm}
\includegraphics[width=0.45\textwidth]{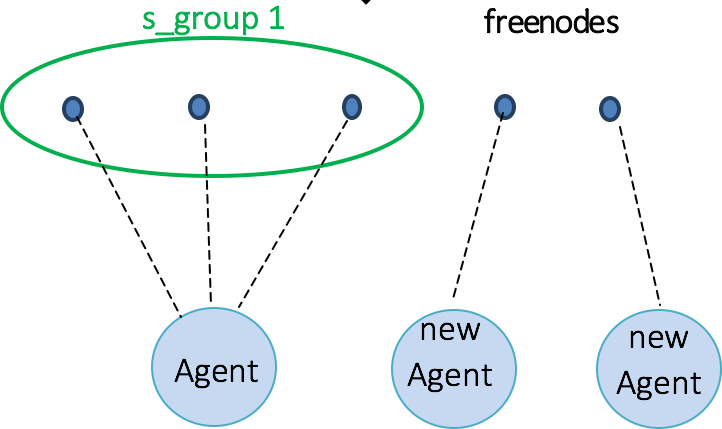}
\caption{SD-Mon evolution. Before (top) and after eliminating s\_group 2 (bottom).}
\label{SD-Mon-evo}
\end{center}
\end{figure}

An agent is started by a master SD-Mon node for each s\_group and for each free node. Configured tracing is applied on every monitored node, and traces are stored in binary format in the agent file system. The shadow network follows system changes so that agents are started and stopped at runtime as required, as shown in Figure~\ref{SD-Mon-evo}. Such changes are persistently stored so that the last configuration can be reproduced after a restart. Of course, the shadow network can be always updated via the User Interface.

Each agent takes care of an s\_group or of a free node. At start-up it tries to get in contact with its nodes and apply the tracing to them as stated by the master. Binary files are stored in the host file system.
Tracing is internally used in order to track s\_group operations happening at runtime. An asynchronous message is sent to the master whenever one of these changes occurs.
Since each process can only be traced by a single process at a time, each node (included those belonging to more than one s\_group) is controlled by only one agent.
When a node is removed from a group or when the group is deleted, another agent takes over, as shown in Figure~\ref{SD-Mon-evo}.
When an agent is stopped, all traces on the controlled nodes are switched off.

The monitoring network is also supervised, in order to take account of network fragility, and when an agent node goes down another node is deployed to play its role; there are also periodic consistency checks for the system as a whole, and when an inconsistency is detected then that part of the system can be restarted.

SD-Mon does more than monitor activities one node at a time. In
particular inter-node and inter-group messages are displayed at
runtime. As soon as an agent is stopped, the related tracing files are
fetched across the network by the master and they are made available
in a readable format in the master file system.

\begin{figure*}[!t]
\begin{center}
\includegraphics[width=0.65\textwidth]{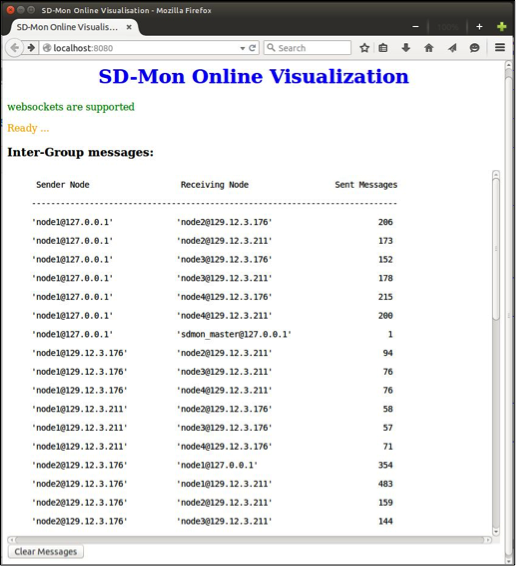}
\caption{SD-Mon: online monitoring.}
\label{SD-Mon-online}
\end{center}
\end{figure*}

SD-Mon provides facilities for online visualisation of this data, as well as \emph{post hoc} offline analysis.  Figure~\ref{SD-Mon-online} shows, in real time, messages that are sent \emph{between} s\_groups. This data can also be used as input to the animated Devo visualisation, as illustrated in the right-hand side of Figure~\ref{Devo-pic}.

\section{Systemic Evaluation}
\label{sec:case-studies}
Preceding sections have investigated the improvements of individual aspects of an Erlang system, e.g. ETS tables in Section~\ref{sec:vm-ets}.
This section  analyses the impact of the new tools and technologies from
Sections~\ref{sec:language-scalability}, \ref{sec:improving-vm-scalability}, \ref{sec:scalable-tools}
in concert. We do so by investigating the deployment, reliability,
and scalability of the Orbit and ACO benchmarks from Section~\ref{sec:benchmarks}.
The experiments reported here are representative. Similar experiments
show consistent results for a range of micro-benchmarks, several
benchmarks, and the very substantial (approximately 150K lines of Erlang) Sim-Diasca
case study~\cite{simdiasca} on several state of the art NUMA architectures, and the four
clusters specified in Appendix~A. A coherent presentation of many of these
results is available in an article by~\cite{evaluating-tpds-16} and in a RELEASE project deliverable\footnote{See Deliverable 6.2, available online. \url{http://www.release-project.eu/documents/D6.2.pdf}}.
The bulk of the experiments reported here are conducted on the Athos cluster
using Erlang/OTP 17.4 and the associated SD Erlang libraries.

\begin{figure*}
\centering
\includegraphics[width=0.8\textwidth]{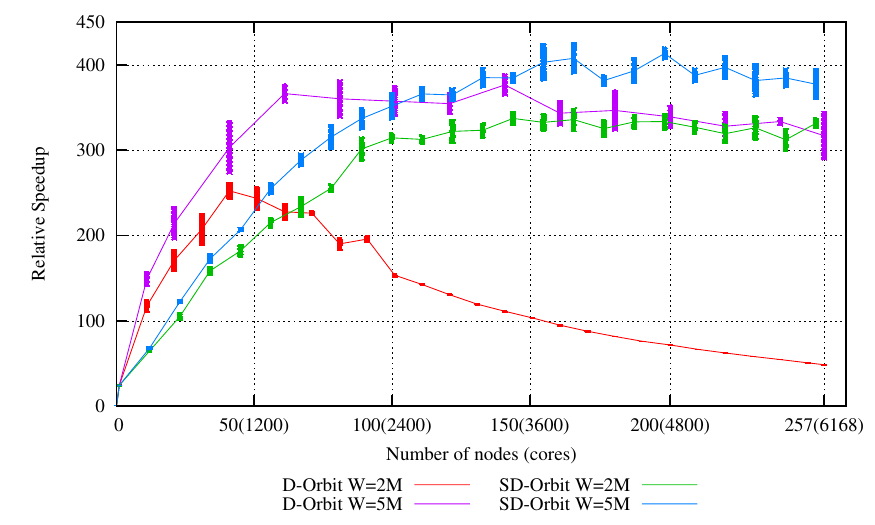}
\caption{Speedup of distributed Erlang vs.\ SD Erlang Orbit (2M and 5M elements) [Strong Scaling].}
\label{fig:orbit-2-5-speedup}
\end{figure*}

The experiments cover two measures of scalability.  As Orbit does a
fixed size computation, the scaling measure is relative speedup (or
strong scaling), i.e.\ speedup relative to execution time on a single
core.  As the work in ACO increases with compute resources, weak
scaling is the appropriate measure. The benchmarks also evaluate different
aspects of s\_groups: Orbit evaluates the scalability impacts of
network connections, while ACO evaluates the impact of both network
connections and the global namespace required for reliability.

\subsection{Orbit}
\label{subsec:case-studies-orbit}

Figure~\ref{fig:orbit-2-5-speedup} shows the speedup of the D-Orbit and SD-Orbit
benchmarks from Section~\ref{subsubsec:design-s-groups}.
The measurements are repeated seven times, and we plot standard deviation.
Results show that D-Orbit performs better on a small number
of nodes as communication is direct, rather than via a gateway node. As the
number of nodes grows, however, SD-Orbit delivers better speedups,
i.e.~beyond 80 nodes in case of 2M orbit elements, and beyond 100
nodes in case of 5M orbit elements. When we increase the size of Orbit
beyond 5M, the D-Orbit version fails due to the fact that some VMs
exceed the available RAM of 64GB. In contrast SD-Orbit experiments run
successfully even for an orbit with 60M elements.


\subsection{Ant Colony Optimisation (ACO)}
\label{subsec:case-studies-aco}

\paragraph{Deployment} The deployment and monitoring of ACO, and of the large (150K lines of Erlang) Sim-Diasca simulation using \wombat (Section~\ref{sec:scalable-deployment}) on the Athos cluster is detailed in~\cite{evaluating-tpds-16}.

\begin{figure*}
\centering
\includegraphics[width=0.8\textwidth]{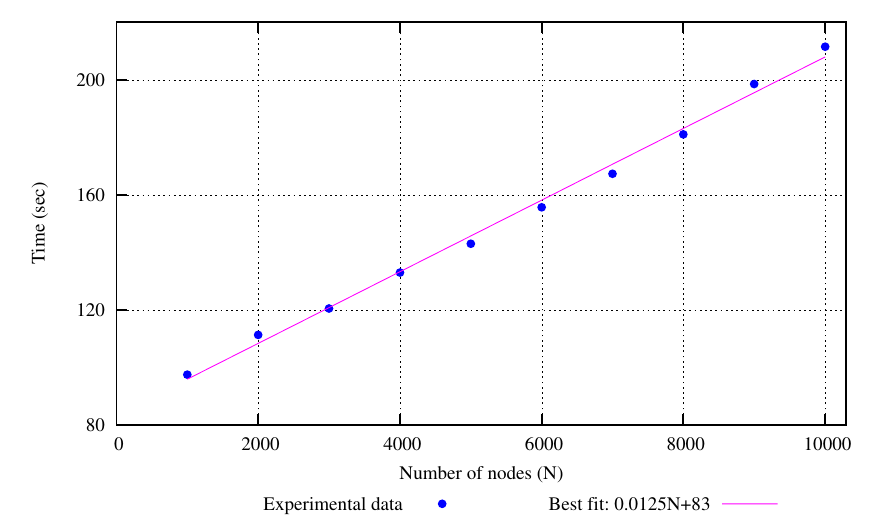}
\caption{Wombat Deployment Time}
\label{fig:wombat-deployment-time}
\end{figure*}

An example experiment deploys 10,000 Erlang nodes without enabling
monitoring, and hence allocates three nodes per core (i.e.\ 72 nodes
on each of 139 24-core Athos hosts).
Figure~\ref{fig:wombat-deployment-time} shows that \wombat
deploys the nodes in 212s, which is approximately 47 nodes per
second. It is more common to have at least one core per Erlang node,
and in a related experiment 5000 nodes are deployed, one per core
(i.e.\ 24 nodes on each of 209 24-core Athos hosts) in 101s, or 50
nodes per second. Crucially in both cases the deployment time is
linear in the number of nodes. The deployment time could be reduced to
be logarithmic in the number of nodes using standard
divide-and-conquer techniques. However there hasn't been the demand to
do so as most Erlang systems are long running servers.

The measurement data shows two important facts: it shows that \wombat scales well (up to a deployment base of 10,000 Erlang nodes), and that \wombat is non-intrusive because its overhead on a monitored node is typically less than 1.5\% of effort on the node.

We conclude that \wombat is capable of deploying and monitoring
substantial distributed Erlang and SD Erlang programs.  The
experiments in the remainder of this section use standard distributed
Erlang configuration file deployment.

\begin{figure*}
\centering
\includegraphics[width=0.8\textwidth]{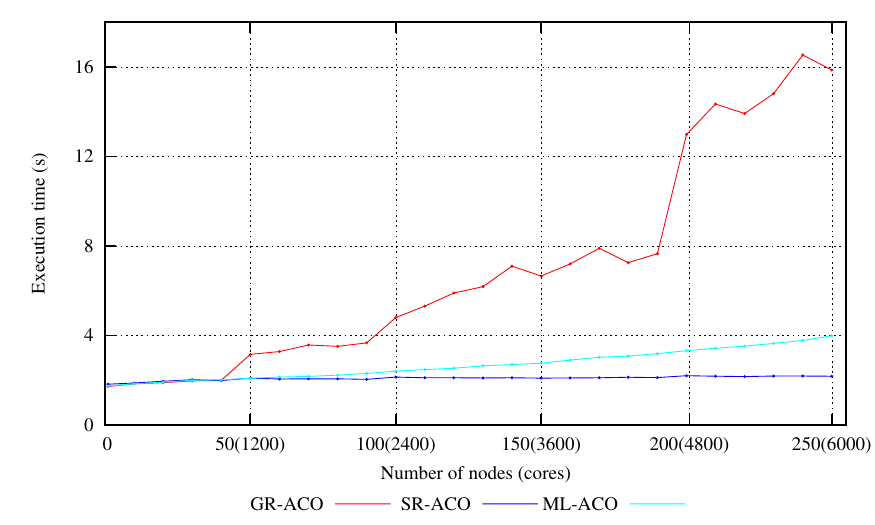}
\caption{ACO execution times, Erlang/OTP 17.4 (RELEASE) [Weak Scaling].}
\label{fig:VRELEASE-times-good}
\end{figure*}

\paragraph{Reliability} SD Erlang changes the organisation of
processes and their recovery data at the language level, so we seek to
show that these changes have not disrupted Erlang's world-class
reliability mechanisms at this level. As we haven't changed them we
don't exercise Erlang's other reliability mechanisms, e.g.\ those for
managing node failures, network congestion, etc. A more detailed
study of SD Erlang reliability, including the use of replicated
databases for recovering Instant Messenger chat sessions, finds
similar results~\cite{chechina-16-scalable}.

We evaluate the reliability of two ACO
versions using a Chaos Monkey service that kills processes in the running
system at random~\cite{bennett-12-chaos}. Recall that GR-ACO provides
reliability by registering the names of critical processes globally,
and SR-ACO registers them only within an s\_group
(Section~\ref{subsec:aco-overview}). 

For both GR-ACO and SR-ACO a Chaos Monkey runs on every Erlang node,
i.e.\ master, submasters, and colony nodes, killing a random Erlang
process every second. Both ACO versions run to completion.  Recovery,
at this failure frequency, has no measurable impact on runtime. This
is because processes are recovered within the Virtual machine using
(globally synchronised) local recovery information. For example, on a
common X86/Ubuntu platform typical Erlang process recovery times are
around 0.3ms, so around 1000x less than the Unix process recovery time
on the same platform~\cite{Lutac:2016}.  We have conducted more
detailed experiments on an Instant Messenger benchmark, and obtained
similar results~\cite{chechina-16-scalable}.

We conclude that both GR-ACO and SR-ACO are reliable, and that SD
Erlang preserves distributed Erlang reliability model. The remainder
of the section outlines the impact of maintaining the recovery
information required for reliability on scalability.

\begin{figure*}[!ht]
\centering
\includegraphics[width=0.8\textwidth]{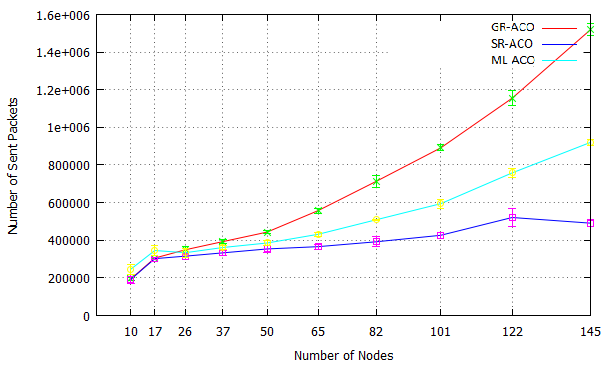}
\caption{Number of sent packets in ML-ACO, GR-ACO, and SR-ACO.}
\label{fig:network-traffic-sent}
\end{figure*}

\paragraph{Scalability} Figure~\ref{fig:VRELEASE-times-good} compares
the runtimes of the ML, GR, and SR versions of ACO
(Section~\ref{subsec:aco-overview}) on Erlang/OTP 17.4(RELEASE).
As outlined in Section~\ref{subsubsec:design-s-groups}, \mbox{GR-ACO} not only
maintains a fully connected graph of nodes, it registers process names
for reliability, and hence scales significantly worse than the
unreliable \mbox{ML-ACO}. We conclude that providing reliability with
standard distributed Erlang process registration dramatically limits
scalability.

While ML-ACO does not provide reliability, and hence doesn't register
process names, it maintains a fully connected graph of nodes which
limits scalability.  SR-ACO, that maintains connections and registers
process names only within s\_groups scales best of all.
Figure~\ref{fig:VRELEASE-times-good} illustrates how maintaining
the process namespace, and fully connected network, impacts
performance. This reinforces the evidence from the Orbit benchmarks,
and others, that partitioning the network of Erlang nodes
significantly improves performance at large scale.

To investigate the impact of SD Erlang on network traffic, we measure
the number of sent and received packets on the GPG cluster for three
versions of ACO: ML-ACO, GR-ACO, and SR-ACO.
Figure~\ref{fig:network-traffic-sent} shows the total number of sent packets.
The highest traffic (the red line) belongs to the GR-ACO and the
lowest traffic belongs to the SR-ACO (dark blue line). This shows that
SD Erlang significantly reduces the network traffic between Erlang
nodes. Even with the s\_group name registration SR-ACO has less
network traffic than ML-ACO that has no global name registration.


\subsection{Evaluation Summary}
\label{subsec:case-studies-discussion}
We have shown that \wombat is capable of deploying and monitoring
substantial distributed Erlang and SD Erlang programs like ACO and Sim-Diasca.
The Chaos Monkey experiments with GR-ACO and SR-ACO show that both are
reliable, and hence that SD Erlang preserves the distributed Erlang language-level reliability
model.

As SD Orbit scales better than D-Orbit, SR-ACO scales better than
ML-ACO, and SR-ACO has significantly less network traffic, we conclude
that, even when global recovery data is not maintained, partitioning
the fully-connected network into s\_groups reduces network traffic and
improves performance. While the distributed Orbit instances (W=2M)
and (W=5M) reach scalability limits at around 40 and 60 nodes, Orbit
scales to 150 nodes on SD Erlang (limited by input size), and SR-ACO is
still scaling well on 256 nodes (6144 cores). Hence not only have we
exceeded the 60 node scaling limits of distributed Erlang identified
in Section~\ref{sec:languagescaling}, we have not reached the scaling
limits of SD Erlang on this architecture.

Comparing GR-ACO and ML-ACO scalability curves
shows that maintaining global recovery data, i.e.\ a process name
space, dramatically limits scalability. Comparing GR-ACO and SR-ACO
scalability curves shows that scalability can be recovered by partitioning the nodes into appropriately-sized s\_groups, and hence maintaining the recovery data only
within a relatively small group of nodes.  These results are consistent
with other experiments. \sjt{ The point here is that the size of the
  s\_groups must be bounded, and the number of groups grow linearly
  with the number of nodes. It is not s\_groups \emph{per se} that
  make everything work: we could, after all, have one s\_group :-)}
\pwtcomment{Is the argument clearer now Simon?}


\section{Discussion}
\label{sec:discussion}
Distributed actor platforms like Erlang, or Scala with Akka, are a
common choice for internet-scale system architects as the model, with
its automatic, and VM-supported reliability mechanisms makes it
\emph{extremely easy to engineer scalable reliable systems}. Targeting
emergent server architectures with hundreds of hosts and tens of
thousands of cores, we report a systematic effort to improve the
scalability of a leading distributed actor language, while preserving
reliability. The work is a \textit{vade mecum} for addressing
scalability of reliable actor languages and frameworks. It is also
high impact, with downloads of our improved Erlang/OTP running at 50K
a month.

We have undertaken the first systematic study of scalability in a
distributed actor language, covering VM, language and persistent
storage levels. We have developed the BenchErl and DE-Bench tools for
this purpose.  Key VM-level scalability issues we identify include
contention for shared ETS tables and for commonly-used shared
resources like timers. Key language scalability issues are the costs
of maintaining a fully-connected network, maintaining global recovery
information, and explicit process placement.  Unsurprisingly the
scaling issues for this distributed actor language are common to other
distributed or parallel languages and frameworks with other paradigms
like CHARM++~\cite{kale1993charm++}, Cilk~\cite{blumofe1995cilk}, or
Legion~\cite{grimshaw1997legion}. We establish scientifically the
folklore limitations of around 60 connected nodes for distributed
Erlang (Section~\ref{sec:erlang-scalability}).

The actor model is no panacea, and there can still be scalability
problems in the algorithms that we write, either within a single actor
or in the way that we structure communicating actors.  A range of
pragmatic issues also impact the performance and scalability of actor
systems, including memory occupied by processes (even when quiescent),
mailboxes filling up, etc. Identifying and resolving these problems is
where tools like Percept2 and \wombat are needed.
However, \emph{many of the scalability issues arise where Erlang
departs from the private state principle of the actor model},
e.g. in maintaining shared state in ETS tables, or a shared global
process namespace for recovery.

We have designed and implemented a set of Scalable Distributed (SD)
Erlang libraries to address \emph{language-level} scalability
issues. The key constructs are s\_groups for partitioning the network and global
process namespace, and semi-explicit process placement for deploying
distributed Erlang applications on large heterogeneous architectures
in a portable way.  We have provided a state transition operational
semantics for the new s\_groups, and validated the library implementation
against the semantics using QuickCheck (Section~\ref{sec:language-scalability}).

To improve the scalability of the Erlang VM and libraries we have
improved the implementation of shared ETS tables, time management and
load balancing between schedulers.  Following a systematic analysis of
ETS tables, the number of fine-grained (bucket) locks and of reader
groups have been increased. We have developed and evaluated four new
techniques for improving ETS scalability:%
\begin{inparaenum}[(i)]
\item programmer control of number of bucket locks;
\item a contention-adapting tree data structure for \texttt{ordered\_set}s;
\item queue delegation locking; and
\item eliminating the locks in the meta table.
\end{inparaenum}
We have introduced a new scheduler utilisation balancing mechanism to
spread work to multiple schedulers (and hence cores), and new
synchronisation mechanisms to reduce contention on the \mbox{widely-used}
time management mechanisms.  By June 2015, with Erlang/OTP 18.0, the
majority of these changes had been included in the primary releases.
In any scalable actor language implementation such thoughtful design
and engineering will be required to schedule large numbers of actors
on hosts with many cores, and to minimise contention on shared VM
resources (Section~\ref{sec:improving-vm-scalability}).

To facilitate the development of large Erlang systems, and to make
them understandable we have developed a range of tools. The
proprietary \wombat tool deploys and monitors large distributed Erlang
systems over multiple, and possibly heterogeneous, clusters or
clouds. We have made open source releases of four concurrency tools:
Percept2 now detects concurrency bad smells; Wrangler provides
enhanced concurrency refactoring; 
the Devo tool is enhanced to provide interactive visualisation of SD
Erlang systems; and the new SD-Mon tool monitors SD Erlang systems.
\pwtcomment{Simon: What about something like:} We anticipate that
these tools will guide the design of tools for other large scale
distributed actor languages and frameworks (Section~\ref{sec:scalable-tools}).

\begin{table*}[!t]
  \newcommand{\sep}{\hspace*{1em}}
  \renewcommand{\arraystretch}{1.4}
\caption{Cluster Specifications (RAM per host in GB).\label{tab:clusters}}{%
  \begin{tabular}{@{}l@{\sep}r@{\sep}r@{\sep}r@{\sep}r@{\sep}p{4.5cm}@{\sep}c@{\sep}p{2.5cm}@{}} \toprule
    & & \multicolumn{3}{c}{\textbf{Cores}} &  &  & \\[-3pt]
    \cmidrule(r){3-5}\\[-2em]
    & & \textbf{per\ } &  & \textbf{max\ \ } & & & \\[-2pt]
    \textbf{Name} & \textbf{Hosts} & \textbf{host} & \textbf{total\ } & \textbf{avail.} & \textbf{Processor} & \textbf{RAM} & \textbf{Inter-connection} \\
      \midrule

GPG    & 20  & 16 & 320    & 320   & Intel Xeon E5-2640v2 8C, 2GHz        & 64      & 10GB Ethernet \\
Kalkyl & 348 & 8  & 2,784  & 1,408 & Intel Xeon 5520v2 4C, 2.26GHz        & 24--72  & InfiniBand 20 Gb/s \\
\multirow{2}{*}{TinTin} & \multirow{2}{*}{160} & \multirow{2}{*}{16} & \multirow{2}{*}{2,560} & \multirow{2}{*}{2,240} & {AMD Opteron 6220v2 Bulldozer 8C, 3.0GHz} & \multirow{2}{*}{64--128} & 2:1 oversubscribed QDR Infiniband \\
Athos  & 776 & 24 & 18,624 & 6,144 & Intel Xeon E5-2697v2 12C, 2.7GHz      & 64      & Infiniband FDR14 \\
\bottomrule
\end{tabular}}
\end{table*}%

We report on the reliability and scalability implications of our new
technologies using a range of benchmarks, 
and consistently use the Orbit and ACO benchmarks throughout the
article.  While we report measurements on a range of NUMA and cluster
architectures, the key scalability experiments are conducted on the
Athos cluster with 256 hosts (6144 cores). Even when global recovery
data is not maintained, partitioning the network into s\_groups
reduces network traffic and improves the performance of the Orbit and
ACO benchmarks above 80 hosts.  Crucially we exceed the 60 node limit
for distributed Erlang and do not reach the scalability limits of SD
Erlang with 256 nodes/VMs and 6144 cores. Chaos Monkey experiments show
that two versions of ACO are reliable, and hence that SD Erlang
preserves the Erlang reliability model. However the ACO results show
that maintaining global recovery data, i.e.\ a global process name
space, dramatically limits scalability in distributed
Erlang. Scalability can, however, be recovered by maintaining recovery
data only within appropriately sized s\_groups.  These results are
consistent with experiments with other benchmarks and on other
architectures (Section~\ref{sec:case-studies}).

In future work we plan to incorporate RELEASE technologies, along with
other technologies in a generic framework for building performant
large scale servers. In addition, preliminary investigations suggest
that some SD Erlang ideas could improve the scalability of other
actor languages.\footnote{See Deliverable 6.7
  (\url{http://www.release-project.eu/documents/D6.7.pdf}),
  available online.}
For example the Akka framework for Scala could benefit from semi-explicit
placement, and Cloud Haskell from partitioning the network.


\appendix
\section*{Appendix A: Architecture Specifications}
\label{sec:platforms}

The specifications of the clusters used for measurement are summarised
in Table~\ref{tab:clusters}. We also use the following NUMA machines.
(1) An AMD Bulldozer with 16M L2/16M L3 cache, 128GB RAM, four AMD
Opteron 6276s at 2.3~GHz, 16~``Bulldozer'' cores each, giving a total
of 64 cores.
(2) An Intel NUMA with 128GB RAM, four Intel Xeon E5-4650s at 2.70GHz,
each with eight hyperthreaded cores, giving a total of 64 cores.


\section*{Acknowledgements}
  We would like to thank the entire RELEASE project team for technical
  insights and administrative support. Roberto Aloi and Enrique Fernandez
  Casado contributed to the development and
  measurement of WombatOAM. This work has been supported by the
  European Union grant RII3-CT-2005-026133 ``SCIEnce: Symbolic
  Computing Infrastructure in Europe'', IST-2011-287510 ``RELEASE: A
  High-Level Paradigm for Reliable Large-scale Server Software'', and
  by the UK's Engineering and Physical Sciences Research Council grant
  EP/G055181/1 ``HPC-GAP: High Performance Computational Algebra and
  Discrete Mathematics''.

\bibliographystyle{plain}


\end{document}